\documentclass[11pt]{article}
\usepackage{graphicx}
\usepackage{amssymb}
\usepackage{amsmath}
\usepackage{graphicx}
\usepackage{amstext}
\usepackage{leftidx}
\usepackage{esint}
\usepackage[utf8]{inputenc}
\usepackage{color}
\usepackage{cite}
\usepackage{hyperref}
\usepackage{multido}

\usepackage{multido}

\makeatletter
\ams@newcommand{\vardot}[2]{%
  {\mathop{#2\kern0pt}\limits^{\vbox to-1.4\ex@{\kern-\tw@\ex@
   \hbox{\normalfont\multido{}{#1}{.}}\vss}}}}
\makeatother
\newcommand{\rL}{\rho_\Lambda}

\newcommand{\CC}{\Lambda}

\newcommand{\rv}{\rho_{\rm vac}}
\newcommand{\Pv}{P_{\rm vac}}
\newcommand{\rvo}{\rho^0_{\rm vac}}

\newcommand{\rco}{\rho^0_{c}}

\newcommand{\nueff}{\nu_{\rm eff}}

\newcommand{\tnu}{\tilde{\nu}}

\newcommand{\bk}{{\bf k}}
\newcommand{\mpl}{m_{\rm Pl}}
\newcommand{\MPl}{{\cal M}_{\rm Pl}}

\newcommand{\be}{\begin{equation}}
\newcommand{\ee}{\end{equation}}

\newcommand{\jtext}[1]{{\textcolor{black}{#1}}}

\newcommand{\ha}{\hat{a}}
\newcommand{\astar}{a_{*}}

\newcommand{\wv}{w_{\rm vac}}

\usepackage{graphicx,calc}
\usepackage{authblk}
\usepackage{simpler-wick}
\usepackage{mathtools}
\usepackage{physics}
\usepackage{float}
\usepackage{csvsimple}
\providecommand{\abs}[1]{\lvert#1\rvert}
\usepackage{amsmath}
\usepackage{amsfonts}
\usepackage{comment}
\usepackage{braket}
\usepackage{slashed}
\usepackage{wrapfig}
\usepackage{array}
\usepackage{cancel}
\usepackage{tensor}
\newcolumntype{P}[1]{>{\centering\arraybackslash}p{#1}}
\newcolumntype{M}[1]{>{\centering\arraybackslash}m{#1}}

\def\plusheigh{-\the\dimexpr\fontdimen22\textfont2\relax}

\DeclareMathOperator{\calH}{\mathcal{H}}
\newcommand{\upperRomannumeral}[1]

\begin{document}

\hyphenation{theo-re-ti-cal gra-vi-ta-tio-nal theo-re-ti-cally mo-dels Not-with-standing pro-blem accor-din-gly pa-ra-me-ters cos-mo-lo-gi-cal con-si-de-ra-tions Ho-we-ver va-cu-um re-gu-la-ri-za-tion vio-la-ting re-nor-ma-li-za-tion e-ma-na-ting re-nor-ma-li-za-bi-li-ty ex-pli-cit-ly e-xact ex-pan-ding}

\begin{center}
%{\bf \LARGE Running Vacuum from QFT in curved spacetime}\\
\vskip 2mm
%{\bf \LARGE Unstable de Sitter vacuum and running vacuum: towards a unified quantum field theory of\\ inflation and dark energy}
{\bf \LARGE Towards a unified quantum field theory of\\ dark energy and inflation: unstable de Sitter vacuum and running vacuum}
%towards a consistent approach to $H^4$-inflation}
%\vskip 2mm
%{\bf \Large The dynamics of $\rv(H)$ from the quantized matter fields}

 \vskip 8mm

\textbf{Joan Sol\`a Peracaula\footnote{Corresponding author: sola@fqa.ub.edu},  Alex Gonz\'alez-Fuentes}

Departament de F\'isica Qu\`antica i Astrof\'isica, \\
and   Institute of Cosmos Sciences,\\ Universitat de Barcelona,
Av. Diagonal 647, E-08028 Barcelona, Catalonia, Spain\\
\vspace{0.5cm}
\textbf{Cristian Moreno-Pulido}\\
Departament d'Inform\`atica, Matem\`atica Aplicada i Estad\'\i stica, \\
Universitat de Girona, \\
Campus Montilivi
17003  Girona, Spain

\vskip0.5cm

%E-mails:   sola@fqa.ub.edu

 \vskip2mm

\end{center}
\vskip 15mm

\begin{quotation}
\noindent {\large\it \underline{Abstract}}.
Inflation is a necessary cosmic mechanism to cure basic inconsistencies of the standard model of cosmology. In its absence, we could not understand the observed spatial flatness,  the homogeneity and isotropy of the CMB, and yet the origin of structure formation, nor the large amount of entropy today. These problems are usually `fixed' by postulating the existence of a scalar field (the ``inflaton''). However, other less {\it ad hoc} options are actually possible.
In the running vacuum model (RVM) framework, the vacuum energy density (VED) is a function of the Hubble rate $H$ and its time derivatives: $\rho_{\rm vac}=\rho_{\rm vac}(H, \dot{H},\ddot{H},\dots)$.  In this context, the VED is dynamical (there is no rigid cosmological term $\Lambda$). In the FLRW epoch,  $\rv$ evolves very slowly with expansion, as befits the  observed $\Lambda\simeq$ const.\,behavior.
In contrast, in the very early universe  the vacuum fluctuations of the quantized matter fields induce higher powers $H^N$ capable of unleashing fast inflation in a short period in which $H\simeq$ const.  We call this mechanism `RVM-inflation'. It does not require an inflaton field since inflation is brought about by pure quantum field theory (QFT)  effects on the dynamical background. It is different from Starobinsky's inflation, in which $H$ is never constant. In this work, we study a closely related scenario: the decay of the exact de Sitter vacuum into FLRW spacetime, in its radiation epoch, and its impact on the current universe, and compare it with the RVM. The two QFT calculations are renormalized using an off-shell adiabatic prescription. We find that in both cases inflation is driven by $H^4$ powers accompanied by subleading contributions of order $H^2$ that ease a graceful-exit transition into the radiation-dominated epoch, where the FLRW regime starts and ultimately develops a mildly evolving VED in the late universe: $\delta\rho_{\rm vac}\sim {\cal O}(\mpl ^2 H^2)$.  The proposal presented here aims at a unified QFT approach to inflation and dark energy (conceived as dynamical vacuum energy) with potentially measurable phenomenological consequences in the present universe, and constitutes a first step toward establishing its full theoretical and phenomenological consistency.

\end{quotation}

\newpage

\tableofcontents

%\tableofcontents

\newpage
\section{Introduction}
The cosmological constant (CC) in Einstein's equations\cite{Einstein1917}, $\CC$, constitutes a cornerstone of the standard or concordance model of cosmology, aka $\CC$CDM\cite{Peebles1993,Turner:2022gvw}. Despite the latter being bestowed as the leading paradigm to account for the cosmological evolution of our universe in the last few decades\,\cite{SupernovaSearchTeam:1998fmf,Planck:2018vyg},  it is not less true that it has by now a rather checkered history sprinkled with several deep theoretical problems and practical dysfunctions that have  undermined its theoretical and phenomenological consistency and have become a handicap to its reputation. These include phenomenological snags (or more than that) persisting in modern observations, such as the so-called cosmological tensions between the $\CC$CDM predictions on the local measurements of the current Hubble parameter $H_0$ and on the growth rate of structure formation; see  e.g. \cite{CosmoVerseNetwork:2025alb,Abdalla:2022yfr,Perivolaropoulos:2021jda} and references therein for a comprehensive review of these tensions, and \cite{Vagnozzi:2023nrq} for further  insights and alternative points of view. Recently, a new kind of (acute) tension has appeared in the already quite battered landscape of the $\CC$CDM. It bears relation with the observations of the  James Webb Space Telescope (JWST)\,\cite{Gardner:2006ky,Labbe:2022ahb}; they  have revealed the existence of an unexpectedly large population of supermassive galaxies at large redshifts in the approximate range $z\gtrsim 7-10$, which is completely at odds with the expectations of the $\CC$CDM model. This discovery has spurred theoretical work to revise the current models of galaxy formation, but also our views on early-universe cosmology to address the observed discrepancies.

Behind the phenomenological hitches and glitches that are currently drawing the most attention of cosmologists, the excruciating issue of the cosmological constant problem (CCP)\cite{Weinberg89}-- a long-standing theoretical conundrum -- has never diminished  by an inch its intensity as of the problem being first pinpointed by Zeldovich more than half a century ago\,\cite{Zeldovich1967}.  The severity of the CCP remains immutable and hence is equally harrowing at present, no matter if we prefer to look in another direction. The mystery, in fact, is waiting patiently for a way out that does not appear anywhere\cite{Weinberg89,Sahni:1999gb,Carroll:2000fy,PeeblesRatra2003,Padmanabhan2003,Copeland2006,Aitchison2009,JSPRev2013,JSPRev2015,JSPRev2022}\footnote{See also \cite{JSPCosmoverse} for a more informal, although fairly detailed, account of the CCP.}. To tackle the problem efficiently, we need to collect more information on the nature of the  dark energy (DE).
Interestingly enough, recent measurements by the DESI collaboration come to rescue and suggest that DE could be a dynamical component of the universe\cite{DESI:2024mwx,DESI:2024aqx,DESI:2025zgx,DESI:2025fii}. If so, this could help alleviate some aspects of the CCP, in particular the cosmic coincidence problem\,\cite{PeeblesRatra2003}. But it may also help in more practical issues, such as improving the phenomenological fits to the overall cosmological data as compared to the $\CC$CDM with a rigid CC term $\CC$. In fact, a number of recent analyzes have demonstrated the effectiveness of dynamical DE in improving the description of cosmological observations beyond the standard $\CC$CDM. The idea that DE may be dynamical has a long pedigree. Already about ten years ago, the large positive impact of dynamical DE was highlighted by several devoted studies which considered a large set of cosmological observations of various sorts; see \cite{Sola:2015wwa,Sola:2016jky,Sola:2017znb,DiValentino:2016hlg,Zhao:2017cud,SolaPeracaula:2016qlq,Sola:2016zeg,SolaPeracaula:2017esw}. For older preliminary studies, see  e.g. \cite{Shafieloo:2005nd,Basilakos:2009wi,Grande:2011xf,Sahni:2014ooa,Gomez-Valent:2014rxa,Gomez-Valent:2014fda},  among others.

When assessing the DE models which provided significant evidence of dynamical DE in the past and continue doing so at present, we find the Running Vacuum Model (RVM) framework, based on quantum field theory (QFT) in curved spacetime; see e.g the reviews \cite{JSPRev2022,JSPRev2015,JSPRev2013}. The RVM advocates for the vacuum energy density (VED) of QFT as the ultimate explanation for the DE (and inflation, see below); and this despite the aforementioned CCP, since the RVM  provides a genuinely new framework that could help mitigate the problem, especially in what concerns the preposterous fine-tuning issue involved in it; see \cite{Fossil2008} and recent work \cite{CristianJoan2020,CristianJoan2022a,CristianJoan2022b,CristianJoanSamira2023,SolaPeracaula:2025yco}, and the review \cite{JSPRev2013}. The RVM  predicts  a cosmic evolution of the VED from fundamental principles. Obviously, not every model with evolving DE belongs to the RVM class, not by a long shot. There are myriad models of ``time-evolving $\CC(t)$''\cite{Overduin:1998zv}  and of time-evolving scalar fields (quintessence and the like) that have no relation whatsoever with the RVM\,\cite{PeeblesRatra2003,Padmanabhan2003,Copeland2006}. Most of these models are {\it ad hoc} since one ascribes some arbitrary cosmic time (or scale factor) dependence to $\CC$, or just replaces the latter with some particular scalar field potential having no fundamental motivation. In the RVM context, in contrast, the DE is once more vacuum energy (density), but it proves dynamical owing to the quantum effects modifying the background evolution of the spacetime. Such an evolution impinges on the QFT description of the VED and as a result the latter acquires a dynamical component which in the current universe is evolving quadratically with the Hubble rate: $\delta\rv\sim \nueff\,\mpl^2 H^2$. The notable property is that the coefficient $\nueff$ of the term $H^2$ is computable in QFT since it plays the role of $\beta$-function coefficient of the running vacuum\cite{Fossil2008,
CristianJoan2020,CristianJoan2022a}. In addition, such a dynamical law of the VED in the current universe has been successfully tested against the global cosmological observations in different studies, e.g. in the recent works \cite{SolaPeracaula:2021gxi,SolaPeracaula:2023swx,deCruzPerez:2025dni} and the older ones previously mentioned\cite{Sola:2015wwa,Sola:2016jky,Sola:2017znb,SolaPeracaula:2016qlq,Sola:2016zeg,SolaPeracaula:2017esw}. In all of them, the global fit to the data can be significantly improved as compared to the $\CC$CDM and the current cosmological tensions can be alleviated. Notable are also the RVM-inspired proposals \cite{Gomez-Valent:2024tdb,Gomez-Valent:2024ejh,BDRVM,BDRVMbis,BDRVM2}, which provide possible solutions to these tensions from different perspectives\cite{SolaPeracaula:2024iil}.
 See also \cite{Montani:2024ejp} for related ideas on running parameters, and \cite{Akarsu:2021fol,Akarsu:2023mfb,Gomez-Valent:2023uof,Giare:2025pzu,Soriano:2025gxd,Anchordoqui:2023woo}  for additional studies on the cosmological tensions and the nature of DE. It is worth mentioning as well that the RVM behavior can  be mimicked by other gravitational frameworks, including $f(T)$ theories \cite{Cai:2015emx,Errahmani:2025vde},  entropic-force and holography\,\cite{Basilakos:2012ra,Basilakos:2014tha,Komatsu:2013qia,Rezaei:2022bkb}, low-energy string theory\cite{Mavromatos:2020kzj,PhantomVacuum2021}, finite-temperature renormalization effects\,\cite{Park:2024kfn} and lattice quantum gravity\,\cite{Dai:2024vjc}.  It may also have implications for the time variation of the constants of Nature\,\cite{Fritzsch:2012qc,Sola:2016our,Fritzsch:2016ewd}.

%%%%%%%%%%%%%%%%%%%%%%%%%

In another vein, inflation\cite{Guth:1980zm,Linde:1981mu} is also a fundamental cornerstone of the current cosmological paradigm.  It is a necessary cosmic phenomenon that must have occurred in the primeval epoch of the universe to explain a number of inconsistencies of the $\CC$CDM, since the latter cannot be retrograded to arbitrarily early times. In the absence of inflation, we could not understand the homogeneity and isotropy of the observed CMB, nor the current degree of spatial flatness of the universe at present without fine-tuning. Structure formation emerging from the tiny primeval vacuum fluctuations could not be understood either if the fluctuations were not magnified by inflation. A related difficulty is to account for the large amount of entropy today, $S\sim 10^{88}$ (in natural units), which in the $\CC$CDM is at odds with the necessary causal connection that must govern all the microscopic phenomena. All of these problems are usually `fixed' by postulating the existence of a cosmic scalar field (called the ``inflaton'') which takes care of the necessary arrangements during the very early stages of the expansion history\cite{KolbTurner,LiddleLyth,RubakovGorbunov} --  see also \cite{Kallosh:2025ijd,Martin:2013tda} for updated reviews with an indispensable list of references.

%%%%%%%%%%%%%%%%%%%%%%%%%%%%%%%%%%%%%%%%%%%%

Although the introduction of the inflationary period and a corresponding inflaton field can be a remedy for the aforementioned problems, it may be looked upon as being not very natural, since after all it appears as an {\it ad hoc} patching up of the very early expansion history of the $\CC$CDM.  In contrast, in the RVM context, one can have a unified cosmic history which involves inflation in the early stages in an integrated way into an overarching picture of the cosmological evolution until our days. In fact,  `RVM-inflation'\,\cite{CristianJoan2022a,SolaPeracaula:2025yco} connects smoothly to the standard radiation epoch of the Friedmann-Lemaître-Robertson-Walker (FLRW) regime and in particular realizes ``graceful exit'' automatically. This is possible because in the RVM quantum effects on the effective action yield different (even) powers of the expansion rate $H$. Inflation can then be driven by higher powers $H^N$ ($N>2$) before the low-energy regime is steered by the subleading $\sim H^2$.  This fact was proven in the aforementioned papers, although it was long recognized on mere phenomenological grounds\,\cite{BLS2013,Perico:2013mna}.

Prompted by the phenomenological success of the RVM in the description of the early and the late universe, in  the present study we compare the expansion history of the RVM  with that of the unstable de Sitter vacuum decaying into radiation. The advantage of de Sitter spacetime is that it is an exactly solvable QFT and hence no adiabatic expansion is needed. However, the exact solution requires a rather demanding QFT calculation, as we shall see. Now, because exact de Sitter can only be admitted as an initial spacetime in the very early universe, we must let it decay. In this work we study its decay into  radiation, as we previously did for the RVM\,\cite{SolaPeracaula:2025yco}. At this point,  semianalytical and numerical methods are required to describe the solution. One finds that in both the RVM and unstable de Sitter vacuum, inflation can be produced by the leading power $H^4$ of the Hubble rate in a short period where $H\simeq$ const. \cite{SolaPeracaula:2025yco}. In the RVM case, a variety of higher order  powers of $H$ could actually be involved,  $H^4$ being however the simplest\,\cite{CristianJoan2022a,CristianJoan2022b,CristianJoanSamira2023,SolaPeracaula:2025yco}.

Needless to say, de Sitter spacetime has been extensively studied in the literature. A very short list of previous investigations can be found, for example, in \cite{Chernikov:1968zm,Candelas:1975du,Dowker:1975tf,Brown:1976wc,Bunch:1978yq,Ford:1984hs,Mottola:1984ar,Dolgov:2005se,Kamenshchik:2021tjh,Landete:2013axa,Firouzjahi:2022xxb,Firouzjahi:2023wbe};  see also the textbooks \cite{BirrellDavies82,ParkerToms09,Fulling89,Mukhanov:2007zz} and references therein.  In some of these papers, it was considered  the possibility that de Sitter spacetime decays e.g. into Minkowski space, and in fact this has been argued as a possible solution to the CCP. In this work, we will also address the decay of de Sitter spacetime, but as previously noted, we consider its decay into the standard radiation epoch of FLRW cosmology. This also leads to a graceful exit of the inflationary period. The two $H^4$ inflationary mechanisms being considered in our study, based respectively on RVM and unstable de Sitter vacuum, have interesting similarities, but also remarkable differences that will be unraveled here. In both cases, we renormalize the energy-momentum tensor (EMT) using the off-shell adiabatic procedure proposed in \cite{CristianJoan2020,CristianJoan2022a}, which allows us to explore the impact of the quantum effects at the different epochs of the cosmological evolution. The adiabatic approach is very convenient in typical cosmological spacetimes for which the comoving angular frequency $\Omega_k$ is slowly varying at early and late times, so that an expansion of the solution in higher order adiabatic corrections is asymptotically convergent (meaning, as usual, only up to a finite order) in the manner of a WKB series. In the absence of this property, the particle physics interpretation becomes too hard, as is well-known in the general context of QFT in curved spacetime\cite{BirrellDavies82,ParkerToms09,Fulling89,Mukhanov:2007zz}. But in its presence the oscillatory Fourier modes of the field  inside the horizon admit a particle description in terms of what is called the adiabatic vacuum.  The adiabatic behavior is very useful, e.g. to estimate particle production in a time-dependent background and, in particular, for the study of gravitational generation of cosmological relics in the very early universe, see e.g.\cite{Ford:2021syk,Kolb:2023ydq}.

We should also stress that the mechanism of $H^4$-inflation involved in RVM and de Sitter inflation follows a rather different pattern as compared to Starobinsky's $R^2$-inflation\,\cite{Starobinsky:1980te}. For a detailed comparison of the two inflationary mechanisms, see\cite{JSPRev2015,Basilakos:2015yoa,Mavromatos:2020kzj}.  Despite having $R^2\sim H^4$ in the action, it turns out that, at the level of the field equations, all of the higher order terms  depend on time derivatives of the Hubble rate $H$, and hence they all vanish for $H=$ const. in the Starobinsky case.
Thus, for large values of $H$, the  $H^4$-power is dominant in the inflationary epoch, in contrast to the Starobinsky inflation, which is characterized by a period where $\dot{H}$ (rather than $H$) remains constant\cite{JSPRev2015}.

The presence of higher powers of the Hubble rate in the early universe can be useful not only to trigger inflation. For instance, as noted in \cite{Mavromatos:2020crd}, they could help eschewing the trouble of string theories with the  `swampland' criteria on the impossibility to construct metastable de Sitter vacua, which if so would hinder or even forbid the existence of de Sitter solutions in the context of a low-energy effective theory of quantum gravity. Related studies can be found in
\cite{Mavromatos:2020kzj,Mavromatos:2020crd,PhantomVacuum2021,BasMavSol,basilakos2,basilakos3,NickPhiloTrans,Gomez-Valent:2023hov,Dorlis2024,Dorlis2024b,Dorlis:2025zzz,Dorlis:2025amf}. For a comprehensive and in depth exposition of the RVM both in QFT and the corresponding stringy formulation, see \cite{NickJoan_PR}.

In summary, the main focus of this paper is the detailed study of the de Sitter spacetime decaying into FLRW spacetime with an eye to the resemblances with its brother scenario, the running vacuum model.   Clearly, the former can only be an ephemeral initial state in cosmic history since $H$ cannot remain constant in realistic cosmology. However, the maximal initial symmetry of de Sitter space allows one to compute the Fourier field modes exactly before it decays into FLRW.  We study in depth this transition  by properly renormalizing the EMT with the help of the aforementioned off-shell adiabatic procedure. To our knowledge, it is the first time that such a renormalization approach has been applied to de Sitter space. We have checked that the classic on-shell result by Bunch \& Davies\,\cite{Bunch:1978yq}, and Dowker \& Critchley \cite{Dowker:1975tf},  is recovered as a particular case of our calculation. As noted previously, our off-shell result for de Sitter allows a direct comparison with existing RVM calculations for FLRW spacetime\cite{SolaPeracaula:2025yco}  for which only an approximate WKB-type solution is possible and which have been addressed with the same renormalization method \cite{CristianJoan2020,CristianJoan2022a,CristianJoan2022b,CristianJoanSamira2023}. We find remarkable similarities hinting towards a unified QFT approach to the cosmological history since in both cases, RVM and de Sitter,  the off-shell feature allows to explore the cosmic evolution of the vacuum energy and pressure.

The structure of the paper can be summarized as follows.  In~Sec.\,\ref{ZPEScalar}, we consider the classical action and EMT of a scalar field non-minimally coupled to gravity in FLRW spacetime. Next, we address the quantum version of this theory in the RVM context. Upon renormalizing the EMT using an off-shell adiabatic prescription, we derive the finite zero-point energy (ZPE) associated with these fluctuations and show that the corresponding VED evolves with the cosmological expansion, i.e. we find $\rv=\rv(H)$, a prominent feature of our QFT approach which, as previously indicated,  enables us to explore the impact of vacuum dynamics on different epochs of the expansion history.  In Sec.\,\ref{sec:deSitter} we deal with de Sitter spacetime and start computing the exact field modes and the unrenormalized ZPE.  In Sec.\,\ref{sec:RendeSitter} we face the renormalization of the exact de Sitter theory and determine the  corresponding VED and vacuum pressure. We let the de Sitter vacuum decay into radiation, and in Sec.\,\ref{sec:Inflation} we show that the running VED in the early de Sitter universe is responsible for inflation and leads to graceful exit without requiring {\it ad hoc} inflaton fields. We also compare unstable de Sitter inflation with  `RVM inflation', both triggered by $\sim$$H^4$ powers emanating from quantum effects.  In Sec.\,\ref{sec:pheno} we explore a variety of phenomenological consequences in the late universe. For instance, for both scenarios we predict a mildly evolving $\rv(H)$ that deviates from the rigid $\rv=\CC_{\rm obs}/(8\pi G_N)$ value of the concordance model by a small term proportional to  $H^2$. This is remarkable since this fact alone could explain the dynamical DE observed at present from pure vacuum QFT effects. Additionally, we show that these quantum corrections lead to an `effective' form of quintessence or phantom DE behavior.
\jtext{Whether quintessence and phantom fields for the current universe can be rendered expendable - or if the role played by the inflaton can be fully replaced in our framework – cannot be asserted at this point.  However our unified framework constitutes at least a serious  attempt to try to  explain inflation and dynamical DE in terms of a fundamental approach based on running vacuum energy in QFT in curved spacetime.}

Finally, we propose a possible solution to the entropy problem based on the existing QFT  link between the early and late universe in this unified framework, which provides a strong liaison between these two maximally distant cosmic epochs. We believe it speaks off positively of the overall consistency of our approach. The conclusions are delivered in Sec.\,\ref{sec:conclusions}. {We relegate to four appendices A,B,C and D at the end for useful formulas and some bulky computational details so as not to interrupt the discussion in the main text. In particular, in Appendix D we expand on the consistency of the off-shell adiabatic renormalization method.} In our presentation, natural units ($\hbar=c=1$) will be used throughout, except when special features should be highlighted.

\section{Energy-momentum tensor for a non-minimally coupled scalar field in cosmological spacetime}\label{ZPEScalar}

A calculation of the vacuum energy density (VED) in flat (Minkowski) spacetime, even after renormalization,  has no impact whatsoever on the physics of the cosmological constant (CC) as the latter cannot be defined in Minkowskian spacetime, since $\CC\neq 0$ is inconsistent with a flat solution to Einstein's equations. Hence, the cosmological constant problem cannot be addressed in flat spacetime; rather, it should be framed in relation to the dynamics of the expanding universe.  It means that to make contact with the physical $\CC$ that has been measured, we need to move to a curved background and compute the VED induced by the quantized matter fields in an appropriate renormalization framework. In a full quantum  gravity (QG) theory, where the gravitational field should also be quantized, all of the contributions from gravitons and quantum matter fields must be taken into account. Nevertheless, a fully consistent QG is still under construction. We know since long ago that GR is non-renormalizable when quantized in the framework of perturbative QFT\,\cite{tHooft:1974toh,Goroff:1985sz}. So one has to move to an effective field theory treatment\cite{Donoghue:1994dn}.  Not even string theory provides a final description of QG that we can use right now to solve the main problems associated with the VED in cosmology. In the meantime, a formulation of gravity quantization in the framework of perturbative QFT has been claimed to be consistent with strict renormalizability since long ago\,\cite{Stelle:1976gc},  despite various criticism\cite{Woodard:2015zca}. Many problems are still open, but a renormalizable theory is vindicated provided that higher derivative (HD) terms
up to quadratic order are included, see e.g. \cite{Buoninfante:2025dgy} and references therein.

All that said, in this work we will still remain in the semiclassical framework of QFT in curved spacetime as our main guide for the study of the VED in cosmology. After all, it is not clear whether QG is necessary since gravity could be an emerging phenomenon\,\cite{Jacobson:1995ab,Padmanabhan:2003gd,Padmanabhan:2007tm} and in that case only the matter fields need to be quantized. This is the minimum scenario that we can secure, and hence the part that in any case must be taken into account.  Therefore, we will explore the implications of the semiclassical QFT approach to construct a unified theory of inflation and DE. We want to show that it is feasible and that inflation and dark energy can be accounted for without invoking {\it ad hoc} fields such as inflatons, quintessence and the like.   The main assumption, which indeed serves as a golden rule of the entire approach based on QFT in curved spacetime is that whatever the ultimate formulation of the quantum theory of gravitation might be, it should still be reasonable to suppose that there is a semiclassical limit to that theory so that the spacetime curvature can be treated classically even in the presence of quantum matter. Such an approximation ought to be satisfactory for systems with small fluctuations in $T_{\mu\nu}$ (the EMT)\,\cite{Kuo:1993if}. This assumption can fail far away from the Planck scale, but it should remain reasonable otherwise and could even be the only possible approach in the absence of QG. It is our contention that it should be possible to alleviate the  CCP  within this semiclassical formulation, see e.g. \cite{JSPRev2022}. For different studies bearing relation with the VED in the semiclassical approach, see e.g.\cite{Antoniadis:2006wq,Maggiore2011,Bilic2011,Capozziello2011,BennieW2013,KohriMatsui2017,Sanchez:2020rqj,Mottola2022,Bass:2023ece} and references therein.

In order to address the renormalization of the VED in QFT in curved spacetime, we start from the EMT of the classical field theory.
For the sake of simplicity in our presentation, we assume that there is only one quantum matter component in the form of a real scalar field $\phi$ which enters quadratically in  the matter Lagrangian.  For a generalization involving an arbitrary number of quantized scalar fields and even an arbitrary number of quantized fermionic fields, see\,\cite{CristianJoanSamira2023}. In practical scenarios, however,  we will sum over  field multiplicities and adapt the formulas conveniently.
Furthermore, we will focus on the computation of the zero-point energy (ZPE) of that field in FLRW spacetime since the ZPE is a pure quantum effect that is present for all fields irrespective of their spins, being a generic component of the VED. For this reason we do not include any classical potential for $\phi$ at this stage, just the mass term of that field. Even that bit entails a significantly complicated calculation in QFT in curved spacetime. Finally, a  non-minimal coupling $\xi$ between $\phi$ and curvature will also be assumed. This coupling is not strictly necessary for renormalization purposes in our calculation, since no self-interactions of the field will be considered, but it will play an important role for making possible a new mechanism of inflation and also for the  phenomenological implications at low energies (i.e. for the current universe).

\subsection{Classical field theory}\label{sec:ClassicalEMT}

The classical field action associated to a scalar field $\phi$ of mass $m$ non-minimally coupled to curvature reads\footnote{Metric and curvature conventions are as in \cite{CristianJoan2022a}, see e.g. Appendix A of that reference. They correspond to $(+,+,+)$ in the standard classification of \cite{MTW}.}
\begin{equation}\label{eq:Sphi}
  S[\phi]=-\int d^4x \sqrt{-g}\left(\frac{1}{2}g^{\mu \nu}\partial_{\nu} \phi \partial_{\mu} \phi+\frac{1}{2}(m^2+\xi R)\phi^2 \right)\,,
\end{equation}
where $\xi$ is the non-minimal curvature coupling.  For the particular value $\xi=1/6$, the massless ($m=0$)  action has conformal symmetry, i.e. is symmetric under simultaneous rescalings of the $g_{\mu\nu}$ and $\phi$ with a local function $\alpha(x)$: $g_{\mu\nu}\to e^{2\alpha(x)}g_{\mu\nu}$ and  $\phi\to e^{-\alpha(x)}\phi$. However, as already noted, we shall keep $\xi$ general, since this parameter will play an important role in our considerations, as we shall see. Actually, minimal coupling ($\xi=0$) will prove an unsuitable option.

The corresponding Klein-Gordon (KG) equation for $\phi$ can be  derived from the above action upon variation with respect to the scalar field:
\begin{equation}\label{eq:KG}
(\Box-m^2-\xi R)\phi=0\,,
\end{equation}
where $\Box\phi=g^{\mu\nu}\nabla_\mu\nabla_\nu\phi=(-g)^{-1/2}\partial_\mu\left(\sqrt{-g}\, g^{\mu\nu}\partial_\nu\phi\right)$. From the standard definition of the EMT through the metric functional variation of the matter action, we find
\begin{equation}
\begin{split}
T_{\mu \nu}^{\phi}=&-\frac{2}{\sqrt{-g}}\frac{\delta S_\phi}{\delta g^{\mu\nu}}= (1-2\xi) \partial_\mu \phi \partial_\nu\phi+\left(2\xi-\frac{1}{2} \right)g_{\mu \nu}\partial^\sigma \phi \partial_\sigma\phi\\
& -2\xi \phi \nabla_\mu \nabla_\nu \phi+2\xi g_{\mu \nu }\phi \Box \phi +\xi G_{\mu \nu}\phi^2-\frac{1}{2}m^2 g_{\mu \nu} \phi^2.
\end{split} \label{EMTScalarField}
\end{equation}
Since our scalar matter  field sits in  FLRW background, we must consider it together with  the corresponding Einstein-Hilbert (EH) action. Therefore, the total action is
\begin{equation}\label{eq:EH}
S_{\rm total}=  S_{\rm EH}+ S[\phi]=\int d^4 x \sqrt{-g}\,\left(\frac{1}{16\pi G}\, R  - \rL\right) + S[\phi]\,.
\end{equation}
We pointed out in the previous section that the cosmological term is  physically meaningful only in curved spacetime.  Its observed value $\CC_{\rm obs}$ becomes naturally linked to the VED, $\rv$, through Einstein's equations: $\rv=\Lambda_{\rm obs}/(8\pi G)$.
However, even in this gravitational context we should not confuse  the bare parameter  $\rL$ in the EH action with the physical $\rv$, where the former is related in a similar way with the bare values of $\CC$  and $G$. The corresponding connection with physical quantities is not immediate at this point, and it will only emerge upon properly renormalizing the theory.   Einstein's equations follow from the standard variation of the total action $S_{\rm total}$  with respect to the metric:
\begin{equation} \label{EinsteinEqs}
\frac{1}{8\pi G}\,G_{\mu \nu}=-\rho_\Lambda g_{\mu \nu}+T_{\mu \nu}^{\phi}\,,
\end{equation}
where  $G_{\mu\nu}=R_{\mu\nu}-(1/2) g_{\mu\nu} R$  is the usual Einstein tensor.  The above are classical field equations that do not incorporate quantum effects yet. This will be done upon quantization.

In our calculation in cosmological spacetime we shall assume a flat three-dimensional metric and for convenience we will use the conformal FLRW line element  $ds^2=a^2(\tau)\eta_{\mu\nu}dx^\mu dx^\nu$, which is conformally related to Minkowski metric  $\eta_{\mu\nu}={\rm diag} (-1, +1, +1, +1)$ through the scale factor $a(\tau)$, given as a function of conformal time $\tau$. Derivatives with respect to $\tau$ will be denoted by primes, so, for instance, $\mathcal{H}\equiv a^\prime/a$ is the corresponding Hubble function in conformal time. In order to render the final results in terms of the usual Hubble function $H(t)=\dot{a}/a$ in cosmic time $t$ (where a dot indicates differentiation with respect to $t$), we recall that  $d\tau=dt/a$. Therefore, we have, for example, $\mathcal{H}=a H$, $\mathcal{H}^\prime=a^2(H^2+\dot{H})$,
$\mathcal{H}^{\prime\prime}=a^3\left(2H^3+4 H\dot{H}+\ddot{H}\right), \dots$ which are useful relations for the mentioned conversion.
\subsection{Quantization and adiabatic expansion}\label{sec:Quantization}

The above equations hold good in classical field theory. However, in QFT, $\phi$ is a quantized matter field contributing also with quantum fluctuations $\delta\phi$.  In this case,  it is convenient to separate the fluctuations from the background part:
 \begin{equation}
 \phi\left({\bf x},\tau \right) = \phi_b (\tau)+\delta\phi \left({\bf x},\tau \right)\,.
 \end{equation}
The Klein-Gordon equation \eqref{eq:KG} can be written more explicitly in the conformal metric as follows:
\begin{equation}\label{eq:phiexpansion}
    \begin{split}
        \phi''+2\calH\phi'-\nabla^2 \phi + a^2(m^2+\xi R)\phi=0\,,
    \end{split}
\end{equation}
where $R=6 a''/a^3$ for spatially flat FLRW
metric. Notice that the background field $\phi_b (\tau)$ is assumed to be spatially homogeneous in cosmology, but not so the fluctuating part, which is why the field can be decomposed in Fourier modes:
\begin{equation}
    \phi(\tau, \mathbf{x})=\int \frac{d^3k}{(2\pi)^{3/2}} \left[ A_\mathbf{k} e^{i\mathbf{k}\mathbf{x}} \phi_k+A^*_\mathbf{k} e^{-i\mathbf{k} \mathbf{x}} \phi^*_k \right] \; .
\end{equation}
Upon quantization, the Fourier expansions are promoted
to operators in the Heisenberg representation, on which we impose the canonical commutation
relations. These are encoded in those satisfied by the creation and annihilation operators, $A_k^\dagger$ and $A_k$:
\begin{equation}
[A_\bk, A_\bk'^\dagger]=\delta({\bf k}-{\bf k'}), \qquad [A_\bk,A_ \bk']=0\,. \label{CommutationRelation}
\end{equation}
In terms of the Fourier field modes, Eq.\,\eqref{eq:phiexpansion} takes on the form
\begin{equation}
    \phi_k''+2\calH\phi_k'+(\omega_k^2(m)+a^2\xi R)\phi_k=0 \; ,
\end{equation}
where we have defined $\omega_k^2(m)=k^2+a^2m^2$, with $k\equiv|{\bf k}|$ the modulus of the
comoving momentum (the physical momentum being $k/a$). With a rescaling of the field $\phi=\varphi/a$ and of the field modes, $\phi_k=\varphi_k/a$, we can get rid of the friction term proportional to $\calH$ in the previous differential equation, and we obtain a simpler one:
\begin{equation}\label{eq:varphimodes}
    \begin{split}
        \varphi_k''+\left(\omega_k^2(m)+a^2\left(\xi-\frac{1}{6}\right)R\right)\varphi_k=0\,.
    \end{split}
\end{equation}
We can also separate the background and the fluctuating part of the rescaled field, $\varphi\left({\bf x},\tau \right) = \varphi_b (\tau)+\delta\varphi\left({\bf x},\tau \right)$, and
the Fourier expansion of the fluctuation can be written
\begin{equation}\label{eq:diffvarphi}
    \delta \varphi=\int \frac{d^3k}{(2\pi)^{3/2}} \left[ A_\mathbf{k}e^{i\mathbf{k}\mathbf{x}} h_k(\tau)+A_\mathbf{k}^\dagger e^{-i\mathbf{k}\mathbf{x}} h_k^*(\tau) \right] \; ,
\end{equation}
where $h_k(\tau)$ are the mode functions of the vacuum fluctuations, which we are looking for. They are no longer $h_k(\tau)\sim e^{\pm ik\tau}$ as in Minkowski spacetime, this being the reason why in a curved background we cannot have a straightforward interpretation in terms of particles with definite frequencies.
The nontrivial mode functions $h_k(\tau)$ satisfy the same differential equation \eqref{eq:varphimodes}:
\begin{equation} \label{KGmodes}
h^{\prime\prime}_{k}+\Omega_k^2(\tau)h_k=0\,,
\end{equation}
where
\begin{equation} \label{eq:Omegak}
\Omega_k^2(\tau)\equiv \omega_k^2(m)+a^2(\tau)\left(\xi-1/6\right)R=k^2+a^2(\tau)m^2+a^2(\tau)\left(\xi-1/6\right)R\,,
\end{equation}
is the comoving squared angular frequency.
Despite it being a linear differential equation, the effective frequency $\Omega_k(\tau)$ is not constant and hence it corresponds to an anharmonic oscillator which does not possess a closed analytic solution, except for very particular cases, such as e.g. the massless limit with minimal coupling ($\xi=0$) -- a situation which is far from our main interests. However,   it has an exact solution  for arbitrary $m$ and $\xi$ for $H=$ const. (de Sitter spacetime), and this one is of great interest in our case. It will be treated in detail in Sec.\ref{sec:deSitter} and following.

In general, one must construct an approximate solution through an adiabatic series expansion, which is essentially a WKB-type solution\cite{BirrellDavies82,ParkerToms09,Fulling89,Mukhanov:2007zz}. To this aim, one introduces a phase-integral ansatz for the mode function:
\begin{equation}\label{eq:phaseintegral}
h_k (\tau)=\frac{1}{\sqrt{2W_k (\tau)}}\exp \left(-i\int^\tau W_k (\tilde{\tau})d\tilde{\tau}\right)\,,
\end{equation}
which is normalized through the Wronskian condition
 $h_k^{} h_k^{*\prime}-h_k^\prime h_k^* = i$. This normalization is necessary for the quantum field to satisfy the canonical commutation relations, given the corresponding relations \eqref{CommutationRelation} for the creation and annihilation operators.
Proceeding in this way, we can trade the original mode functions $h_k$ for the new functions $W_k$. It is easy to check that the $W_k$ modes satisfy a nonlinear (WKB-type) differential equation:
\begin{equation}\label{Non-LinDiffEq}
W_k^2=\Omega_k^2 -\frac{1}{2}\frac{W_k^{\prime \prime}}{W_k}+\frac{3}{4}\left( \frac{W_k^\prime}{W_k}\right)^2\,.
\end{equation}
The template \eqref{eq:phaseintegral} is motivated by the fact that for constant $\Omega_k$ (i.e. independent of time), it provides the exact solution for positive frequency modes by taking $W_k=\Omega_k$. If, however,  $\Omega_k(\tau)$ is not constant but is  a slowly varying function, one can still solve the above equation perturbatively with the help of an asymptotic series which can be organized through adiabatic orders. When the expansion rate of the universe is sufficiently slow so that the derivative terms in Eq.\,\eqref{Non-LinDiffEq} may be ignored, we recover the standard WKB approximation
\begin{equation}\label{eq:phaseintegralWKB}
h_k (\tau)=\frac{1}{\sqrt{2\Omega_k (\tau)}}\exp \left(-i\int^\tau \Omega_k (\tilde{\tau})d\tilde{\tau}\right)\,.
\end{equation}
This leading-order WKB approximation is valid when $\left(a'(\tau)\right)^2$ and $a''(\tau)$ are both small compared to $\Omega^2(\tau)$. In the light of \eqref{eq:Omegak}, this solution can be viewed
as a slow expansion or, equivalently,  a high frequency approximation. It is therefore applicable for large $k$, hence short wave lengths (as e.g. in geometrical Optics), and weak gravitational fields. Furthermore, in view of the fact that the mode \eqref{eq:phaseintegralWKB} is a positive frequency solution, with a slowly varying frequency,  it may be taken as an approximate definition of the vacuum state, and hence no particle creation occurs and particle number remains constant. This fact provides a clue to solving  Eq.\eqref{Non-LinDiffEq} beyond the crude approximation \eqref{eq:phaseintegralWKB} through an adiabatic expansion to higher orders, and this leads to the definition of the adiabatic vacuum\cite{BirrellDavies82,ParkerToms09,Fulling89,Mukhanov:2007zz}, as we discuss briefly below.

In practice, the suitability of the adiabatic expansion is related to the smallness of the adiabaticity ratio $|\Omega'_k (\tau)/\Omega^2_k (\tau)|\ll1$. This property does not hold for general spacetimes, but for typical cosmological spacetimes it does, i.e. for them the comoving angular frequency $\Omega_k(\tau)$ slowly varies at both early and late times. This becomes manifest through the adiabatic expansion itself. The adiabaticity counting rule is that each time derivative increases by one unit the adiabatic order and the solution for $W_k$ can be written as
\begin{equation}\label{eq:WKBseries}
W_k=\omega_k^{(0)}+\omega_k^{(2)}+\omega_k^{(4)}+\omega_k^{(6)}+\cdots,
\end{equation}
in which the superscript indicates the adiabatic order. If the expansion is carried out up to the $N$th adiabatic order, the vacuum state annihilated by all ladder operators $A_\bk$ satisfying \eqref{CommutationRelation} is known as the $N$th-order adiabatic vacuum.  The vacuum expectation values (VEV's) computed in the adiabatic approach always refer to that vacuum state at the given order. It is, of course, an approximate vacuum state. This approach constitutes the basis for the so-called adiabatic regularization\cite{BirrellDavies82,ParkerToms09,Fulling89,Mukhanov:2007zz}, which is essentially an expansion in the number of time derivatives. On physical grounds, this is tantamount to saying that we place ourselves in a slowly varying spacetime, for which the particle number is an adiabatic invariant that remains essentially constant.

One can see immediately that the adiabatic expansion in the cosmological context ends up as an expansion in powers of $H$ and its time derivatives\cite{CristianJoan2020,CristianJoan2022a,CristianJoan2022b,CristianJoanSamira2023}. The adiabaticity condition is satisfied both at early and late times, since the time derivatives become increasingly suppressed. In particular, for the early universe the mechanism of RVM-inflation also preserves this condition since it is based on a period for which $H\simeq$ const.  Only even adiabatic orders ($N=0,2,4, \dots$) are permitted by the requirement of general covariance of the theory.  From explicit calculation, one verifies that all odd adiabatic orders are indeed absent (see the aforementioned papers). Finally, it is important to remember that the WKB expansion is an asymptotic series, and therefore it should be truncated to relatively low adiabatic orders, beyond which its convergence gets degraded. The explicit expansion \eqref{eq:WKBseries} has been computed in detail up to $6$th adiabatic order in \cite{CristianJoan2022a}, we refer the reader to this reference for details. However, in practice, we will need only the results up to $4$th adiabatic order, which were presented previously in\cite{CristianJoan2020}.

As noted before, QFT in curved spacetime is not a theory of particles with definite frequencies but rather a quantum theory of fields.  Since physics resides in the fields, the main actor carrying physical information of the theory is the EMT. We restrict our considerations to the matter fields represented by the scalar field action \eqref{eq:Sphi}. Furthermore, since we are specifically addressing the effect of the quantum fluctuations of the quantized matter fields, we need to focus on computing just the VEV of the EMT, which we may call the `vacuum EMT' for short. This quantity depends on bilinears of the fluctuation and its time derivatives since $\braket{\delta\phi}=0$ and hence the linear terms do not contribute. For example, let us focus on the 00th component of the vacuum EMT (a quantity which defines the unrenormalized ZPE). We expand it in terms of the field modes $h_k$ defined previously. The result follows from substituting the above Fourier expansion of the field into the classical EMT \eqref{EMTScalarField} and applying the commutation relations \eqref{CommutationRelation} between the ladder operators, with the following result\cite{CristianJoan2020,CristianJoan2022a}\footnote{ We shall define the quantity $\Delta^2$ in the next section. For $\Delta^2=0$, the above result reduces to that of \cite{Anderson:1987yt}, as it should. However,  the presence of $\Delta^2$ is crucial in our framework in order to generate an off-shell adiabatic expansion  of the modes and ultimately of the EMT.}:
\begin{equation*}
    \begin{split}
        \braket{T_{00}^{\delta\phi}}=&\frac{1}{4\pi^2 a^2} \int dk k^2 \left[ \abs{h_k'}^2
        +(\omega_k^2+a^2\Delta^2)\abs{h_k}^2\right.
        \end{split}
        \end{equation*}
        \begin{equation}\label{eq:vacuumEMT}
        \begin{split}
       & \left.+\left( \xi-\frac{1}{6} \right)\left( -6\calH^2\abs{h_k}^2+6\calH(h_k'h_k^*+{h_k^*}'h_k)\right) \right]\,.
    \end{split}
\end{equation}
Since this expression comes from a composite operator (the EMT) made out of quadratic products of field operators, they do not define a smooth distribution, so we expect the VEV to be divergent in QFT.  In fact, in momentum space the vacuum EMT involves integration over all modes,  $\int\frac{d^3k}{(2\pi)^3}( \dots) $, which prove to be UV-divergent integrals up to fourth adiabatic order (in $n=4$ spacetime dimensions).  Hence, to extract physical results on the vacuum effects, renormalization is mandatory. Adiabatic orders higher than $4$ are finite in $4$-dimensional spacetime\cite{BirrellDavies82}.

\subsection{Off-shell adiabatic renormalization}\label{sec:Renormalization}

The vacuum EMT being UV-divergent must be subject  to renormalization. In the present context, this is done by appropriately subtracting the first four (UV-divergent) adiabatic orders.  In the works \cite{CristianJoan2020,CristianJoan2022a} the following ``off-shell subtraction prescription'' was proposed to renormalize the vacuum EMT:
\begin{equation}\label{RenormalizedEMTScalar}
\left\langle T_{\mu\nu}^{\delta\phi} \right\rangle_{\rm ren} (M)=\left\langle T_{\mu\nu}^{\delta\phi} \right\rangle (m)-\left\langle T_{\mu\nu}^{\delta\phi}\right\rangle^{(0-4)}(M).
\end{equation}
This specific prescription will be referred to as `off-shell ARP' (adiabatic regularization prescription)\footnote{Despite that the subtraction scheme is not manifestly covariant, adiabatic regularization is known to be equivalent to covariant point splitting, in which the spacetime points are split only in the spacelike hypersurface of constant conformal time $\tau$. This was proven for zero spatial curvature in\,\cite{Birrell78} and generalized for any curvature in \cite{Anderson:1987yt}. As a result,  the value of the adiabatically renormalized EMT is the same as that from such a covariant method.}. %The off-shell feature introduced in our case does not alter this fact.}.
Notice that we understand that in the off-shell piece of \eqref{RenormalizedEMTScalar} (viz. the one evaluated at the arbitrary scale $M$) we are to replace $\omega_k^2(m)=k^2+a^2m^2$ with $\omega_k^2(M)=k^2+a^2M^2$. The recipe above also plays the role of a renormalization prescription, since it renders the superficially UV-divergent integrals globally finite when the integrands are worked out. This method constitutes an extension of the original (on-shell) adiabatic regularization procedure\cite{BirrellDavies82,ParkerToms09,Fulling89}. See also \cite{{FerreiroNavarroSalas2019}} for an application to coupling renormalization.
The superscript $(0-4)$  in the second term on the \textit{rhs} of \eqref{RenormalizedEMTScalar} denotes the (UV-divergent) orders being subtracted, whereas the first term is the on-shell value. The on-shell value can be computed, in principle, to any desired adiabatic order. However, keeping the subtraction scale $M$ generic, it plays the role of renormalization scale. In practice this means that we can test the evolution of the VED with $M$, and this proves useful in cosmology, as we shall see.  In fact, having a floating scale $M$ in QFT is a characteristic of the renormalization group (RG)  analysis, as is well known.

Ideally, to implement the off-shell ARP recipe \eqref{RenormalizedEMTScalar}, we  must first calculate the on-shell term $\langle T_{\mu\nu}^{\delta\phi} \rangle (m)$ through an exact result and then subtract the first four adiabatic orders involved in the off-shell term $\langle T_{\mu\nu}^{\delta\phi} \rangle (M)$.  When we look, for example, at Eq.\,\eqref{eq:vacuumEMT}, this means that we need to know the exact modes $h_k$ for the computation of $\langle T_{\mu\nu}^{\delta\phi} \rangle (m)$ and the approximate modes $h_k$ up to fourth adiabatic order for the computation of $\langle T_{\mu\nu}^{\delta\phi} \rangle (M)$. Unfortunately, an exact solution for $\langle T_{\mu\nu}^{\delta\phi} \rangle (m)$ is not always available.  For example, in \cite{CristianJoan2020,CristianJoan2022a} the computation of the VED for the non-minimally coupled scalar action \eqref{eq:Sphi} was performed within the (FLRW) background, but since an exact result for $\langle T_{\mu\nu}^{\delta\phi} \rangle (m)$ cannot be found, it was computed up to $4$th and $6$th adiabatic orders. Higher orders are extremely cumbersome and moreover the adiabatic expansion is an asymptotic expansion, so its truncation is mandatory after a few orders. Fortunately, an exact solution is available in de Sitter spacetime. This will be addressed in Sec.\,\ref{sec:deSitter}.

The off-shell contribution $\langle T_{\mu\nu}^{\delta\phi}\rangle^{(0-4)}(M)$ in \eqref{RenormalizedEMTScalar} can be computed
from the ordinary adiabatic expansion up to 4th order, with the mentioned proviso that we have to replace  $\omega_k(m)$ with $\omega_k(M)$.  In doing this, we must  take into account the contribution of the term $\Delta^2\equiv m^2-M^2$  in Eq.\,\eqref{eq:vacuumEMT}, which is to be treated as being of adiabatic order 2, see \cite{CristianJoan2020,CristianJoan2022a}. The corresponding adiabatic expansion of the field modes $h_k$ is obtained from  Eq.\eqref{eq:phaseintegral}  by  using the WKB expansion \eqref{eq:WKBseries}. The explicit calculation is a bit of a laborious task, but it is straightforward. The final result before integrating over momentum reads as follows,\cite{CristianJoan2020,CristianJoan2022a}:
\begin{equation*}
    \begin{split}
        \braket{T_{00}^{\delta\phi}}^{(0-4)}(M)&=\frac{1}{8\pi^2a^2}\int dk k^2 \left[2\omega_k+\frac{a^4M^4\calH^2}{4\omega_k^5}-\frac{a^4M^4}{16\omega_k^7}(2\calH''\calH-{\calH'}^2+8\calH'\calH^2+4\calH^4) \right. \\
        &+\frac{7a^6M^6}{8\omega_k^9}(\calH'\calH^2+2\calH^4)-\frac{105a^8M^8\calH^4}{64\omega_k^{11}} \\
        &+\left(\xi-\frac{1}{6}\right)\left( -\frac{6\calH^2}{\omega_k}-\frac{6a^2M^2\calH^2}{\omega_k^3}+\frac{a^2M^2}{2\omega_k^5}(6\calH''\calH-3{\calH'}^2+12\calH'\calH^2)\right. \\
        \end{split}
        \end{equation*}
\begin{equation}\label{eq:fourthorderT00}
        \begin{split}
         \phantom{aaaaaaaaaaaaa}&-\left.\frac{a^4M^4}{8\omega_k^7}(120\calH'\calH^2+210\calH^4)+\frac{105a^6M^6\calH^4}{4\omega_k^9}\right)\\
        &+\left.\left(\xi-\frac{1}{6}\right)^2\left( -\frac{1}{4\omega_k^3}(72\calH''\calH-36{\calH'}^2-108\calH^4)+\frac{54a^2M^2}{\omega_k^5}(\calH'\calH^2+\calH^4) \right)\right] \\
        &+\frac{1}{8\pi^2a^2}\int dk k^2 \left[ \frac{a^2\Delta^2}{\omega_k}-\frac{a^4\Delta^4}{4\omega_k^3}+\frac{a^4\calH^2M^2\Delta^2}{2\omega_k^5}-\frac{5}{8}\frac{a^6\calH^2M^4\Delta^2}{\omega_k^7} \right. \\
        &+\left.\left(\xi-\frac{1}{6}\right)\left(-\frac{3a^2\Delta^2\calH^2}{\omega_k^3}+\frac{9a^4M^2\Delta^2\calH^2}{\omega_k^5}  \right) \right] \; .
    \end{split}
\end{equation}
We should recall and emphasize that since this expression is evaluated at the scale $M$ (rather than at $m$), it must be understood that $\omega_k$ in \eqref{eq:fourthorderT00} is $\omega_k(M)=\sqrt{k^2+a^2M^2}$. The on-shell result is recovered for $M=m$ for which $\Delta^2=0$.

In the above formula we can easily recognize  the very particular case of  Minkowskian spacetime, which  is obtained for $a=1$ ($\mathcal{H}=0)$ as well as setting $\Delta^2=0$ and $\xi=0$:
\begin{equation}\label{eq:Minkoski}
 \langle T_{00}^{\delta \phi}\rangle^{\rm Mink}=\frac{1}{4\pi^2}\int dk k^2 \omega_k =  \int\frac{d^3k}{(2\pi)^3}\,\left(\frac12\hbar \sqrt{k^2+m^2}\right)\,,
\end{equation}
where $\hbar$ has been temporarily restored in the trailing term for a better identification of the result. Eq.\,\eqref{eq:Minkoski} gives the ZPE associated with the quantum fluctuations in flat spacetime. This is the expression traditionally quoted in the literature\cite{Brown:1992db,Akhmedov:2002ts,Ossola:2003ku,Martin:2012bt}, which tends to an oversimplification of the problem. It involves quartic (as well as subleading) UV-divergences whose proper renormalization treatment and physical interpretation in a gravity context cannot be performed with a simple cutoff, not even with minimal subtraction with  dimensional regularization. A realistic analysis can only be carried out in curved spacetime, where the result is clearly much more complicated. Indeed,  Eq.\,\eqref{eq:fourthorderT00}  involves again quartic, quadratic and logarithmically divergent terms,  as in the naive result in Minkowski space, but now some of the subleading divergences are intrinsically tied to the curvature of spacetime since they are proportional to powers of the Hubble function. The Minkowskian ZPE result \eqref{eq:Minkoski} knows nothing about the spacetime curvature since $a=1$ is assumed, which is why it is vastly insufficient for a realistic discussion of the VED in cosmology and its connection with the CCP. See \cite{JSPRev2022} for a detailed discussion.

While the unrenormalized ZPE is, of course, quartically divergent  -- even in the simplest flat spacetime situation \eqref{eq:Minkoski} ---the adiabatically renormalized ZPE is finite. It ensues from applying the off-shell ARP \eqref{RenormalizedEMTScalar} and can be obtained by working out explicitly the integrals in Eq.\,\eqref{eq:fourthorderT00} using e.g. the formulae in Appendix \ref{sec:DRintegrals}. The computation is relatively lengthy, but the final result up to fourth adiabatic order takes on a very compact form:
\begin{equation}\label{RenormalizedEMT}
\begin{split}
&\langle T_{00}^{\delta \phi}\rangle_{\rm ren}(M)
=\frac{a^2}{128\pi^2 }\left(-M^4+4m^2M^2-3m^4+2m^4 \ln \frac{m^2}{M^2}\right)\\
&+\left(\xi-\frac{1}{6}\right)\frac{3 a^2 {H}^2 }{16 \pi^2 }\left(M^2-m^2+m^2\ln \frac{m^2}{M^2} \right)-\left(\xi-\frac{1}{6}\right)^2\, \frac{9 a^2\left(\dot{H}^2 -2{H}\ddot{H} - 6 H^2\dot{H}\right)}{16\pi^2}\ln \frac{m^2}{M^2}+\cdots
\end{split}
\end{equation}
where the dots stand for higher adiabatic terms beyond 4th order.

 {The expression \eqref{RenormalizedEMT} gives the $00$th component of the renormalized vacuum EMT, therefore the renormalized ZPE for the non-minimally coupled scalar field in the RVM context. This result was obtained in \cite{CristianJoan2020,CristianJoan2022a} and it will be very helpful to compute the off-shell renormalized ZPE in the de Sitter case (cf. Secs. \ref{sec:deSitter} and \ref{sec:RendeSitter}). See also Appendix \ref{sec:RunninGandVED} for more details. }

\subsection{Renormalized Einstein equations in QFT in curved spacetime}\label{sec:GeneralizedEqs}

In order to implement renormalization of QFT in curved spacetime, we must extend the classical vacuum action (or Einstein-Hilbert action) by considering the higher derivative (HD) gravity terms up to the second adiabatic order. This will lead us to the renormalized Einstein field equations, which incorporate the quantum effects from the quantized matter fields\footnote{As noted, we emphasize that since we work with a fixed gravitational background (FLRW) there are no other renormalization effects apart from those associated with the fluctuations of the quantized matter fields. In a QG context one would also consider graviton contributions, but this leads to its own complications.}. The corresponding renormalized form of Einstein's equations (compare with the original form \eqref{EinsteinEqs}) reads
\begin{equation}\label{eq:MEEs}
\frac{1}{8\pi G(M)} G_{\mu \nu}+\rho_\Lambda(M) g_{\mu \nu}+\alpha(M)\ \leftidx{^{(1)}}{\!H}_{\mu \nu}= \langle T_{\mu \nu}^{\delta \phi}\rangle_{\rm ren}(M)\,,
\end{equation}
where $M$ is the renormalization scale introduced before.
Here  $\leftidx{^{(1)}}{\!H}_{\mu \nu}$ is the (covariantly conserved)  HD tensor which appears from the metric functional variation of  the $R^2$-term in the  higher derivative vacuum action for FLRW spacetime, and $\alpha$ is the corresponding bare coupling.  We remind the reader that  the term emerging from  the variation of $R_{\mu\nu}R^{\mu\nu}$ (the square of the Ricci tensor), called  $\leftidx{^{(2)}}{\!H}_{\mu \nu}$, is not necessary in our case since it is not independent of $\leftidx{^{(1)}}{\!H}_{\mu \nu}$  for conformally flat spacetimes  (in particular, FLRW)\,\cite{BirrellDavies82}.

Parameters  $G(M),\rL(M)$ and $\alpha(M)$ are the renormalized couplings at the scale $M$, and $\langle T_{\mu \nu}^{\delta \phi}\rangle_{\rm ren}(M)$ is the (adiabatically) renormalized vacuum EMT computed from Eq.\,\eqref{RenormalizedEMTScalar}, whose $00$th component is given by \eqref{RenormalizedEMT}. Using this result, we can subtract the renormalized Einstein equations \eqref{eq:MEEs}, term by term,  at two scales $M$ and $M_0$, and we find:
\begin{equation}\label{eq:SubtractedMEEs}
\frac{1}{8\pi}\delta G^{-1}(m,M,M_0) G_{\mu \nu}+\delta\rL(m,M,M_0) g_{\mu \nu}+ \delta\alpha(M,M_0)\ \leftidx{^{(1)}}{\!H}_{\mu \nu}= \delta\langle T_{00}^{\delta \phi}\rangle^{\rm (0-4)}_{\rm ren}(M,M_0)\,,
\end{equation}
where we have defined
$\delta\langle T_{00}^{\delta \phi}\rangle^{\rm (0-4)}_{\rm ren}(M,M_0)\equiv\langle T_{00}^{\delta \phi}\rangle^{\rm (0-4)}_{\rm ren}(M)-\langle T_{00}^{\delta \phi}\rangle^{\rm (0-4)}_{\rm ren}(M_0)$ and similarly for $\delta\rL$, $\delta G^{-1}$ and $\delta\alpha$.  Upon using the known form of $G_{00}$ and  $\leftidx{^{(1)}}{\!H}_{00}$ in the conformal metric  (given e.g. in Appendix A of  \cite{CristianJoan2022a}), we obtain
\begin{equation}\label{SubtractionrL}
\delta\rL(m,M,M_0)\equiv\rL(M)-\rL(M_0)=\frac{1}{128\pi^2}\left(M^4-M_0^{4}-4m^2(M^2-M_0^{2})+2m^4\ln  \frac{M^{2}}{M_0^2}\right)\,,
\end{equation}
\begin{equation} \label{SubtractionG}
\delta G^{-1}(m,M,M_0)\equiv G^{-1}(M)-G^{-1}(M_0)=\left(\xi-\frac{1}{6}\right)\frac{1}{2\pi}\left[M^2 - M_0^{2} -m^2\ln \frac{M^{2}}{M_0^2}\right]
\end{equation}
and
\begin{equation} \label{Subtractionalpha}
\delta\alpha(M,M_0)\equiv\alpha(M)-\alpha(M_0)= -\frac{1}{32\pi^2}\left(\xi-\frac{1}{6}\right)^2 \ln \frac{M^2}{M_0^{2}}\,.
\end{equation}
These relations are important since they furnish the scaling laws obeyed by the couplings $\rL(M)$, $G(M)$, $\alpha(M)$ in the renormalized Einstein's equations and display their respective (finite) shifts when we perform a change of scale  (renormalization point)  from $M_0$ to $M$.
We shall encounter these same equations when we perform the adiabatic renormalization of the vacuum EMT in de Sitter spacetime, which we treat in Sec.\,\ref{sec:deSitter}.  {It is interesting to note that the above relations can also be obtained using the effective action method. In Appendix \ref{sec:RunninGandVED} we recall briefly this alternative path and use the mentioned relations to show the consistency of the running quantities $\rv$ and $G$ with the scale $M=H$.}

\subsection{Renormalized vacuum energy and scaling evolution}\label{sec:RenVEDRVM}

We may now tackle the renormalization of the full energy-momentum tensor of vacuum. The latter must involve the ZPE but also the contribution from the renormalized parameter $\rL(M)$ in the EH action. This is necessary since in the classical limit (absence of vacuum fluctuations) the vacuum EMT must reduce to just $-\rho_\Lambda g_{\mu\nu}$. Thus,  in covariant form, the full renormalized vacuum EMT reads
\begin{equation}\label{eq:FullvacEMT}
    \braket{T_{\mu\nu}^{\mathrm{vac}}}_{\rm ren}(M)=-\rho_\Lambda(M) g_{\mu\nu}+\braket{T_{\mu\nu}^{\delta\phi}}_{\rm ren}(M)\,.
\end{equation}
We can extract the renormalized VED by just considering the  $00$th component of this expression. The VED as measured by an observer with 4-velocity $U^\mu$ is $\rv=\langle  T_{\mu\nu}^{\rm vac} U^\mu U^\nu\rangle$.
In the rest frame of the observer, the $4$-velocity is $U^\mu=\left(1/\sqrt{-g_{00}},0,0,0\right)=\left(1/a,0,0,0\right)$ and satisfies $g_{\mu\nu}U^\mu U^\nu=-1$. Thus, the renormalized VED is given by
\begin{equation}\label{RenVDE}
\rv(M)= \frac{\langle T_{00}^{\rm vac}\rangle_{\rm ren}(M)}{a^2}=\rho_\Lambda(M) +\frac{\langle T_{00}^{\delta \phi}\rangle_{\rm ren}(M)}{a^2}\,.
\end{equation}
Combining equations \eqref{RenVDE} and \eqref{RenormalizedEMT} and making explicit the dependence of the renormalized VED on both $M$ and $H$, we have
\begin{equation}\label{RenVDEexplicit}
\begin{split}
\rv(M,H)&= \rho_\Lambda (M)+\frac{1}{128\pi^2 }\left(-M^4+4m^2M^2-3m^4+2m^4 \ln \frac{m^2}{M^2}\right)\\
&+\left(\xi-\frac{1}{6}\right)\frac{3 {H}^2 }{16 \pi^2 }\left(M^2-m^2+m^2\ln \frac{m^2}{M^2} \right)+\left(\xi-\frac{1}{6}\right)^2 \frac{9\left(6H^2\dot{H}+2H\ddot{H}-\dot{H}^2\right)}{16\pi^2}\ln \frac{m^2}{M^2}+\cdots
\end{split}
\end{equation}
where, again,  the dots stand for higher adiabatic terms beyond 4th order. Notice that the on-shell value ($M=m$) of this expression is simply $\rv(m,H)=\rL(m)$, which is independent of $H$, as it should since the vacuum EMT has been subtracted at the point $M$, so for $M=m$ only the first term of \eqref{RenVDE} should remain up to $4$th adiabatic order. This confirms the correct normalization of the expression obtained for the VED.

However, for a physical interpretation of the theory insofar as concerns the cosmological context under discussion, an appropriate choice of $M$ is required. According to the standard practice in ordinary gauge theories,  the choice of the scale should be made near the typical energy scale of the process. Here, the ``process'' is, of course,  the cosmic expansion, and it has been suggested to make the choice of $M$ at the value of the Hubble rate $H$  for each cosmic epoch under consideration, since $H$ gives the characteristic energy scale of the FLRW spacetime at any given moment; see\cite{CristianJoan2020,CristianJoan2022a,CristianJoan2022b,CristianJoanSamira2023}  and \cite{JSPRev2022} for a review\footnote{For more details, see Appendix \ref{sec:RunninGandVED} where we provide a detailed discussion of the off-shell adiabatic renormalization procedure and demonstrate the consistency of our scale setting procedure.}.
Therefore, from Eq.\,\eqref{RenVDEexplicit} we find the evolution of the VED between two cosmic epochs $M=H$ and $M_0=H_0$:
\begin{equation}\label{DiffVEDphys}
\begin{split}
\rv(H)&=\rv(H_0)+\frac{3\left(\xi-\frac{1}{6}\right)}{16\pi^2}\left[H^2\left(H^2-m^2+m^2\ln\frac{m^2}{H^2}\right)-H_0^2\left(H_0^2-m^2+m^2\ln\frac{m^2}{H_0^2}\right)\right]\\
&+\left(\xi-\frac{1}{6}\right)^2\frac{9a^2}{16\pi^2}\left(6H^2\dot{H}+2H\ddot{H}-\dot{H}^2\right)\ln \frac{m^2}{H^2}\\
&-\left(\xi-\frac{1}{6}\right)^2\frac{9a^2}{16\pi^2}\left(6H_0^2\dot{H}_0+2H_0\ddot{H}_0-\dot{H}_0^2\right)\ln \frac{m^2}{H_0^2}+\cdots
\end{split}
\end{equation}
where we have defined  $\rho_{\rm vac}(H)\equiv\rho_{\rm vac}(M=H,H)$ and similarly $\rho_{\rm vac}(H_0)\equiv\rho_{\rm vac}(M_0=H_0,H_0)$.
This equation provides the VED at the scale $M=H$ in terms of the VED at another renormalization scale $M_0=H_0$, and hence it  expresses the `running' of the VED between the two scales. In obtaining this relation from \eqref{RenVDEexplicit} we should remark that  we used  Eq.\,\eqref{SubtractionrL} for $\rL(M)-\rL(M_0)$. This difference cancels identically against corresponding terms in \eqref{RenVDEexplicit}. As a result the VED evolution has the remarkable property that it is free of quartic mass contributions $\sim m^4$. An interpretation of this fact in terms of the $\beta$-function of the running VED is given in Sec. \ref{sec:betaVED}.

\subsection{Dynamical VED in the RVM: present universe versus
early universe}\label{sec:VEDinRVM}

In the late universe the ${\cal O}(H^4)$ terms of the VED given by Eq.\,\eqref{DiffVEDphys} are negligible and one can easily show that the leading evolution of the vacuum energy  can be expressed as follows \cite{CristianJoan2020,CristianJoan2022a}:
\begin{equation}\label{eq:RVMform}
\rv(H)\simeq \rvo+\frac{3\nu(H)}{8\pi}\,(H^2-H_0^2)\,\mpl^2\,,
\end{equation}
where $\rvo\equiv\rv(H_0)$ is identified with today's value of the VED through the observed CC, i.e. $\rvo=\CC_{\rm obs}/(8\pi G)$.
We have introduced the effective coupling
\begin{equation}\label{eq:nueff2}
\nu(H)\equiv\frac{1}{2\pi}\,\left(\xi-\frac16\right)\,\frac{m^2}{\mpl^2}\left(-1+\ln \frac{m^2}{H^{2}}-\frac{H_0^2}{H^2-H_0^2}\ln \frac{H^2}{H_0^2}\right)\,.
\end{equation}
Owing to the log behavior, this coupling changes very slowly with the Hubble rate and the effect from the last term becomes quickly suppressed for higher values of $H$ above $H_0$. In addition, $\ln \frac{m^{2}}{H^2}\simeq \ln \frac{m^{2}}{H_0^2}\gg1$ for the late universe, and hence $\nu(H)$ can be approximated by letting $H\to H_0$, i.e. by the effective parameter
\begin{equation}\label{eq:nueffAprox2}
\nueff\equiv\nueff(H_0)=\frac{1}{2\pi} \left( \xi-\frac{1}{6}\right) \frac{m^2}{m_\mathrm{Pl}^2}\left(-2+\ln\frac{m^2}{H_0^2}\right)\simeq\epsilon\ln\frac{m^2}{H_0^2}\,,
\end{equation}
where
\begin{equation}\label{eq:epsilonparameter}
\epsilon=\frac{1}{2\pi}\,\left(\xi-\frac{1}{6}\right)\,\frac{m^2}{\mpl^2}\,.
\end{equation}
Parameters $\nueff$ and $\epsilon$ are both small ($|\nueff|, |\epsilon|\ll 1$), but $\epsilon\ll\nueff\ll1$ since $\ln\frac{m^2}{H_0^2}={\cal O}(100)$ for virtually any known particle mass (recall that $H_0\sim 10^{-42}$ GeV).
For practical purposes, we can write \eqref{eq:RVMform} as follows:
\begin{equation}\label{eq:RVMcanonical}
\rv(H)=\rvo+\frac{3\nueff}{8\pi G_N}\,(H^2-H_0^2)\,.
\end{equation}
Here $G_N\equiv 1/\mpl^2$ is the currently measured value of the gravitational constant, expressed in terms of the Planck mass (in natural units).
We note that the expression obtained for the VED in the late universe, Eq.\eqref{eq:RVMcanonical}, turns out to adopt the canonical RVM form; see \cite{JSPRev2022}. Therefore, $\nueff$ is subject to the known phenomenological bounds $\nueff\lesssim 10^{-4}-10^{-3}$\,\cite{SolaPeracaula:2021gxi,SolaPeracaula:2023swx,deCruzPerez:2025dni}.

On the other hand, in the very early universe, the dominant terms are the higher powers ${\cal O}(H^4)$ in  Eq.\,\eqref{DiffVEDphys}. The terms with time derivatives of $H$ do not contribute significantly for $H\simeq$ const.  and hence RVM-inflation is driven by
 \begin{equation}\label{eq:VEDinfl}
\rho_\mathrm{vac}(H)\simeq\frac{3\left( \xi-\frac{1}{6}\right)}{16\pi^2}\left[ H^4+H^2m^2\left(\ln\frac{m^2}{H^2}-1\right)\right]\,.
\end{equation}
The ${\cal O}(H^2)$ term represents a subdominant correction and we have neglected the contribution from the current universe (the terms evaluated at $H_0$). The above result is just the canonical RVM prediction for the VED in the early universe\cite{SolaPeracaula:2025yco}. In the course of our analysis, we shall compare these RVM formulas  for the current and for the early universe with those emerging from the unstable de Sitter vacuum scenario, which is the main focus of the present work.

\subsection{$\beta$-function of the renormalized VED in the RVM: physical running}\label{sec:betaVED}

In contrast to the usual renormalization schemes for the VED, the ARP prescription that we use in the RVM context does not produce a super-fast running of the VED and hence it does not induce a large enhancement of the cosmological term by quantum effects. This is because there is an exact cancellation of the huge $\sim m^4$ contributions, as mentioned at the end of Section \ref{sec:RenVEDRVM}. This can be more formally understood by computing the $\beta$-function corresponding to the physical running of the VED, which we call $\beta_{\rv}$. A straightforward calculation from Eq.\eqref{RenVDEexplicit} renders:
\begin{equation*}
\begin{split}
\beta_{\rv}(M,H)=&M\frac{\partial\rv(M,H)}{\partial M}=M\frac{d\rL(M)}{dM}-\frac{1}{32\pi^2}\left(M^2-m^2\right)^2\\
&+\left(\xi-\frac{1}{6}\right)\frac{3 {H}^2 }{8 \pi^2}\left(M^2-m^2\right)-\left(\xi-\frac{1}{6}\right)^2 \frac{9\left(6H^2\dot{H}+2H\ddot{H}-\dot{H}^2\right)}{8\pi^2}
\end{split}
\end{equation*}
\begin{equation}\label{eq:RGEVED1}
\phantom{aaaaaaaaaaaaa}=\left(\xi-\frac{1}{6}\right)\frac{3 {H}^2 }{8 \pi^2}\left(M^2-m^2\right)-\left(\xi-\frac{1}{6}\right)^2 \frac{9\left(6H^2\dot{H}+2H\ddot{H}-\dot{H}^2\right)}{8\pi^2}\,,
\end{equation}
where
\begin{equation}\label{eq:RGErL}
M\frac{d\rL(M)}{dM}= M\lim_{M\to M_o} \frac{\delta\rL(m,M,M_0)}{M-M_0}=\frac{1}{32\pi^2}\left(M^2-m^2\right)^2\,,
\end{equation}
a result that follows after using Eq.\,\eqref{SubtractionrL}. The latter is the renormalization group equation (RGE) for the coupling $\rL$, but not the RGE for $\rv$.
In fact, the first two terms in the first line of Eq.\,\eqref{eq:RGEVED1} cancel out exactly and as a result the RGE for the physical $\rv$ is free of quartic mass contributions $\sim m^4$. This confirms that the running of the VED, $\rv$, does not depend on dangerous terms  which would inordinately boost the VED to extremely large values. Recall that to explore the running of the VED at a particular epoch $H$, we set $M=H$. Therefore, if the relevant epoch under consideration is the late universe, the last term of \eqref{eq:RGEVED1} -- being of ${\cal O}(H^4)$ -- is irrelevant and we are left with
\begin{equation}\label{eq:RGEVEDlate}
\left.\beta_{\rv}(M,H)\right|_{M=H}\simeq -\left(\xi-\frac{1}{6}\right)\frac{3 {H}^2 m^2}{8 \pi^2}\,,
\end{equation}
with $H^2\ll m^2$ for any known particle mass. In other words, we get  $\beta_{\rv}\propto H^2 m^2$ rather than the conventional quartic mass effects $\sim m^4$\cite{Brown:1992db,Akhmedov:2002ts,Ossola:2003ku,Martin:2012bt}. Obviously, a slope proportional to $H^2 m^2$ is extremely soft compared to $\sim m^4$ and therefore does not increase the speed of the VED running in any significant way. This explains why $\rv$ (and the corresponding cosmological term) can behave as an approximate cosmological constant at each epoch of the cosmological expansion. In stark contrast,  the parameter $\rL$ runs as $\sim m^4$, as seen from \eqref{eq:RGErL} when we set $M=H$, but it has no impact on the physical running of the VED since the effect on $\rL$ has cancelled exactly in Eq.\,\eqref{eq:RGEVED1}, as previously noted. The unacceptable behavior $\sim m^4$  is nonetheless the usual one that is attributed to the physical cosmological term in most  of the literature. But, as we have just seen from our considerations, the real running of the VED is much softer $\sim H^2m^2$.

The ultimate physical running, however, is determined by the full evolution in terms of $H$. To better understand this point, assume that $M=M(H)$ is some arbitrary (although sufficiently smooth) function of $H$, with an inverse $H=H(M)$. The total derivative of the VED with respect to $M$ defines the $\beta$-function of the full running with $H$:
\begin{equation}\label{eq:DerivTotal}
\begin{split}
\beta_{\rm vac}(M,H)\equiv M\frac{d\rv(M,H(M))}{dM}=&M\left(\frac{\partial\rv}{\partial M}+\frac{\partial\rv}{\partial H}\frac{dH}{dM}\right)=\beta_{\rv}+M\frac{\partial\rv}{\partial H}\frac{dH}{dM}\,.
%=&\beta_{\rho_{vac}}+M(\xi-\frac16)\frac{6H}{16\pi^2}\left(M^2-m^2+m^2\ln\frac{m^2}{M^2}\right)\frac{\partial H}{\partial M}\,.
\end{split}
\end{equation}
From equations \eqref{RenVDEexplicit}  and \eqref{eq:RGEVED1}, and considering only the ${\cal O}(H^2)$ contributions which dominate in the present universe, we find:
\begin{equation}\label{eq:DerivTotal2}
\begin{split}
\beta_{\rm vac}(M,H)=&\left(\xi-\frac{1}{6}\right)\frac{3 H^2 }{8 \pi^2 }\left(M^2-m^2\right)+M\left(\xi-\frac16\right)\frac{3H}{8\pi^2}\left(M^2-m^2+m^2\ln\frac{m^2}{M^2}\right)\frac{\partial H}{\partial M}\, ,
\end{split}
\end{equation}
At this point the above result holds for any function $M=M(H)$. Finally, if we  specialize to $M=H$ as the canonical setting to explore a given cosmic epoch, we obtain
\begin{equation}\label{eq:DerivTotal3}
\begin{split}
\beta_{\rm vac}(H)=&\left(\xi-\frac{1}{6}\right)\frac{3 m^2H^2 }{8 \pi^2 }\left(-2+\ln\frac{m^2}{H^2}\right)\simeq \left(\xi-\frac{1}{6}\right)\frac{3 m^2H^2 }{8 \pi^2 }\ln\frac{m^2}{H_0^2} \,,
\end{split}
\end{equation}
where again we have neglected the ${\cal O}(H^4)$ contributions in the late universe and in the log we have used the current value of the Hubble rate, $H_0$. Again, since for any known particle $\ln\frac{m^2}{H_0^2}\gg 1$, we have neglected the additive numerical term. The above result can be alternatively derived by computing $H d\rv(H)/dH$ directly from Eq.\eqref{DiffVEDphys} within the same approximation. Combining equations \eqref{eq:nueffAprox2} and \eqref{eq:DerivTotal3}, we find the relation  between $\beta_{\rm vac}$ and the parameter $\nueff$:
\begin{equation}\label{eq:betavnueff}
\beta_{\rm vac}(H)=\frac{3\nueff}{4\pi}\mpl^2 H^2\,.
\end{equation}
This $\beta$ function has different sign as compared to that in Eq.\eqref{eq:RGEVEDlate}, since the former explores the full evolution (total derivative) with respect to $M$ and not just the variation of $\rv$ with $M$ at fixed $H$ (as in the latter), i.e. the partial derivative. The above equation  tells us that if $\nueff>0$ (resp. $\nueff<0$) the VED decreases (resp. increases) with expansion, and therefore the VED effectively behaves as quintessence (resp. phantom DE) -- see Sec.\,\ref{sec:EosVacuum} for more details. Interestingly enough, we find that the quantum vacuum mimics dynamical DE\cite{CristianJoan2022b}. This is a very interesting property of the quantum vacuum in the RVM context, which we shall further explore here and compare with the de Sitter scenario in the next section.

\section{Inflation and de Sitter spacetime}\label{sec:deSitter}

De Sitter spacetime\,\cite{deSitter1917} is well-known to be the maximally symmetric solution to the vacuum Einstein field equations $R_{\mu\nu}=\Lambda g_{\mu\nu}$, $R=4\Lambda$,  with positive curvature and hence positive cosmological constant, $\CC>0$. Its symmetry group is $SO(1,n)$ for n spacetime dimensions, therefore enjoying the same degree of symmetry as  Minkowski spacetime ($10$ Killing vectors for $n=4$). Because of this, analytical QFT methods can be applied with relative simplicity. Geometrically, for $n=4$ it can be viewed as a 4-dimensional hyperboloid with positive curvature embedded in five-dimensional Minkowski spacetime\, \cite{BirrellDavies82,ParkerToms09}.  De Sitter space is particularly useful for modeling the inflationary universe both early on and at very late times of cosmic history.  In fact, under an appropriate coordinatization it can be brought into an exponentially expanding FLRW spacetime (see below). Traditionally, inflation is usually described with the help of a scalar field\cite{KolbTurner,LiddleLyth,RubakovGorbunov,Kallosh:2025ijd,Martin:2013tda}, which is sometimes called the ``inflaton''.  During the inflationary stage,  the growth of quantum fluctuations of the scalar field can explain the observed large-scale structure of the universe. Not surprisingly, the study of QFT in the de Sitter background is of central importance in cosmology. For the late universe, however,  a new scalar field is usually introduced, which is called quintessence (and the like), although in this case it is usually a classical scalar field with some effective potential.  An alternative description of the early inflation period is obtained by using higher derivative terms, such as in the case of Starobinsky's $R^2$-inflation\cite{Starobinsky:1980te}. In both  approaches, the description is very useful, but it proves insufficient, since the inflationary period on top of the standard FLRW regime is imposed totally {\it ad hoc}, somehow sewed at hand in the early stage of the cosmic evolution. In fact, there is no liaison between the very early epoch and the current epoch in the context of these inflationary scenarios. This is obviously a drawback. Here we wish to propose a unification scenario based on quantum field theory in curved spacetime in which the very early universe and the current universe are described uninterruptedly by the same unified QFT framework.

In the previous sections, we have seen that quantum effects from quantized matter fields can trigger appropriate powers of the Hubble rate $H^4$ and $H^2$. The combined effect of these powers acting in sequence can describe (within a single unified framework) the period of very early inflation and late-time dark energy (DE) dominance, which eventually leads to an ultimate de Sitter epoch in the remote future. This can be achieved in the context of the running vacuum model (RVM) framework, see \cite{JSPRev2013,JSPRev2015,JSPRev2022}. In particular, the $H^4$-inflation mechanism has recently been studied in \cite{SolaPeracaula:2025yco} within the strict RVM context and was reviewed in the previous sections. Here we shall compare RVM-inflation with de Sitter inflation, which is also triggered by the power $H^4$ of the Hubble rate and includes subleading $H^2$ powers as well.  In both cases, the leitmotif of inflation is the presence of a very short period where $H$ remains constant (or approximately constant) and very large in the early universe: $H=H_I\equiv\sqrt{8\pi G\rho_I/3}=\sqrt{\CC_I/3}$.
It originates from a huge value of the VED, $\rho_I=\rv\sim M_X^4$, typical of some GUT scale $M_X$. Here $\CC_I$ is an effective cosmological term in the early universe. Thus, we have
\begin{equation}\label{eq:Hexp}
    \left(\frac{\dot{a}}{a}\right)^2=\frac{8\pi G\rho_I}{3}=\frac{\CC_I}{3}\equiv H_I^2\,.
\end{equation}
It follows that the scale factor increases exponentially with the cosmic time,
\begin{equation}\label{eq:aexp}
a(t)\propto e^{H_It}\,,
\end{equation}
and therefore it leads indeed to an inflationary epoch. The metric of such epoch takes on the spatially flat FLRW form with an exponential scale factor:
\begin{equation}\label{eq:metricdeSittert}
ds^2=-dt^2+e^{2H_It}\left(dx^2+ dy^2+dz^2\right)\,.
\end{equation}
Interestingly, this metric can describe inflation both in the very early universe and in the very late universe (with a much smaller value of $H\simeq\sqrt{\Lambda_{\rm obs}/3}$ in the last case, of course) since the vacuum energy eventually dominates the cosmic expansion.
Equation \eqref{eq:Hexp} holds good provided that the VED is either strictly constant,  $\rv=\rho_I$,  or mildly evolving with the expansion, $\rv=\rv(H)$, but in any case dominant over any other form of energy in the universe during some period in which $\rv(H)\simeq \rho_I$.

Now a fixed de Sitter background cannot furnish an acceptable description of cosmology, as it would be unrealistic for the subsequent cosmological expansion, and hence it can only be an approximation, valid for a short primordial period. To  achieve  successful inflation, the constant value $H_I$ must be large enough, presumably connected to the typical scale $M_X$ of a Grand Unified Theory (GUT) around the Planck scale, $\mpl$.  At this point, we cannot be more precise.
But we do know that for this picture to be minimally realistic, the universe must eventually exit the de Sitter exponential period \eqref{eq:aexp} and enter a much more temperate stage, since the universe must connect with the standard radiation-dominated epoch\footnote{The origin of the initial de Sitter state might be connected  to stringy versions of the RVM. For instance, in \cite{Dorlis:2025gvb} it is discussed the appearance of exact de Sitter vacua as stable solutions of a dynamical system approach
to cosmology, based on specific potentials motivated by  stringy RVM formulations\cite{Mavromatos:2020kzj,Mavromatos:2020crd,PhantomVacuum2021}. The universe tunnels between these vacua, and this behavior can capture current eras.}.  The ordinary FLRW regime then begins and we must recover the ordinary expansion law for relativistic matter $a\sim t^{1/2}$. This transition can be implemented if the huge VED that is stored in the initial de Sitter background, $\rv\lesssim \mpl^4$, decays into radiation.  But before coming to grips with these details, we need to obtain the renormalized EMT corresponding to de Sitter spacetime and extract the physical VED out of it.

\subsection{Exact mode functions}\label{sec:ExactModes1}

Let us consider the exact solution to the field equations for $H=$ const. This is possible in de Sitter spacetime thanks to its highest degree of symmetry\,\cite{BirrellDavies82,ParkerToms09}\footnote{See also \cite{Kamenshchik:2021tjh} for self-interacting scalar fields in de Sitter spacetime with minimal coupling ($\xi=0$).}.  By differentiating on both sides of Eq.\eqref{eq:aexp} and using the relation $dt=ad\tau$ between cosmic time and conformal time,  we find $da/a^2=H d\tau$, and hence upon integration
\begin{equation}\label{eq:attau}
a=-\frac{1}{H\tau}\,,
\end{equation}
where the conformal time is defined in the negative interval $\tau \in (-\infty , 0)$, in which $0<a<\infty$, and we understand that $H=H_I$ remains constant in the period considered. The de Sitter metric \eqref{eq:metricdeSittert} in conformal time then takes the conformally flat form,
\begin{equation}\label{eq:metricdeSittertau}
ds^2=\frac{1}{H_I^2\tau^2}\left(-d\tau^2+dx^2+ dy^2+dz^2\right)\,,
\end{equation}
with a coordinate singularity at $\tau=0$. The Ricci scalar for de Sitter spacetime gets simplified:
\begin{equation}\label{eq:RdeSitter}
    R=6(2H^2+\dot{H})=12H^2 \; .
\end{equation}
Using \eqref{eq:attau}, it is easy to see that the differential equation \eqref{eq:varphimodes} for the field modes can be put in the form
\begin{equation}\label{eq:varphiequation}
       \varphi_k''+\left[k^2+\frac{m^2+\left(\xi-1/6\right) R}{\tau^2H^2}\right]\varphi_k=0\,.
\end{equation}
This equation admits an exact solution in terms of Bessel functions or, more conveniently in this case, in terms of Hankel functions $\mathbb{H}_\varsigma$ of appropriate order $\varsigma$. Recall that the Hankel functions of first and second kind are related to the Bessel functions of the first and second kind as follows: $\mathbb{H}_\varsigma^{(1)} = J_\varsigma+iY_\varsigma$ and $\mathbb{H}_\varsigma^{(2)} = J_\varsigma-iY_\varsigma$.  In general these are independent since the Bessel functions can be complex functions. The suitable change of variables in Eq.\,\eqref{eq:varphiequation} bringing it into Bessel's canonical form are the following: $\varphi_k=\sqrt{-k\tau}\, u(z)=\sqrt{k\abs{\tau}}\,u(z)$ and $z=-k\tau=k\abs{\tau}$. One can easily check that the new mode function $u(z)$ indeed satisfies $z^2u''(z)+zu'(z)+(z^2-\varsigma^2) u(z)=0$, which is precisely the standard Bessel equation  of order $\varsigma$,
where in this case
\begin{equation}\label{eq:nudefinition}
  \varsigma^2=\frac{1}{4}-\frac{m^2+\left(\xi-1/6\right) R}{H^2}=
 \frac{9}{4}-\frac{m^2+12H^2\xi}{H^2}=\frac94-\mu^2-12\xi\,,
\end{equation}
upon using \eqref{eq:RdeSitter}. Here, for convenience and for further use, we have defined $\mu\equiv m/H$. Thus the order of the Bessel equation is a function of $H$ through $\mu$ and depends on the parameter $\xi$, i.e. $\varsigma=\varsigma(\mu; \xi)$.  This fact will play a role in future considerations.
Notice that in our case $\varsigma$ can only be real or pure imaginary.

The solution $u(z)$ to the Bessel equation can be taken as a linear combination of $J_\varsigma$ and $Y_\varsigma$ or, alternatively, of Hankel functions of first and second kind:
\begin{equation}\label{eq:solHankel}
    \varphi_k=\sqrt{\abs{\tau}} \left[ \alpha_k \mathbb{H}_\varsigma^{(1)} (k\abs{\tau})+\beta_k \mathbb{H}_\varsigma^{(2)} (k\abs{\tau})\right]\,.
\end{equation}
The coefficients $\alpha_k$ and $\beta_k$ in the above solution are dimensionless since $\varphi_k$ must have power dimension $-1/2$ of energy in natural units, according to the the Fourier expansions in Sec. \ref{sec:Quantization}. Using the original field, $\phi_k=\varphi_k/a=H\abs{\tau} \varphi_k$, the exact solution reads
\begin{equation}
    \phi_k=H\abs{\tau}^{3/2} \left[\alpha_k \mathbb{H}_\varsigma^{(1)} (k\abs{\tau})+\beta_k \mathbb{H}_\varsigma^{(2)} (k\abs{\tau}) \right]\,.
\end{equation}
We now make a choice of the coefficients by imposing the Bunch-Davies vacuum limit for $\tau\xrightarrow{}-\infty$ (corresponding to $a\to 0$ in the very early epoch\,\cite{Bunch:1978yq}), so that $\alpha_k=\frac{\sqrt{\pi}}{2} e^{\frac{i\pi}{2}\left( \varsigma+\frac{1}{2}\right)}$ and $\beta_k=0$, and hence the Hankel function of the second kind does not participate.  The field modes with appropriate normalization are then
\begin{equation}\label{eq:Phisolution}
    \phi_k=\frac{\sqrt{\pi}}{2} H \abs{\tau}^{3/2}  e^{\frac{i\pi}{2}\left( \varsigma+\frac{1}{2}\right)}\ \mathbb{H}_\varsigma^{(1)} (k\abs{\tau})\,.
\end{equation}
The corresponding fluctuation modes $h_k=\phi_k a=\phi_k/(H\abs{\tau})$ are
\begin{equation}\label{eq:hkmodesdeSitter}
    h_k=\frac{\sqrt{\pi}}{2} \abs{\tau}^{1/2}  e^{\frac{i\pi}{2}\left( \varsigma+\frac{1}{2}\right)} \ \mathbb{H}_\varsigma^{(1)} (k\abs{\tau}) \; .
\end{equation}
The Bunch-Davies boundary condition is usually imposed deep inside the horizon. It guaranties that for very short physical wavelengths $k/a=k\abs{\tau} H\gg H$ the solution behaves as a positive-definite energy wave in Minkowski space. For a given mode $k$ this condition ensures $k|\tau|\gg1$ and hence the modes can be thought of as being essentially insensitive to curvature effects  for $\tau\xrightarrow{}-\infty$, as then  $a^2R=12/\tau^2\to 0$ -- cf. Eq.\,\eqref{eq:Omegak} -- which is why they are Minkowskian in this limit. We can check that the desired condition is fulfilled by the Hankel function of the first kind $\mathbb{H}_\varsigma^{(1)}(z)$, with $z=k\abs{\tau}=-k\tau>0$ (see Appendix \ref{sec:AppendixHankel}):
\begin{equation}\label{eq:asymptoticHankel}
    \mathbb{H}_\varsigma^{(1)}(|z|\xrightarrow{}\infty) \sim \sqrt{\frac{2}{\pi z}}e^{i\left(z-\frac{1}{2}\varsigma \pi -\frac{\pi}{4} \right)} \,.
\end{equation}
In fact, in this limit, the mode function $h_k$ adopts the desired form\footnote{Had we used, instead,  $k\tau=-z<0$ ($-\infty<\tau<0$) as the argument of the Hankel functions, then it would be $\mathbb{H}_\varsigma^{(2)}(k\tau)$, rather than $\mathbb{H}_\varsigma^{(1)}(k\tau)$, the one satisfying the Bunch-Davies boundary condition, cf. Appendix \ref{sec:AppendixHankel}.}
\begin{equation}
h_k(\tau\xrightarrow{}-\infty)=-\frac{1}{H\tau} \frac{\sqrt{\pi}}{2} H\abs{\tau}^{3/2}  e^{i\frac{\pi}{2}\left( \varsigma+\frac12 \right)}\mathbb{H}_\varsigma^{(1)} (-k\tau) \simeq \frac{\sqrt{\abs{\tau}\pi}}{2} \sqrt{\frac{2}{\pi k \abs{\tau}}} e^{-ik\tau}=\frac{e^{-ik\tau}}{\sqrt{2k}}\,.
 \end{equation}
It corresponds to a positive-energy solution  since $i\frac{\partial h_k}{\partial\tau}=\omega_k h_k$ (the mass term being neglected at very short distances, $\omega_k\simeq k>0$) and it also fulfills the Wronskian normalization condition  $h_k^{} h_k^{*\prime}-h_k^\prime h_k^* = i$, as it should.

\subsection{Zero-point vacuum energy}\label{sec:ExactModes2}

Equipped with the exact and appropriately normalized field modes \eqref{eq:hkmodesdeSitter}, we may now compute the vacuum EMT \eqref{eq:vacuumEMT} for de Sitter spacetime, specifically its $00$th component or ZPE, although later we shall provide the full EMT as well.
To simplify the notation, from now on we will generally omit the argument $z\equiv k\abs{\tau}= -k\tau>0$ in the Hankel functions. Moreover, when computing the derivative with respect to $\tau$, we apply the chain rule and leave the prime to indicate derivative with respect to the argument (for example, $\frac{d}{d\tau}\mathbb{H}_\varsigma (-k\tau)= -k\,\left(d\mathbb{H}_\varsigma(z)/dz\right)\equiv -k\, \mathbb{H}_\varsigma'$).
Therefore, the mode functions and derivatives can be written as follows:\footnote{From the context of the calculations the prime notation for derivatives should not cause any confusion. As previously mentioned, in all cases a prime denotes derivative with respect to the argument of the corresponding function. Thus, for the Hubble function and mode functions $\calH'=d\calH/d\tau$,  $h'_k=d h_k/d\tau$, respectively, whereas $\mathbb{H}'=d\mathbb{H}/dz$, etc. for the Hankel functions.}
\begin{equation}
\begin{split}
    h_k'&=-\frac{\sqrt{\pi}}{2}  e^{\frac{i\pi}{2}\left( \varsigma+\frac{1}{2}\right)} \left( \frac{\abs{\tau}^{-1/2}}{2} \mathbb{H}_\varsigma^{(1)} + \abs{\tau}^{1/2} k {\mathbb{H}_\varsigma^{(1)}}'\right)\,, \\
    h_k^*&=\frac{\sqrt{\pi}}{2}\abs{\tau}^{1/2}  e^{-\frac{i\pi}{2}\left( \varsigma^*+\frac{1}{2}\right)} \mathbb{H}_\varsigma^{(1)*}\,, \\
    {h_k^*}'&=-\frac{\sqrt{\pi}}{2}  e^{-\frac{i\pi}{2}\left( \varsigma^*+\frac{1}{2}\right)} \left( \frac{\abs{\tau}^{-1/2}}{2} \mathbb{H}_\varsigma^{(1)*} + \abs{\tau}^{1/2} k {\mathbb{H}_\varsigma^{(1)*}}'\right)\,.
\end{split}
\end{equation}
The expressions that appear in the 00th component of the vacuum EMT -- cf. Eq.\,\eqref{eq:vacuumEMT} -- are then
\begin{equation}
    \begin{split}
        \abs{h_k}^2&=\frac{\pi}{4}\abs{\tau}  e^{-\pi\Im \varsigma}\abs{\mathbb{H}_\varsigma^{(1)}}^2\,,\\
        \abs{h_k'}^2&=\frac{\pi}{4}e^{-\pi\Im \varsigma} \left[ \frac{1}{4\abs{\tau}}\abs{\mathbb{H}_\varsigma^{(1)}}^2+\abs{\tau} k^2 \abs{{\mathbb{H}_\varsigma^{(1)}}'}^2+\frac{k}{2} \left( \mathbb{H}_\varsigma^{(1)} {\mathbb{H}_\varsigma^{(1)*}}'+\mathbb{H}_\varsigma^{(1)*}{\mathbb{H}_\varsigma^{(1)}}' \right) \right]\,, \\
        (h_k' h_k^*+{h_k^*}'h_k)&=-\frac{\pi}{4}e^{-\pi\Im \varsigma}  \left[ \abs{\mathbb{H}_\varsigma^{(1)}}^2+k\abs{\tau}\left( \mathbb{H}_\varsigma^{(1)} {\mathbb{H}_\varsigma^{(1)*}}'+\mathbb{H}_\varsigma^{(1)*}{\mathbb{H}_\varsigma^{(1)}}' \right) \right]\,.
    \end{split}
\end{equation}

After inserting these expressions into Eq.\,\eqref{eq:vacuumEMT} we are led to the exact result of the ZPE of a non-minimally coupled scalar field in de Sitter spacetime:
\begin{equation}\label{eq:T00mexactk}
    \begin{split}
        \braket{T_{00}^{\delta\phi}}^{\rm dS}(m)&=\frac{\pi}{4}\frac{1}{4\pi^2a^2}e^{-\pi\Im \varsigma} \int dk k^2 \left\{ \frac{1}{4\abs{\tau}}\abs{\mathbb{H}_\varsigma^{(1)}}^2+\abs{\tau} k^2 \abs{{\mathbb{H}_\varsigma^{(1)}}'}^2+\frac{k}{2} \left( \mathbb{H}_\varsigma^{(1)} {\mathbb{H}_\varsigma^{(1)*}}'+\mathbb{H}_\varsigma^{(1)*}{\mathbb{H}_\varsigma^{(1)}}' \right) \right. \\
        &+\left.\omega_k^2 \abs{\tau} \abs{\mathbb{H}_\varsigma^{(1)}}^2+\left(\xi-\frac{1}{6}\right)\left[-\frac{6}{\abs{\tau}} \abs{\mathbb{H}_\varsigma^{(1)}}^2-\frac{6}{\abs{\tau}} \left(\abs{\mathbb{H}_\varsigma^{(1)}}^2+k\abs{\tau}\left( \mathbb{H}_\varsigma^{(1)} {\mathbb{H}_\varsigma^{(1)*}}'+\mathbb{H}_\varsigma^{(1)*}{\mathbb{H}_\varsigma^{(1)}}' \right)\right) \right] \right\}\,.
    \end{split}
\end{equation}
In the above equation we have explicitly indicated that the vacuum EMT is evaluated on-shell, which means that we have used $\omega_k^2=\omega_k^2(m)=k^2+a^2m^2$.
%For the off-shell case, to be considered later on,  we will use $\omega_k^2(M)=k^2+a^2M^2$.
In terms of the previously defined  (dimensionless) variable $z$, the above integration can be rephrased as
\begin{equation}\label{eq:T00mexact}
\begin{split}
    \braket{T_{00}^{\delta\phi}}^{\rm dS}(m)&=\frac{H^2}{16\pi \abs{\tau}^2} e^{-\pi\Im \varsigma} \int dz\,z^2 \left\{ \left( \frac{9}{4}+\frac{m^2}{H^2}-12\xi\right)\abs{\mathbb{H}_\varsigma^{(1)}}^2+z^2\abs{\mathbb{H}_\varsigma^{(1)}}^2+z^2 \abs{{\mathbb{H}_\varsigma^{(1)}}'}^2 \right. \\
    &+\left.z\left( \frac{3}{2}-6\xi\right) \left( \mathbb{H}_\varsigma^{(1)} {\mathbb{H}_\varsigma^{(1)*}}'+\mathbb{H}_\varsigma^{(1)*}{\mathbb{H}_\varsigma^{(1)}}' \right) \right\}\,.
\end{split}
\end{equation}

Using the explicit formulas for the asymptotic expansions of the Hankel functions in powers of $1/z$ given in Appendix \ref{sec:AppendixHankel}, it is easy to convince oneself that the integral \eqref{eq:T00mexact} is UV-divergent (viz. for $z\to\infty$) and one may easily recognize the presence of quartic, quadratic, and logarithmic divergences, which we will have to deal with.

\subsection{Adiabatic expansion of the ZPE }\label{sec:RenEMTdeSitter}

Since the  ZPE obtained in the previous section is UV-divergent, it requires renormalization, which we implement through  the off-shell ARP \eqref{RenormalizedEMTScalar}. This means that we have to subtract the fourth-order adiabatic expansion \eqref{eq:fourthorderT00} evaluated at the arbitrary renormalization scale $M$ from the exact on-shell result found in the previous section. That expansion is valid for any $H(t)$ in FLRW spacetime with flat three-dimensional space\,\cite{CristianJoan2020,CristianJoan2022a}. However, in the particular case of de Sitter spacetime,  Eq. \eqref{eq:fourthorderT00} becomes simplified when written in terms of the usual Hubble function, since $H$ is constant and hence $\dot{H}$ and other higher order derivatives vanish. With the help of Eq.\,\eqref{eq:attau}, the fourth-order adiabatic expansion can be further worked out and brought into the following simpler expression:
\begin{equation}\label{eq:4thorderdeSitter}
    \begin{split}
        \braket{T_{00}^{\delta\phi}}^{(0-4)}(M)&=\frac{H^2\tau^2}{8\pi^2}\int dk k^2 \left[ 2\omega_k+\frac{M^4}{4\tau^6H^4\omega_k^5}-\frac{15}{16}\frac{M^4}{\tau^8H^4\omega_k^7}+\frac{21}{8}\frac{M^6}{\tau^{10}H^6\omega_k^9}-\frac{105}{64} \frac{M^8}{\tau^{12}H^8 \omega_k^{11}} \right. \\
        &+\left(\xi-\frac{1}{6}\right)\left(-\frac{6}{\tau^2\omega_k}-6\frac{M^2}{\tau^4H^2\omega_k^3}+\frac{21}{2}\frac{M^2}{\tau^6H^2\omega_k^5}-\frac{165}{4}\frac{M^4}{\tau^8H^4\omega_k^7}+\frac{105}{4}\frac{M^6}{\tau^{10}H^6\omega_k^9}  \right) \\
        &+\left. \left( \xi-\frac{1}{6}\right)^2\left( 108\frac{M^2}{\tau^6H^2\omega_k^5}\right) \right] \\
        &+\frac{H^2\tau^2}{8\pi^2}\int dk k^2\left[ \frac{\Delta^2}{\tau^2H^2\omega_k}-\frac{1}{4}\frac{\Delta^4}{\tau^4H^4\omega_k^3}+\frac{1}{2} \frac{M^2\Delta^2}{\tau^6H^4\omega_k^5}-\frac{5}{8} \frac{M^4\Delta^2}{\tau^8H^6\omega_k^7} \right. \\
        &+\left.\left(\xi-\frac{1}{6}\right)\left( -3\frac{\Delta^2}{\tau^4H^2\omega_k^3}+9\frac{\Delta^2M^2}{\tau^6H^4\omega_k^5} \right)  \right] \; .
    \end{split}
\end{equation}
As before, it is convenient to change the integration variable from $k$ to dimensionless $z \equiv k\abs{\tau}$,  which we already used for the Hankel functions in the previous section. Furthermore,  $\omega_k(m)=\sqrt{k^2+a^2m^2}$  must now be replaced with $\omega_k(M)=\sqrt{k^2+a^2M^2}$, and it is convenient to introduce the dimensionless counterpart of the latter:
\begin{equation}\label{eq:defwz}
    \omega_z(M)\equiv \sqrt{z^2+\frac{M^2}{H^2}}=\abs{\tau} \omega_k(M)\,.
\end{equation}
Therefore, Eq.\,\eqref{eq:4thorderdeSitter} becomes
\begin{equation}\label{eq:4thorderdeSitterz}
    \begin{split}
        \braket{T_{00}^{\delta\phi}}^{(0-4)}(M)&=\frac{H^2}{8\pi^2\abs{\tau}^2}\int dz z^2 \left[ 2\omega_z+\frac{M^4}{4H^4\omega_z^5}-\frac{15}{16}\frac{M^4}{H^4\omega_z^7}+\frac{21}{8}\frac{M^6}{H^6\omega_z^9}-\frac{105}{64} \frac{M^8}{H^8 \omega_z^{11}} \right. \\
        &+\left(\xi-\frac{1}{6}\right)\left(-\frac{6}{\omega_z}-6\frac{M^2}{H^2\omega_z^3}+\frac{21}{2}\frac{M^2}{H^2\omega_z^5}-\frac{165}{4}\frac{M^4}{H^4\omega_z^7}+\frac{105}{4}\frac{M^6}{H^6\omega_z^9}  \right) \\
        &+\left. \left( \xi-\frac{1}{6}\right)^2\left( 108\frac{M^2}{H^2\omega_z^5}\right) \right] \\
        &+\frac{H^2}{8\pi^2\abs{\tau}^2}\int dz z^2\left[ \frac{\Delta^2}{H^2\omega_z}-\frac{1}{4}\frac{\Delta^4}{H^4\omega_z^3}+\frac{1}{2} \frac{M^2\Delta^2}{H^4\omega_z^5}-\frac{5}{8} \frac{M^4\Delta^2}{H^6\omega_z^7} \right. \\
        &+\left.\left(\xi-\frac{1}{6}\right)\left( -3\frac{\Delta^2}{H^2\omega_z^3}+9\frac{\Delta^2M^2}{H^4\omega_z^5} \right)  \right]\,,
    \end{split}
\end{equation}
where we should emphasize that $\omega_z$ in the above equation is $\omega_z(M)$ as defined in \eqref{eq:defwz}.
The divergent terms of the previous integral for $M=m$ ($\Delta=0$) are easily identified:
\begin{equation}\label{eq:divpartonshell}
\begin{split}
    \braket{T_{00}^{\delta\phi}}^{(0-4)}_{\mathrm{Div}}(m)&=\frac{H^2}{8\pi^2\abs{\tau}^2}\int dz z^2 \left[ 2\omega_z+\left(\xi-\frac{1}{6}\right)\left(-\frac{6}{\omega_z}-6\frac{m^2}{H^2\omega_z^3} \right) \right]\,.
\end{split}
\end{equation}
Once more we can immediately recognize the presence of quartic, quadratic and logarithmically divergent terms, that is to say, the same type of UV-divergences that we have identified in the exact ZPE form \eqref{eq:T00mexact} -- cf. Appendix \ref{sec:exactintegrals}. Therefore, with the above results we are ready to apply the off-shell ARP \eqref{RenormalizedEMTScalar} in order to cancel the divergent terms and produce a renormalized expression for the ZPE.  However, notice that the $\Delta$-dependent terms in the last two rows of Eq.\,\eqref{eq:4thorderdeSitterz} must also be taken into account and here we have also to separate the UV-divergent ones from the finite contributions. One can show by explicit calculation that the divergent part of $\braket{T_{00}^{\delta\phi}}^{(0-4)}(M)$, i.e. of Eq.\,\eqref{eq:4thorderdeSitterz}, is the same as the divergent part of $\braket{T_{00}^{\delta\phi}}^{(0-4)}(m)$ -- given by  Eq.\,\eqref{eq:divpartonshell}. In other words, the UV-divergent terms of these expressions (namely the specific terms carrying poles in dimensional regularization) are exactly equal.
%We denote this feature as follows:
%\begin{equation}\label{eq:DivMm}
%    \left.\braket{T_{00}^{\delta\phi}}^{(0-4)}_{\mathrm{Div}}(M)\right|_{\rm \textcolor{red}{UV}}=
%   \left.\braket{T_{00}^{\delta\phi}}^{(0-4)}_{\mathrm{Div}}(m)\right|_{\rm \textcolor{red}{UV}}\,.
%\end{equation}
Therefore, the quantity
\begin{equation}\label{eq:finite}
\braket{T_{00}^{\delta\phi}}^{(0-4)}_{\mathrm{Div}}(m)-
   \braket{T_{00}^{\delta\phi}}^{(0-4)}_{\mathrm{Div}}(M)
\end{equation}
is finite, as it consists of finite terms that explicitly depend on the scale $M$. This result should be expected since the off-shell EMT cannot involve additional UV-divergences beyond those already contained in the on-shell EMT as this would alter the consistency of the renormalization procedure.  Details of the proof of this fact
%\begin{comment}{\,\eqref{eq:DivMm}} \end{comment}
are provided in Appendix \ref{sec:FiniteEMTdeSitter}.
%However, we should emphasize that
% $\braket{T_{00}^{\delta\phi}}^{(0-4)}_{\mathrm{Div}}(m)$ and $\braket{T_{00}^{\delta\phi}}^{(0-4)}_{\mathrm{Div}}(M)$ differ in finite terms which depend explicitly on the scale $M$.%
Such a residual, finite,  part  will be, of course,  of foremost importance for our calculation.

This also holds for the exact de Sitter scenario. From another straightforward calculation, using the asymptotic expansion of the exact ZPE
and operating the ARP subtraction procedure \eqref{RenormalizedEMTScalar}, the divergent parts of the exact on-shell solution and of the fourth-order adiabatic expansion cancel each other out in full. Once more, we refer the reader to Appendix \ref{sec:FiniteEMTdeSitter}. So,
\begin{equation}\label{eq:DivMm2}
    \braket{T_{00}^{\delta\phi}}_\mathrm{Div}^{\rm dS}(m)- \braket{T_{00}^{\delta\phi}}^{(0-4)}_{\mathrm{Div}}(M)\,, \qquad \braket{T_{00}^{\delta\phi}}_\mathrm{Div}^{\rm dS}(m)- \braket{T_{00}^{\delta\phi}}^{(0-4)}_{\mathrm{Div}}(m)
\end{equation}
are perfectly finite.
%We express this result as follows:
%\begin{equation}\label{eq:DivMm2}
    %\left.\braket{T_{00}^{\delta\phi}}_\mathrm{Div}^{\rm exact}(m)\right|_{\rm \textcolor{red}{UV}}-\left.\braket{T_{00}^{\delta\phi}}^{(0-4)}_{\mathrm{Div}}(M)\right|_{\rm \textcolor{red}{UV}}=\left.\braket{T_{00}^{\delta\phi}}_\mathrm{Div}^{\rm exact}(m)\right|_{\rm \textcolor{red}{UV}}-\left.\braket{T_{00}^{\delta\phi}}^{(0-4)}_{\mathrm{Div}}(m)\right|_{\rm \textcolor{red}{UV}}=0\,,
To show this result, we use the aforementioned fact that the UV-divergent parts of $\braket{T_{00}^{\delta\phi}}^{(0-4)}_{\mathrm{Div}}(m)$ and $\braket{T_{00}^{\delta\phi}}^{(0-4)}_{\mathrm{Div}}(M)$ are exactly the same. The off-shell adiabatic prescription provides indeed the finite renormalized value of the ZPE. All that said,  we still need to extract the finite result left out from such a subtraction when one keeps the finite
parts of the calculation and not just the UV-divergent terms. After all this finite reminder is the physical result we are after, of course.  However, such a calculation is much more cumbersome and will be considered in the next section.

\section{Off-shell adiabatic renormalization in de Sitter spacetime}\label{sec:RendeSitter}

In the previous section, we have shown the finiteness of the adiabatically renormalized ZPE of the vacuum fluctuations in de Sitter spacetime.  However, we need to explicitly determine the finite renormalized result.   An approximate treatment can be attempted, which is based on asymptotically expanding the Hankel functions before performing the integrations, see Appendix \ref{sec:AppendixHankel}. This approach is certainly useful to check the cancellation of UV-divergences in a relatively simple manner, and it provides the kind of expected leading terms in the final result. However, the full result requires a more precise  treatment and can only be obtained by computing the integrals in an exact way and then expanding the result in the appropriate limit. This can be done with the help of special functions, see Appendix \ref{sec:exactintegrals}.  The exact treatment is certainly more cumbersome, but since it is possible, we have followed it.  With these ingredients at hand, we are in position to compute the renormalized VED and vacuum pressure, and hence eventually find the equation of state of the quantum vacuum as well.

\subsection{Renormalized ZPE and VED}\label{sec:RenZPEandVED}

\
Our starting point is the exact result for the unrenormalized ZPE provided either by equation\,\eqref{eq:T00mexactk} or\,\eqref{eq:T00mexact}. The involved integrals are UV-divergent, so the integration measure $d^3k$ is promoted to $d^Nk$ in the context of dimensional regularization (DR)  with $N\equiv 3-2\varepsilon$, and the limit $\varepsilon\to 0$ is taken for granted. The integrals are of the type \eqref{eq:WeberSchafheitlin} in the Appendix \ref{sec:exactintegrals}. We consider their computation using the formulas \eqref{eq:WeberSchafheitlin2}. The corresponding limit $\varepsilon\to 0$ of these results is given by the expansions \eqref{eq:WeberSchafheitlin3}.
After some bulky calculations (see a summary in Appendix \ref{sec:calculationHankelIntegrals}), it can eventually be delivered in a rather compact form as follows (we omit hereafter the superscript dS for de Sitter):
\begin{equation}\label{eq:unrenorT00}
\begin{split}
    \braket{T_{00}^{\delta\phi}}(m)&=\frac{a^2}{64\pi^2 }\left\{ -\frac{m^2(m^2+12H^2\left(\xi-\frac{1}{6}\right))}{\varepsilon}-\frac{3}{2}\left[m^4-144H^4\left(\xi-\frac{1}{6}\right)^2\right]\right.  \\ &+ \left.m^2\left(m^2+12H^2\left(\xi-\frac{1}{6}\right)\right)\left(\gamma_E+\psi\left[\frac{3}{2}+\varsigma(\mu)\right]+\psi\left[\frac{3}{2}-\varsigma(\mu)\right]+\ln\left(\frac{H^2}{4\pi\Tilde{\mu}^2}\right)\right)\right\}\,.
\end{split}
\end{equation}
Recall that $\varsigma(\mu)$, in the argument of the digamma functions $\psi$,  is the order of the Bessel equation defined by \eqref{eq:nudefinition}. It is therefore a function of $H$ through the ratio $\mu\equiv m/H$, and depends on the parameters $m$ and $\xi$, i.e. $\varsigma=\varsigma({m}/{H};\xi)$, but for simplicity we just use $\varsigma(\mu)$. On the other hand,  parameter $\Tilde{\mu}$ in the log term of \eqref{eq:unrenorT00} is the 't Hooft mass unit in DR \cite{Collins84} (carrying here a tilde so as not to be confused with the previously defined $\mu$). However, we should emphasize that, in our renormalization scheme, this auxiliary  mass  plays no role since it cancels out completely at the level of the final results. It is not at all related to the renormalization scheme being used; recall that we do not use minimal subtraction, but off-shell adiabatic renormalization. Therefore, the only renormalization scale that remains in our calculation is $M$ (see subsequent formulas), until it is ascribed an appropriate physical meaning. The above result is manifestly UV-divergent since it contains an obvious pole at $\varepsilon=0$ (i.e. at $N=3$). However, $\braket{T_{00}^{\delta\phi}}(m)$ carries a finite part, which we are interested in,  of course, but the net finite contribution at the end of the calculation cannot be known until we perform renormalization by off-shell adiabatic subtraction \eqref{RenormalizedEMTScalar}, which will remove the poles and will contribute additional finite terms. Therefore, our next step is to compute
%$\langle T_{\mu\nu}^{\delta\phi}\rangle^{(0-4)}(M)$.
%It can be identified from the UV-divergent terms of Eq.\,\eqref{eq:4thorderdeSitterz} upon expanding $\omega_z(M)$ asymptotically in $z$ (cf. Appendix \ref{sec:FiniteEMTdeSitter}). One finds
  $\braket{T_{00}^{\delta\phi}}^{(0-4)}(M)=\braket{T_{00}^{\delta\phi}}_\mathrm{Div}^{(0-4)}(M)+\braket{T_{00}^{\delta\phi}}_\mathrm{Non-Div}^{(0-4)}(M)$, which we have split into divergent and non-divergent pieces.  In this way we can write Eq.\,\eqref{RenormalizedEMTScalar} as follows:
\begin{equation}\label{eq:subtractionFRandNonDiv}
    \braket{T_{00}^{\delta\phi}}_\mathrm{ren}(M)=\braket{T_{00}^{\delta\phi}}^{\mathrm{dS}}(m)-\braket{T_{00}^{\delta\phi}}^{(0-4)}(M)=\braket{T_{00}^{\delta\phi}}_{\mathrm{FR}}(M)-\braket{T_{00}^{\delta\phi}}^{(0-4)}_\mathrm{Non-Div}(M)\,,
\end{equation}
where
\begin{equation}\label{eq:FR}
    \braket{T_{00}^{\delta\phi}}_{\mathrm{FR}}(M)\equiv\braket{T_{00}^{\delta\phi}}^{\mathrm{dS}}(m)-\braket{T_{00}^{\delta\phi}}^{(0-4)}_{\mathrm{Div}}(M)
\end{equation}
is the finite remainder (FR) after the cancellation of the poles (see below) between these two UV-divergent terms.
Concerning $\braket{T_{00}^{\delta\phi}}_\mathrm{Div}^{(0-4)}$ it can be identified from the UV-divergent integrals in Eq.\,\eqref{eq:4thorderdeSitterz}.
%upon expanding $\omega_z(M)$ asymptotically in $z$ (cf. Appendix \ref{sec:FiniteEMTdeSitter}).
We find
\begin{equation}\label{eq:TMDiv}
\begin{split}
    \braket{T_{00}^{\delta\phi}}_\mathrm{Div}^{(0-4)}(M)&=\frac{H^2}{16\pi^2\tau^2}\int dz z^2 \left[ 4\omega_z-12\left(\xi-\frac{1}{6}\right)\left(\frac{1}{\omega_z}+\frac{M^2}{H^2\omega_z^3}\right)\right. \\
    &+ \left.\frac{2\Delta^2}{H^2\omega_z}-\frac{1}{2}\frac{\Delta^4}{H^4\omega_z^3}-6\left(\xi-\frac{1}{6}\right)\frac{\Delta^2}{H^2\omega_z^3}\right]\\
    &=\frac{H^2}{16\pi^2\tau^2}\left\{ \left[-\frac{1}{2}\frac{M^4}{H^4}\right] \Gamma_2 +\left[ -\frac{1}{2}(6\xi-1)\frac{M^2}{H^2}+\frac{\Delta^2M^2}{2H^4} \right]\Gamma_1\right.\\
    &+\left.\left[ -(6\xi-1)\frac{M^2}{H^2}-\frac{1}{4}\frac{\Delta^4}{H^4}-3\left(\xi-\frac{1}{6}\right)\frac{\Delta^2}{H^2} \right] \Gamma_0 \right\}(1-\varepsilon\ln\frac{M^2}{\Tilde{\mu}^2})\,,
\end{split}
\end{equation}
where the quantities $\Gamma_{0,1,2}$ contain poles for $\varepsilon\to 0$ and are defined in Appendix \ref{sec:DRintegrals}. The above expression acts as an overall counterterm to the unrenormalized ZPE given by \eqref{eq:unrenorT00}.
Regarding $\braket{T_{00}^{\delta\phi}}_\mathrm{Non-Div}^{(0-4)}(M)$ in \eqref{eq:subtractionFRandNonDiv}, it
collects  the manifestly convergent integrals appearing in  \eqref{eq:4thorderdeSitterz},  that is, those with a denominator $\omega_z^n$ with $n>3$. They can also be computed from the master formulas given in Appendix \ref{sec:DRintegrals}.

The explicit computation of \eqref{eq:FR} following the above procedure renders
\begin{equation}
\begin{split}
    \braket{T_{00}^{\delta\phi}}_\mathrm{FR}(M)=
    &\frac{a^2}{64\pi^2}\left\{ - \frac{1}{2}\left(6H^2(1-6\xi)+3m^2-M^2\right)\left(m^2-M^2+12H^2\left(\xi-\frac{1}{6}\right)\right)\right.\\
    &+\left.m^2\left[m^2+12H^2\left(\xi-\frac{1}{6}\right)\right]\left(\psi\left[\frac{3}{2}-\varsigma(\mu)\right]+\psi\left[\frac{3}{2}+\varsigma(\mu)\right]+\ln \frac{H^2}{M^2}\right)\right\}\,.
\end{split}
\end{equation}
We confirm its finiteness since the poles carried by the quantities $\Gamma_{0,1,2}$ have canceled out between the two subtracted expressions.  As for the non-divergent piece participating in Eq.\,\eqref{eq:subtractionFRandNonDiv}, we find:
\begin{equation}\label{eq:minusAnomaly}
\begin{split}
    \braket{T_{00}^{\delta\phi}}^{(0-4)}_\mathrm{Non-Div}(M)&=-\frac{a^2H^4}{960\pi^2}+\frac{9a^2H^4}{2\pi^2}\left(\xi-\frac{1}{6}\right)^2+\frac{m^2a^2H^2}{96\pi^2}+\left(\xi-\frac{1}{6}\right)\frac{3(m^2-M^2)a^2H^2}{8\pi^2}\,.
\end{split}
\end{equation}
%\begin{equation}
%\begin{split}
    %\braket{T_{00}^{\delta\phi}}^{(0-4)}_\mathrm{Non-Div}(H)&=-\frac{a^2H^4}{960\pi^2}+\frac{9a^2H^4}{2\pi^2}\left(\xi-\frac{1}{6}\right)^2+\frac{m^2a^2H^2}{96\pi^2}+\left(\xi-\frac{1}{6}\right)\frac{3(m^2-H^2)a^2H^2}{8\pi^2}\,.
%\end{split}
%\end{equation}
 Subtraction \eqref{RenormalizedEMTScalar} within the off-shell ARP can now be performed to produce the renormalized ZPE in de Sitter spacetime that we are after. Putting the pieces of \eqref{eq:subtractionFRandNonDiv} together, we reach the following result:
\begin{equation}\label{eq:RenormZPEdeSitter}
\begin{split}
    \braket{T_{00}^{\delta\phi}}_\mathrm{Ren}(M)&=\frac{a^2}{128\pi^2}\left\{ \left(M^2-3m^2+36H^2\left(\xi-\frac16\right)\right)\left(m^2-M^2+12H^2\left(\xi-\frac{1}{6}\right)\right)\right.\\
    &+\left.2m^2\left(m^2+12H^2\left(\xi-\frac{1}{6}\right)\right)\left(\psi\left[\frac{3}{2}-\varsigma(\mu)\right]+\psi\left[\frac{3}{2}+\varsigma(\mu)\right]-\ln \frac{M^2}{H^2}\right)\right\}\\
    &+\frac{a^2H^4}{960\pi^2}-\frac{a^2H^2m^2}{96\pi^2}-\frac{3a^2H^2(m^2-M^2)\left(\xi-\frac{1}{6}\right)}{8\pi^2}-\frac{9a^2H^4\left(\xi-\frac{1}{6}\right)^2}{2\pi^2}\,.
\end{split}
\end{equation}
This expression can be compared with its RVM counterpart, Eq.\,\eqref{RenormalizedEMT}, which was obtained in \cite{CristianJoan2020,CristianJoan2022a}. Note that the last term of \eqref{RenormalizedEMT} vanishes for de Sitter spacetime since $H=$ const. The comparison of the remaining terms is not totally straightforward due to the presence of the digamma functions in Eq\,.\eqref{eq:RenormZPEdeSitter}. While the above result corresponds to exact de Sitter space, in the RVM case the on-shell value was computed only up to fourth adiabatic order since it corresponds to  (spatially flat)  FLRW spacetime for which an exact solution is not available since $H$ is not constant.

To obtain the renormalized VED we utilize once more Eq.\,\eqref{RenVDE}. Inserting Eq.\eqref{eq:RenormZPEdeSitter} in it, we arrive at the desired result:
\begin{equation}\label{eq:VED_Ren}
\begin{split}
    \rho_\mathrm{vac}(M,H)&=\rho_\Lambda(M)+\frac{1}{128\pi^2}\left\{  \left(M^2-3m^2+36H^2\left(\xi-\frac{1}{6}\right)\right)\left(m^2-M^2+12H^2\left(\xi-\frac{1}{6}\right)\right)\right.\\
    &+\left.2m^2\left(m^2+12H^2\left(\xi-\frac{1}{6}\right)\right)\left(\psi\left[\frac{3}{2}-\varsigma(\mu)\right]+\psi\left[\frac{3}{2}+\varsigma(\mu)\right]-\ln \frac{M^2}{H^2}\right)\right\}\\
    &+\frac{H^4}{960\pi^2}-\frac{H^2m^2}{96\pi^2}-\frac{3H^2(m^2-M^2)\left(\xi-\frac{1}{6}\right)}{8\pi^2}-\frac{9H^4\left(\xi-\frac{1}{6}\right)^2}{2\pi^2}\,.
\end{split}
\end{equation}
The renormalized VED in de Sitter space depends on $M$ and $H$ as independent variables at this point. In particular, we can identify explicit powers $H^4$ and $H^2$ of the expansion rate. This is in contradistinction to its  RVM counterpart, Eq.\,\eqref{RenVDEexplicit}, where only the power $H^2$ appears explicitly before the setting $M=H$ is implemented.  In the de Sitter case, there is also a less trivial dependence on $H$ through the digamma functions, whose argument depends on the order of the Bessel function, $\varsigma=\varsigma(\mu)$, Eq.\,\eqref{eq:nudefinition}.

Despite $H$ being assumed to be constant as an initial primeval condition,  this ceases to be so as soon as the de Sitter vacuum starts to decay into relativistic matter and  the expansion rate  becomes time-evolving $H=H(t)$. In point of fact, we need to have an unstable de Sitter vacuum, as otherwise the inflationary epoch would be eternal. The transit between the two epochs is indeed possible, and therefore this inflationary mechanism successfully implements ``graceful exit'' (as we shall verify in Sect.\ref{sec:Inflation}). Because the powers $H^4$ and $H^2$ will  become dynamical,  the VED itself will be dynamical.   The role played by these powers is not the same in different epochs.  Power $H^4$ is dominant during inflation and is indeed the trigger of the early inflationary period, whereas power $H^2$ is the ruling hand over the post-inflationary regime dynamics, in particular for the late-time universe.  It is apparent that the VED appears in this unification scenario as a dynamical quantity for the entire cosmic history, spanning from early inflation to the late universe.  To summarize: in our context, inflation is caused by an unstable de Sitter period in which $H$ remains approximately constant for a very short time.  After that  the vacuum decays into radiation and the standard FLRW regime starts. In what follows, we presume this dynamical scenario caused by the unstable de Sitter vacuum decaying into  FLRW spacetime.  The details of such a decay and corresponding solution of the cosmological equations will be furnished in Sec.\ref{sec:Inflation}.

Being $H$ dynamical, let us estimate  the change of the VED  between different epochs of the cosmic expansion, which are tracked by the renormalization scale $M$. As noted previously, relating the value  of an effective quantity at different renormalization scales is, in fact, the main task of the renormalization group.  Following the RG prescription that we applied for the RVM in Sec.\,\ref{sec:RenVEDRVM}, we set $M$ at the value of the Hubble rate $H$  for each cosmic epoch under consideration\cite{CristianJoan2020,CristianJoan2022a,CristianJoan2022b,CristianJoanSamira2023}. Therefore, given the VED at one scale, say $M_0=H_0$ (which could represent e.g. the current universe), we can provide the VED at a different scale $M=H$ (representing another expansion history epoch, typically in our past) using Eq.\eqref{eq:VED_Ren}. Denoting once more $\rv(M,H)$ at $M=H$ simply as $\rv(H)$, and similarly for $\rv(H_0)$, the VED values at the two scales become related as follows:
\begin{equation}\label{eq:rvMMo}
\begin{split}
    &\rho_\mathrm{vac}(H)=\rho_\mathrm{vac}(H_0)+\frac{m^2}{64\pi^2}\left[m^2+12H^2\left(\xi-\frac{1}{6}\right)\right]\left(\psi\left[\frac{3}{2}-\varsigma(\mu)\right]+\psi\left[\frac{3}{2}+\varsigma(\mu)\right]\right)\\
    &-\frac{m^2}{64\pi^2}\left[m^2+12H_0^2\left(\xi-\frac{1}{6}\right)\right]\left(\psi\left[\frac{3}{2}-\varsigma(\mu_0)\right]+\psi\left[\frac{3}{2}+\varsigma(\mu_0)\right]\right)+\frac{m^4}{64\pi^2}\ln \left(\frac{H^2}{H_0^2}\right)\\
    &+\frac{1}{128\pi^2}\left\{ (H^4-H_0^4)\left[\frac{2}{15}+24\left(\xi-\frac{1}{6}\right)-144\left(\xi-\frac{1}{6}\right)^2\right]+m^2(H^2-H_0^2)\left[-\frac{4}{3}-48\left(\xi-\frac{1}{6}\right)  \right]\right\}\,,
\end{split}
\end{equation}
where $\mu=m/H$ and $\mu_0=m/H_0$. Notice that this expression is free of the isolated  quartic mass powers $\sim m^4$ appearing in Eq.\,\eqref{eq:VED_Ren} as they cancel in the difference. In fact, to derive the above equation, we have made use of the result  \eqref{SubtractionrL} for the difference\footnote{One can verify that the relations \eqref{SubtractionrL}-\eqref{Subtractionalpha}(which relate the couplings $G$, $\rL$ and $\alpha$ at two different scales) retain exactly the same form. The reason being that the terms in \eqref{eq:RenormZPEdeSitter} that are not in \eqref{RenormalizedEMT} are scale independent, i.e. do not depend on $M$, and hence cancel out in the subtraction. } $\rL(M)-\rL(M_0)$, which still depends on the  $\sim m^4\ln H^2$  contribution for $M=H$. However, this additional type of quartic mass effect is also unharmful. To see this, it is convenient to define the auxiliary function
\begin{equation}\label{eq:Psifun}
    \Psi(\mu) = \psi\left(\frac{3}{2}-\varsigma(\mu)\right)+\psi\left(\frac{3}{2}+\varsigma(\mu)\right)-\ln \mu^2
\end{equation}
and the corresponding expression for $\Psi(\mu_0)$.
In this way,  Eq.\,\eqref{eq:rvMMo} can be conveniently recast  as follows:
\begin{equation}\label{eq:rvMMoPsi}
\begin{split}
    &\rho_\mathrm{vac}(H)=\rho_\mathrm{vac}(H_0)+\frac{m^2}{64\pi^2}\left[12H^2\left(\xi-\frac{1}{6}\right)\right]\left(\psi\left[\frac{3}{2}-\varsigma(\mu)\right]+\psi\left[\frac{3}{2}+\varsigma(\mu)\right]\right)\\
    &-\frac{m^2}{64\pi^2}\left[12H_0^2\left(\xi-\frac{1}{6}\right)\right]\left(\psi\left[\frac{3}{2}-\varsigma(\mu_0)\right]+\psi\left[\frac{3}{2}+\varsigma(\mu_0)\right]\right)\\
    &+\frac{1}{128\pi^2}\left\{ (H^4-H_0^4)\left[\frac{2}{15}+24\left(\xi-\frac{1}{6}\right)-144\left(\xi-\frac{1}{6}\right)^2\right]-m^2(H^2-H_0^2)\left[\frac{4}{3}+48\left(\xi-\frac{1}{6}\right)  \right]\right\}\\
    &+\frac{m^4}{64\pi^2}\left\{\Psi(\mu)-\Psi(\mu_0)\right\}\,.
\end{split}
\end{equation}
Despite naive appearances, the last term $\sim m^4$ in this equation is not worrisome at all at low energies, since it is suppressed by $\Psi$, which brings in powers of the Hubble rate and inverse powers of $m$. To be quantitatively precise, let us expand the function \eqref{eq:Psifun} in the low energy limit $\mu\to\infty$, which obviously means $\mu=m/H\gg1$.  This expansion entails using the asymptotic limit of the corresponding digamma functions (cf. Appendix \ref{sec:exactintegrals}).  We find
\begin{equation}\label{eq:psiexpansion1}
    \Psi(\mu) = \frac{4(-1+9\xi)}{3\mu^2} + \frac{-11+240\xi-1080\xi^2}{15\mu^4} + \mathcal{O}\left(\frac{1}{\mu^5}\right)\,,\ \ \ \ \ \ \ \ \ \ \ \ \   (\mu\gg1)\,.
\end{equation}
Hence by making explicit the dependence on $H$ in the last term of \eqref{eq:rvMMoPsi}, we obtain
\begin{equation}\label{eq:psiexpansion2}
    m^4\Psi(\mu) = \frac43(-1+9\xi)\,m^2 H^2 + \frac{-11+240\xi-1080\xi^2}{15}\,H^4 + \mathcal{O}\left(\frac{H}{m}\,H^4\right) \ \ \ \ \ \  (H\ll m)\,.
\end{equation}
It follows that the  vacuum effects from  $\sim m^4 \Psi$ are not of  order $m^4$ at low energies, but at most  of order  $\sim m^2H^2\ll m^4$.
%\sim H^4$ at very high energies (where inflation occurs and is dominated by the power $H^4$ over the $m^2H^2$ terms, as we shall discuss in Sec.\ref{sec:Inflation}).
In the opposite limit $\mu\to 0$, i.e. at very high energies, the $m^4\Psi$ term is not dominant either, since then $H^4\gg m^4$.
Of the two limits, the most sensitive one insofar as concerns the cosmological constant problem is the low energy limit $H\ll m$ since it corresponds to the current universe, and here we could not afford contributions of order  $\sim m^4$ in the evolution of the VED.
%%%%%%%%%%%%%%%%%%%%%%%%%%%%%%%%%%%%%
\begin{figure}[t]
  \begin{center}
      \includegraphics[width=0.7\linewidth]{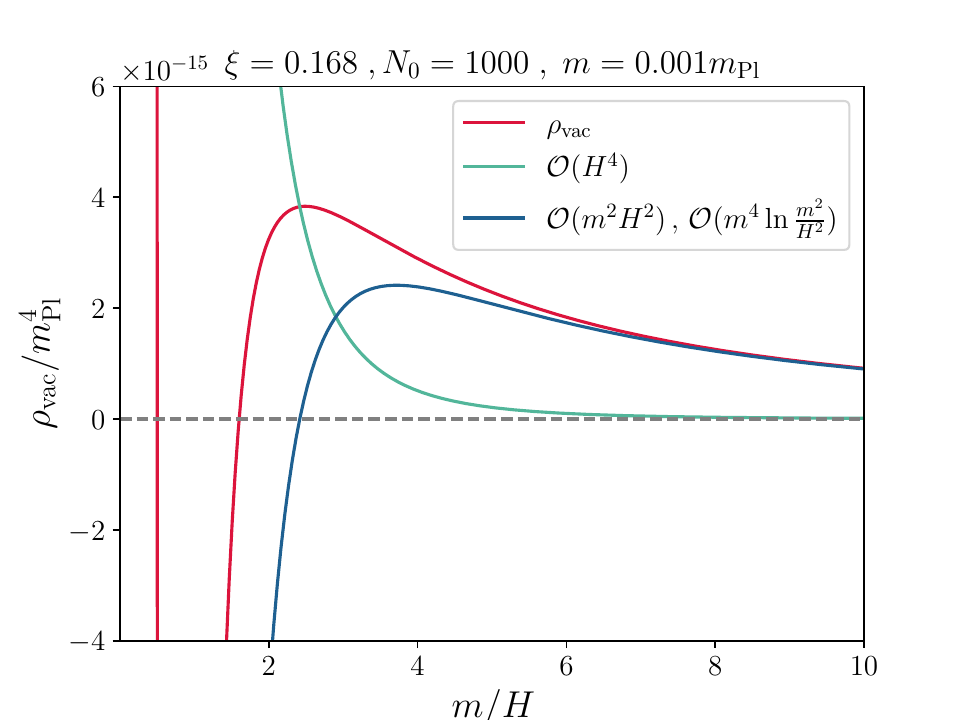}
\caption{Dependence of the VED, $\rho_\mathrm{vac}$, normalized to the Planck density in the inflationary period for the unstable Sitter case as a function of $m/H$. We plot specifically Eq. \eqref{eq:VED_Ren_final2}. The various contributions from the terms of $\mathcal{O}(H^4)$ and $\mathcal{O}(H^2)$ which add up to the total $\rho_\mathrm{vac}$ (red curve) are also displayed. We can take into account the contribution of more scalar fields with the same non-minimal coupling and mass by including a multiplicity factor $N_0$. For this numerical example, we take $N_0=1000$ with $\xi=0.168$ (for all scalar fields) and $m=0.001m_\mathrm{Pl}$ (see Sec.\ref{sec:Inflation} for the numerical analysis and notation).}
\label{fig:negative_rho}
\end{center}
\end{figure}
%%%%%%%%%%%%%%%%%%%%%%%%%%%%%%%%%%%%
Thus,  we find that the $m^4\Psi$ effects on the VED are  of the order of other existing contributions in \eqref{eq:rvMMoPsi}, which at low energy are just of the softer form $m^2 H^2$. This fact is particularly notable in our framework, as it obviously alleviates the fine-tuning problem in the CCP, as first observed in \cite{CristianJoan2020,CristianJoan2022a} in the case of the RVM, where a similar situation occurs, see Sec.\ref{sec:RenVEDRVM}.  So both unified QFT approaches to cosmic history prove free from fine tuning of the VED evolution.

As another consistency check at low energies between the de Sitter scenario and the RVM one, we can verify from Eq.\eqref{eq:VED_Ren}  that the on-shell value $M=m$ of the VED in the current universe ($H\ll m$) is $\rv (H,m)=\rho_\Lambda(m)$, i.e. it is independent of $H$ and coincides exactly with the on-shell value of the VED in the RVM, given by Eq.\,\eqref{RenVDEexplicit} at $M=m$. To check this, we need to use the low-energy expansion \eqref{eq:psiexpansion1}. Of course, in the de Sitter case the result is not exact since the $\sim H^4$ terms remain, but these can be ignored for the present universe.

During the inflationary epoch in the early universe,  the constant terms in Eq. \eqref{eq:rvMMoPsi} evaluated in the current universe (those depending on $H_0$) can be neglected, and therefore in practice the relevant expression of the VED for the study of inflation reads as follows:
\begin{equation}\label{eq:VED_Ren_final2}
\begin{split}
    &\rho_\mathrm{vac}^{\rm infl.}(H)=\frac{3m^2 H^2}{16\pi^2}\left(\xi-\frac{1}{6}\right)\left(\psi\left[\frac{3}{2}-\varsigma(\mu)\right]+\psi\left[\frac{3}{2}+\varsigma(\mu)\right]\right)\\
    &+\frac{1}{128\pi^2}\left\{ H^4\left[\frac{2}{15}+24\left(\xi-\frac{1}{6}\right)-144\left(\xi-\frac{1}{6}\right)^2\right]-m^2H^2\left[\frac{4}{3}+48\left(\xi-\frac{1}{6}\right)  \right]\right\}\\
    &+\frac{m^4}{64\pi^2}\,\left[\psi\left(\frac{3}{2}-\varsigma(\mu)\right)+\psi\left(\frac{3}{2}+\varsigma(\mu)\right)-\ln \mu^2 \right]\,.
\end{split}
\end{equation}
For future convenience, in  Fig.\ref{fig:negative_rho} we plot the above vacuum energy density as a function of $\mu=m/H$. The relative importance of the terms $H^4$, $m^2H^2$ and $m^4\ln(m^2/H^2)$ is clearly highlighted.  It can be seen that the VED can have both signs during its evolution before it stabilizes towards the FLRW regime. We have used typical numerical inputs which we have picked up from the physical region of parameter space, see Sec.\,\ref{sec:PhysicalRegion_xi}  for a detailed discussion of the numerical results. We shall come back to this plot later on.

On comparing the renormalized VED in de Sitter spacetime with the RVM, we observe that they are closely related but exhibit some differences. In particular,  the RVM result \eqref{DiffVEDphys} vanishes identically for $\xi=1/6$ whereas the de Sitter result \eqref{eq:VED_Ren_final2} does not vanish in the conformal limit ($\xi=1/6$ and $m=0$) and yields $\rv\propto H^4$ when we neglect the contributions from the current universe. The result obtained agrees with the analysis of \cite{Bunch:1978yq}. We provide a detailed explanation in Sect.\ref{sec:FullEMT}.

\subsection{Trace renormalization and vacuum pressure}\label{sec:RenPressure}

As indicated, apart from the VED we need the (renormalized) vacuum pressure from the quantized matter fields. The first step in its computation is to determine  the vacuum expectation value of the EMT trace. Using the EMT given by Eq.\,\eqref{EMTScalarField} and denoting the vacuum EMT trace by  $\langle T^{\delta \phi} \rangle\equiv  \langle 0| g^{\mu\nu} T^{\delta \phi}_{\mu\nu}|0\rangle$, we find\,\cite{CristianJoan2022a}
\begin{equation}\label{eq:QuantumTrace}
\langle T^{\delta \phi} \rangle=\left\langle \left(6\xi-1 \right)g^{\mu\nu}\nabla_\mu\delta \phi\nabla_\mu \delta \phi +2(3\xi-1)m^2\delta \phi^2+6\left(\xi-\frac{1}{6} \right)^2R\delta \phi^2+\left(\xi-\frac{1}{6} \right)R\delta \phi^2\right\rangle\,,
\end{equation}
where again only the local quadratic fluctuations of the fields are involved. Since these quadratic terms do not define regular distributions, the above trace is UV-divergent and requires renormalization.
To proceed, we need a  more explicit form for this expression, which  follows by writing the above result in terms of the mode functions $h_k(\tau)$ and corresponding derivatives $h'_k(\tau)$ defined in Sec.\ref{sec:Quantization}. We find:
\begin{equation}\label{eq:EMTtracemodes}
\begin{split}
    \braket{T^{\delta\phi}}&=\frac{1}{(2\pi)^3a^2} \int d^3k\left(-(6\xi-1)\frac{\calH^2}{a^2}+(6\xi-1)\frac{k^2}{a^2}+2(3\xi-1)m^2+6\left(\xi-\frac{1}{6}\right)^2R+\left(\xi-\frac{1}{6}\right)R\right)\abs{h_k}^2 \\
    &-\frac{(6\xi-1)}{a^2}\frac{1}{(2\pi)^3a^2}\int d^3k \abs{h_k'}^2+\frac{(6\xi-1)}{a^2}\frac{\calH}{(2\pi)^3a^2}\int d^3k(h_k{h_k'}^*+h_k'h_k^*) \,.
\end{split}
\end{equation}
Trading  the integration variable $k$ for the previously used $z=k|\tau|$ and at the same time borrowing the explicit form of the exact modes $h_k$ and $h'_k$ in terms of the Hankel functions  -- cf.  Sections \ref{sec:ExactModes1} and  \ref{sec:ExactModes2} --one finds after some calculations:
\begin{equation}\label{eq:EMTtraceHankel}
\begin{split}
    \braket{T^{\delta\phi}}^{\rm dS}(m)&=\frac{H^4}{8\pi}e^{-\pi\Im \varsigma} \int dz z^2 \left\{\left[(6\xi-1)z^2+\frac{9}{4}-\frac{51}{2}\xi+\frac{m^2}{H^2}(-2+6\xi)+72\xi^2\right]\abs{\mathbb{H}_\varsigma^{(1)}}^2\right. \\
    &\left.-(6\xi-1) z^2\abs{{\mathbb{H}_\varsigma^{(1)}}'}^2-\frac{3(6\xi-1)}{2}z\left( \mathbb{H}_\varsigma^{(1)} {\mathbb{H}_\varsigma^{(1)*}}'+\mathbb{H}_\varsigma^{(1)*}{\mathbb{H}_\varsigma^{(1)}}' \right) \right\}\,.
\end{split}
\end{equation}
Here we emphasize through the argument of the trace that it is evaluated at the particles' mass $m$, i.e. that it corresponds to the on-shell value, and we have expressed the result in terms of the ordinary Hubble rate $H$ defined from cosmic time.

Analogously to the computation of the ZPE, the renormalized trace of the EMT follows the same subtraction recipe \eqref{RenormalizedEMTScalar} as for the EMT itself. Adopting the same splitting structure \eqref{eq:subtractionFRandNonDiv} between finite and UV-divergent pieces, we have
\begin{equation}\label{eq:RentraceEMT}
\begin{split}
    \braket{T^{\delta\phi}}_\mathrm{Ren}(M)&=\braket{T^{\delta\phi}}_\mathrm{FR}(M)-\braket{T^{\delta\phi}}^{(0-4)}_\mathrm{Non-Div}(M)\,,
\end{split}
\end{equation}
where
\begin{equation}\label{eq:FREMTtrace}
    \braket{T^{\delta\phi}}_\mathrm{FR}(M)\equiv\braket{T^{\delta\phi}}^\mathrm{dS}(m)-\braket{T^{\delta\phi}}_\mathrm{Div}^{(0-4)}(M)
\end{equation}
is the  finite remainder of the off-shell subtraction of divergent terms in de Sitter space.  As in the ZPE case, $\braket{T^{\delta\phi}}^\mathrm{dS}(m)$ is computed from the exact integral formulae involving products of Hankel functions given in  Appendix \ref{sec:exactintegrals},  and on the other hand, $\braket{T^{\delta\phi}}_\mathrm{Div}^{(0-4)}(M)$ is obtained from the divergent part of the 4th-order adiabatic expansion of \eqref{eq:EMTtracemodes}.  We skip bulky computational details -- (cf. Eq.\eqref{eq:traceappendix} in Appendix \ref{sec:calculationHankelIntegrals}) and simply provide the respective final results after expanding them in the limit $\varepsilon\to 0$. On the one hand,
\begin{equation}\label{eq:tracedesitterexact}
\begin{split}
    \braket{T^{\delta\phi}}^{\rm dS}(m)&=\frac{H^4}{16\pi^2}\left\{ \frac{\mu^2(-2+\mu^2+12\xi)}{\varepsilon}\right\} - \frac{H^4(-2+\mu^2+12\xi)}{16\pi^2} \Bigg(-3-\mu^2+18\xi\\
    &+\mu^2\left[ \ln\left(\frac{H^2}{4\pi\Tilde{\mu}^2}\right)+\gamma_E+\psi\left(\frac{3}{2}-\varsigma\right)+\psi\left(\frac{3}{2}+\varsigma\right) \right]\Bigg)\,,
\end{split}
\end{equation}
where we recall that $\mu=m/H$. On the other hand, the adiabatic expansion up to $4th$ order renders:
\begin{equation}\label{eq:TraceDiv}
\begin{split}
    \braket{T^{\delta\phi}}^{(0-4)}_\mathrm{Div}(M)&=\frac{H^2}{4\pi^2}\left\{ -\frac{m^2M^2}{4H^2}\Gamma_1+\frac{\Delta^2m^2}{4H^2}\Gamma_0+\left(\xi-\frac{1}{6}\right)\left[\frac{3M^2}{2}\Gamma_1+\frac{9M^2+6\Delta^2}{2}\Gamma_0\right]\right\}\left(1-\epsilon\ln\frac{M^2}{\Tilde{\mu}^2}\right)\,,
\end{split}
\end{equation}
in which $\Gamma_i$ are the same quantities which appeared before containing poles for $\varepsilon\to 0$. From the last two results, the subtraction \eqref{eq:FREMTtrace} yields
\begin{equation}
\begin{split}
    \braket{T^{\delta\phi}}_\mathrm{FR}(M)&=\frac{1}{16\pi^2} \left(m^4-216H^4\left(\xi-\frac{1}{6}\right)+\frac{M^2}{H^2}(6\xi-1)-m^2H^2(-1+\frac{M^2}{H^2}+6\xi)\right.\\
    &-\left.m^2H^2(-2+\frac{m^2}{H^2}+12\xi)\left[ \ln\left(\frac{H^2}{M^2}\right)+\psi\left(\frac{3}{2}-\varsigma\right)+\psi\left(\frac{3}{2}+\varsigma\right) \right]\right)\,.
\end{split}
\end{equation}
As expected, it proves finite since the poles have canceled.  Finally, the non-divergent part of $\braket{T^{\delta\phi}}^{(0-4)}(M)$ can also be worked out:
\begin{equation}\label{eq:TNonDiv}
\begin{split}
    \braket{T^{\delta\phi}}_\mathrm{Non-Div}^{(0-4)}(M)&=\frac{1}{4\pi^2}\left\{ \frac{1}{120}\left[ 2H^4-15m^4-15M^4-5m^2(4H^2-6M^2)\right] \right.\\
    &+\left(\xi-\frac{1}{6}\right)\left[ -3M^2H^2-\frac{15}{2}(m^2-M^2)H^2\right]-\left.  72\left(\xi-\frac{1}{6}\right)^2H^4\right\} \, .
\end{split}
\end{equation}
The renormalized trace of the EMT is finally in reach and ensues from inserting the previous results into Eq.\eqref{eq:RentraceEMT}:
%\begin{equation}
%\begin{split}
    %\braket{T^{\delta\phi}}_\mathrm{Ren}(M)&=\frac{1}{16\pi^2}\left\{ -\frac{1}{15}H^4+\frac{3}{2}m^4+m^4\ln \frac{M^2}{m^2}+\frac{1}{2}M^4+\frac{2}{3}m^2H^2-2m^2M^2+ \right.\\
    %&+\left(\xi-\frac{1}{6}\right)\left[-12M^2H^2+12m^4+12m^2H^2\ln \frac{M^2}{m^2}+30M^4-30m^2M^2+18m^2H^2  \right]+\\
   % &+\left.\left(\xi-\frac{1}{6}\right)^2\left[ 144m^2H^2+288H^4 \right] \right\}
%\end{split}
%\end{equation}
\begin{equation}
\begin{split}
    \braket{T^{\delta\phi}}_\mathrm{Ren}(M)&=\frac{H^4}{16\pi^2} \left(\mu^4+(6\xi-1)(6+\frac{M^2}{H^2}-36\xi)-\mu^2(-1+\frac{M^2}{H^2}+6\xi)\right.\\
    &-\left.\mu^2(-2+\mu^2+12\xi)\left[ \ln\left(\frac{H^2}{M^2}\right)+\psi\left(\frac{3}{2}-\varsigma\right)+\psi\left(\frac{3}{2}+\varsigma\right) \right]\right) -  \braket{T^{\delta\phi}}_\mathrm{Non-Div}(M)\\
    &=\frac{1}{16\pi^2}\left\{ -\frac{1}{30}[2H^4-15m^4-15M^4-5m^2(4H^2-6M^2)]\right.\\
    &+\left(\xi-\frac{1}{6}\right)\left[ 12M^2H^2+30(m^2-M^2)H^2\right]+288H^4\left(\xi-\frac{1}{6}\right)^2\\
    &+m^4-216H^4\left(\xi-\frac{1}{6}\right)^2+6M^2H^2\left(\xi-\frac{1}{6}\right)-6m^2H^2\left(\xi-\frac{1}{6}\right)-M^2m^2\\
    &\left.-m^2\left[ m^2+12H^2\left(\xi-\frac{1}{6}\right)\right]\left[ \ln\left(\frac{H^2}{M^2}\right)+\psi\left(\frac{3}{2}-\varsigma\right)+\psi\left(\frac{3}{2}+\varsigma\right)\right]\right\}\,.
\end{split}
\end{equation}
After some rearrangements, the final result can be presented in a more compact form:
\begin{equation}\label{eq:trace_ren_exact}
\begin{split}
    \braket{T^{\delta\phi}}_\mathrm{Ren}(M)&=\frac{1}{16\pi^2}\left\{ -m^2\left[ m^2+12H^2\left(\xi-\frac{1}{6}\right)\right]\left[ \ln\left(\frac{H^2}{M^2}\right)+\psi\left(\frac{3}{2}-\varsigma\right)+\psi\left(\frac{3}{2}+\varsigma\right)\right]\right.\\
    &-\frac{1}{15}H^4+\frac{3}{2}m^4+\frac{1}{2}M^4+\frac{2}{3}m^2H^2-2m^2M^2\\
    &+\left.\left(\xi-\frac{1}{6}\right)\left[ -12M^2H^2+24m^2H^2\right]+72H^4\left(\xi-\frac{1}{6}\right)^2 \right\}\,.
\end{split}
\end{equation}
Following \cite{CristianJoan2022a,SolaPeracaula:2025yco}, the vacuum pressure can be obtained from the expression of the renormalized trace given above and that of the renormalized $00$th component (ZPE) of the vacuum EMT, Eq.\,\eqref{eq:RenormZPEdeSitter}, assuming the perfect fluid form for the vacuum EMT. An observer of $4$-velocity $ U^\mu$ measures a pressure $\Pv=\frac13\left(g^{\mu\nu}+U^\mu U^\nu\right) \langle T_{\mu\nu}^{\rm vac}\rangle$. In the rest frame of the observer, the $4$-velocity is $U^\mu=\left(1/\sqrt{-g_{00}},0,0,0\right)=(1/a,0,0,0)$ in the conformal metric and we have  $g^{00}+U^0 U^0=0$. Thus, the vacuum pressure is
\begin{equation}\label{eq:VacuumPressureDef}
\Pv(M)=\frac13 g^{kk}\langle T_{kk}^{\rm vac}\rangle_{\rm ren}(M)= \frac{\langle T_{ii}^{\rm vac}\rangle_{\rm ren}(M)}{a^2}= -\rho_\Lambda (M)+ \frac{\langle T_{ii}^{\delta \phi} \rangle_{\rm ren}(M)}{a^2}\,,
\end{equation}
(for any fixed $i=1,2,3$, and summation over $k=1,2,3$ is understood),
where use has been made of Eq.\,\eqref{eq:FullvacEMT}.
The vacuum isotropy also implies
\begin{equation}\label{eq:VacuumT11}
\frac{\langle T_{ii}^{\delta \phi} \rangle_{\rm ren}(M)}{a^2}=\frac{1}{3}\left(\langle T^{\delta \phi} \rangle_{\rm ren}(M)+\frac{\langle T_{00}^{\delta \phi} \rangle_{\rm ren}(M)}{a^2}\right)\,,
\end{equation}
since $\langle T^{\delta \phi} \rangle_{\rm ren}(M)=g^{\mu\nu}\langle T^{\delta \phi}_{\mu\nu} \rangle_{\rm ren}(M)=\left(-\langle T_{00}^{\delta \phi} \rangle_{\rm ren}(M)+3\langle T_{ii}^{\delta \phi} \rangle_{\rm ren}(M)\right)\frac{1}{a^2}$.
Finally,  using our definition \eqref{RenVDE} of VED to eliminate $\rL(M)$ in favor of $\rv(M)$ in the above equation, we arrive at the result\cite{CristianJoan2022a}:
\begin{equation}\label{eq:VacuunPressure}
    P_\mathrm{vac}(M)=-\rho_\mathrm{vac}(M)+\frac{1}{3}\left(\braket{T^{\delta\phi}}_\mathrm{ren}(M)+4\frac{\braket{T_{00}^{\delta\phi}}_\mathrm{ren}(M)}{a^2}\right) \,,
\end{equation}
which shows our contention mentioned above.
Clearly, the quantity in parentheses in the above expression is a quantum effect since it is absent in the classical theory. We need to check its possible impact in the QFT context. If it were non-vanishing, quantum effects would imply a departure from the canonical form of the vacuum EoS:
\begin{equation}\label{eq:renormEoS}
    P_\mathrm{vac}(M)=-\rho_\mathrm{vac}(M)\,.
\end{equation}
The expression in parentheses in Eq.\,\eqref{eq:VacuunPressure} can
be evaluated with the help of equations \eqref{eq:RenormZPEdeSitter} and \eqref{eq:trace_ren_exact}.  The result is amazingly simple as it turns out to be identically zero for de Sitter spacetime:
\begin{equation}\label{eq:identitynul}
    \braket{T^{\delta\phi}}^\mathrm{dS}_\mathrm{Ren}(M)+4\frac{\braket{T_{00}^{\delta\phi}}_\mathrm{Ren}^\mathrm{dS}(M)}{a^2}=0 \, .
\end{equation}
As a consequence, the relation between the renormalized vacuum pressure and the renormalized vacuum energy density in the de Sitter case adopts the canonical vacuum form \eqref{eq:renormEoS}, exactly as in the classical case.
Notice, however,  that this holds true only for the renormalized quantities, not for the unrenormalized ones. Indeed, to better assess this fact, let us repeat the above calculation before renormalization, i.e. before subtracting the corresponding off-shell pieces up to $4$th adiabatic order in equations \eqref{RenormalizedEMTScalar} and \eqref{eq:RentraceEMT}. In other words, let us directly use the on-shell results  \eqref{eq:unrenorT00} and \eqref{eq:tracedesitterexact}. We still meet a finite result, but a non-vanishing one, which we may call the quantum anomaly:
\begin{equation}\label{eq:anomaly1}
\begin{split}
    \mathcal{A}=\left( \braket{T^{\delta\phi}}^\mathrm{dS}(m)+4\frac{\braket{T_{00}^{\delta\phi}}^\mathrm{dS}(m)}{a^2} \right) = -\frac{H^2m^2}{32\pi^2} \left(-2+\frac{m^2}{H^2} +12\xi\right) \, .
\end{split}
\end{equation}
On the other hand, the adiabatic computation up to 4th adiabatic order with $H=$ const. yields the same anomaly
\begin{equation}
\begin{split}\label{eq:anomaly2}
    \mathcal{A}=\left( \braket{T^{\delta\phi}}^{(0-4)}(M)+4\frac{\braket{T_{00}^{\delta\phi}}^{(0-4)}(M)}{a^2} \right) = -\frac{H^2m^2}{32\pi^2} \left(-2+\frac{m^2}{H^2} +12\xi\right) \, ,
\end{split}
\end{equation}
where we have used \eqref{eq:TMDiv} and \eqref{eq:TraceDiv}.  As a consequence, the two effects cancel each other when we operate the off-shell subtraction, leaving  $0$ for the net quantum anomaly\footnote{The result  \eqref{eq:anomaly1} is consistent with the analysis of\cite{Firouzjahi:2023wbe}. However, the cancellation of $\mathcal{A}$ by subtracting the adiabatic contribution \eqref{eq:anomaly2}, which is responsible for the exact vacuum  EoS  $\wv=-1$,  is characteristic of our renormalization framework.}.
%This situation is similar to that in the RVM. In Ref.\,\cite{CristianJoan2022a} it was shown that before operating the off-shell subtraction \eqref{eq:RentraceEMT} the standard trace anomaly of the EMT is correctly reproduced. However, after subtraction the renormalized quantities do not depend on it, since the on-shell result  in Eq.\,\eqref{eq:RentraceEMT} is computed at $4$th adiabatic order exactly as the off-shell result.}.
This implies the result \eqref{eq:renormEoS}, which says that the renormalized equation of state (EoS) of the de Sitter vacuum is permanently fixed at $\wv=\Pv/\rv=-1$. It suggests that the symmetries of de Sitter spacetime are imprinted in the renormalized energy-momentum tensor.

This result is in contradistinction to the situation in the RVM for FLRW spacetime, where quantum effects break the canonical EoS of vacuum. In fact, the quantity in parentheses in Eq.\eqref{eq:VacuunPressure} is non-vanishing in that case, and hence the renormalized EoS of the quantum vacuum for the RVM departs slightly from $-1$, as shown in \cite{CristianJoan2022a} -- see also our subsequent discussion in Sec.\eqref{sec:EosVacuum}. However, all quantum effects breaking the canonical result \eqref{eq:renormEoS} are proportional to time derivatives of the Hubble rate and therefore vanish for $H=$ const. This means that during the inflationary regime at constant $H$,  the RVM vacuum remains strictly canonical, as in the de Sitter case, which demonstrates the consistency of the two approaches.  Now, away from the inflationary phase, the RVM vacuum deviates from $\wv=-1$, whereas the EoS of the de Sitter vacuum still remains at the canonical value.

A more detailed study of the evolution of the VED and the effective  EoS of the quantum vacuum will be considered in
Sections \ref{sec:VEDindS} and \ref{sec:EosVacuum}.  In particular, the consequences for the current universe can be rather significant and may constitute a smoking gun of this theoretical framework.
However, before being able to categorize the EoS behavior,  we need to elucidate what is the physical region operating in our unified scenarios, and this means particularly identifying the physical region for inflation in our parameter space.  In Sec. \ref{sec:Inflation},  we undertake the study of inflation within this class of dynamical vacuum scenarios. But before that let us consider the determination of the full renormalized vacuum EMT, not just the ZPE.

\subsection{Full renormalized EMT and Bunch-Davies result}\label{sec:FullEMT}

The full renormalized vacuum EMT in our off-shell ARP can be obtained from the sole knowledge of its $00$th component (ZPE) or of its trace. This is due to the maximal symmetry of de Sitter space and the fact, which we have proven in the previous section, that the de Sitter vacuum has an exact EoS $\wv=-1$.  Indeed, let us come back to Eq.\,\eqref{eq:VacuumPressureDef} for the vacuum pressure. On the other hand, the VED is given by Eq.\,\eqref{RenVDE}, which we take in renormalized form and hence VED and pressure are explicitly depending on the scale $M$.  Therefore, using \eqref{eq:renormEoS} it follows that $\langle T_{ii}^{\delta \phi}\rangle_\mathrm{Ren}(M)=-\langle T_{00}^{\delta \phi}\rangle_\mathrm{Ren}(M)$.  Overall, this implies that
\begin{equation}\label{eq:full EMT1}
\langle T_{\mu\nu}^{\delta \phi}\rangle_\mathrm{Ren}(M)=-g_{\mu\nu}\,\frac{\langle T_{00}^{\delta \phi}\rangle_\mathrm{Ren}(M)}{a^2}\,,
\end{equation}
where we have emphasized here that this relation is valid only for the renormalized quantities since \eqref{eq:renormEoS} is only valid for renormalized vacuum energy and pressure.
Taking the trace on both sides, this relation can also be recast as follows:
\begin{equation}\label{eq:full EMTtrace}
\langle T_{\mu\nu}^{\delta \phi}\rangle_\mathrm{Ren}(M)=\frac14\,g_{\mu\nu}\, \braket{T^{\delta\phi}}_\mathrm{Ren}(M)\,.
\end{equation}
Note, in particular,  that  Eq.\,\eqref{eq:identitynul} is the $00$th component of the previous relation.
It is easy to see that the VED in the de Sitter case can also be written in terms of the vacuum EMT trace:
\begin{equation}\label{eq:VacuuumTraceform}
\rv(M)=\rho_\Lambda (M)+\frac{\langle T_{00}^{\delta \phi}\rangle_\mathrm{Ren}}{a^2}=\rho_\Lambda (M)-\frac14  \braket{T^{\delta\phi}}_\mathrm{Ren}(M)\,.
\end{equation}
From the explicit results obtained in the previous sections we can verify that all of the above relations are consistently satisfied.
Therefore, the full expression for the renormalized vacuum EMT of de Sitter spacetime in our off-shell ARP scheme reads:
%\begin{equation}\label{eq:fullRenARP}
%\begin{split}
%    \braket{T_{\mu\nu}^{\delta\phi}}_\mathrm{Ren}(M)&=-\frac{g_{\mu\nu}}{128\pi^2}\left\{\left[M^2-3m^2+36H^2\left(\xi-\frac16\right)\right]\left[m^2-M^2+12H^2\left(\xi-\frac{1}{6}\right)\right]\right.\\
%    &+\left.2m^2\left[m^2+12H^2\left(\xi-\frac{1}{6}\right)\right]\left[\psi\left(\frac{3}{2}-\varsigma(\mu)\right)+\psi\left(\frac{3}{2}+\varsigma(\mu)\right)-\ln \frac{M^2}{H^2}\right]\right\}\\
%    &-g_{\mu\nu}\left\{\frac{H^4}{960\pi^2}-\frac{H^2m^2}{96\pi^2}-\frac{3H^2(m^2-M^2)\left(\xi-\frac{1}{6}\right)}{8\pi^2}-\frac{9H^4\left(\xi-\frac{1}{6}\right)^2}{2\pi^2} \right\}\,.
%\end{split}
%\end{equation}
\begin{equation}\label{eq:fullRenARP}
\begin{split}
    \braket{T_{\mu\nu}^{\delta\phi}}_\mathrm{Ren}(M)&=-\frac{g_{\mu\nu}}{64\pi^2}\left\{m^2\left[ m^2+\left(\xi-\frac{1}{6}\right) 12H^2\right]\left[\psi\left(\frac{3}{2}+\varsigma\right)+\psi\left(\frac{3}{2}-\varsigma\right)- \ln\left(\frac{M^2}{H^2}\right)\right]\right.\\
    &-\frac{2}{3}m^2H^2-\frac{3}{2}m^4-\frac{1}{2}M^4+2m^2M^2\\
    &+\left.\left(\xi-\frac{1}{6}\right)\left[ 12M^2H^2-24m^2H^2\right]-72H^4\left(\xi-\frac{1}{6}\right)^2 +\frac{H^4}{15} \right\}\,.
\end{split}
\end{equation}
The reader can check that in the on-shell case ($M=m$) the previous formula boils down to the following:
\begin{equation}\label{eq:BDresult}
\begin{split}
    \braket{T_{\mu\nu}}_\mathrm{BD}&=-\frac{g_{\mu\nu}}{64\pi^2}\left\{ m^2\left[m^2+\left(\xi-\frac{1}{6}\right)12H^2\right]\left[\psi\left(\frac{3}{2}+\varsigma\right)+\psi\left(\frac{3}{2}-\varsigma\right)-\ln \left(\frac{m^2}{H^2}\right)\right] \right.\\
    &-\left.m^2\left(\xi-\frac{1}{6}\right)12H^2-\frac{2m^2H^2}{3}-72H^4\left(\xi-\frac{1}{6}\right)^2+\frac{H^4}{15}\right\}\,.
\end{split}
\end{equation}
This is precisely the well-known Bunch-Davies result that was  obtained in \cite{Bunch:1978yq} using the manifestly covariant point-splitting regularization procedure, which, as previously indicated, is equivalent to using adiabatic regularization and hence it is the closest to our approach. The same expression was previously found as well by Dowker \& Critchley in \cite{Dowker:1975tf} using dimensional regularization and zeta-function techniques.
In the conformal massless limit ($\xi=1/6, m=0$) one expects, on simple dimensional grounds, that the trace of \eqref{eq:BDresult} should be proportional to $H^4$. In fact, we can verify it to be true, as the trace is then simply $\braket{T^{\delta\phi}}_\mathrm{\rm BD}=-\frac{1}{240\pi^2}\,H^4$. This trace, as could also be expected,  is nothing but the conformal anomaly, i.e. the anomalous trace of the EMT  in de Sitter spacetime\,\cite{BirrellDavies82}\footnote{See also \cite{CristianJoan2022a} for a detailed discussion in the RVM context. The de Sitter trace anomaly follows also from the general conformal anomaly formula (D.10) in the Appendix D of that reference, as can be easily checked.}.  As it is well known, the trace anomaly is a finite contribution that does not depend on any mass scale. It is easy to check that it ultimately comes from the finite piece \eqref{eq:TNonDiv}, after the latter is inserted into the renormalized trace \eqref{eq:RentraceEMT} and the conformal massless limit is applied for both scales $m$ and $M$.

The matching of \eqref{eq:fullRenARP} with \eqref{eq:BDresult} in the on-shell case is, of course, a highly nontrivial consistency check of our calculation.
However, the off-shell renormalized vacuum EMT that we have determined in our calculation is more general than the Bunch-Davies result \eqref{eq:BDresult}, since it contains an explicit dependence on the scale $M$ for arbitrary values of it, not just on-shell.  The dependence on such a floating scale $M$ is crucial in our framework and is characteristic of the renormalization group approach in QFT. It allows us to explore the evolution of the EMT, and in particular of the VED,  throughout  the cosmic expansion.

Finally, we point out that  the full renormalized vacuum EMT given by \eqref{eq:fullRenARP} depends on quartic  mass powers $\sim m^4$. However, one can check that the difference of EMT values at any two points $H$ and $H_0$ of the cosmic expansion is free of them,  in a way similar to that we have proven for  the VED in Eq.\,\eqref{eq:rvMMoPsi}  using \eqref {eq:psiexpansion2}. Once more this reflects the advantage of the off-shell renormalized theory, since the physical quantities can be compared at different points of the cosmic expansion and the relation is completely smooth, i.e. free of $\sim m^4$ effects that otherwise would require fine tuning. This feature is not possible if one just remains in the on-shell renormalized theory.

\section{$H^4$-inflation: decay of RVM and de Sitter vacuum}\label{sec:Inflation}
Inflation is a necessary stage of the very early cosmic evolution. In its absence, we could not understand the homogeneity and isotropy of the observed CMB, nor the high-level of spatial flatness of the universe at present without fine-tuning; and we would not understand its large amount of entropy today, an issue intimately connected with the horizon problem and hence with causality. Moreover, without inflation, we could not even figure out the origin of the structure formation that we see today. There is a wide variety of inflationary mechanisms in the literature, mostly based on the hypothetical existence of a scalar field called inflaton \cite{KolbTurner,LiddleLyth,RubakovGorbunov,Kallosh:2025ijd,Martin:2013tda}. In this context, which has been the traditional line of approach to the subject since its very inception\,\cite{Guth:1980zm,Linde:1981mu}, inflation is implemented by literally `sewing up' an exponentially expanding period in the very early history of the universe before the ordinary FLRW regime starts. Such a period is usually attributed to {\it ad hoc} inflaton dynamics of various sorts.  Nevertheless, we shall not adopt this approach here, since we aim at a unified QFT picture that involves all stages of the cosmic history from inflation to our days.  One such mechanism is provided by the RVM and is described in detail in the recent study \cite{SolaPeracaula:2025yco}, see also previous considerations along these lines in  \cite{{Fossil2008},JSPRev2015,JSPRev2022} and \cite{CristianJoan2022a,CristianJoan2022b,CristianJoanSamira2023}. On the other hand, it turns out that the time-honored de Sitter spacetime when decaying into relativistic matter can also lead to a very similar unified  QFT scenario, as we shall demonstrate here in detail. To better exemplify the analogies and differences between the two unification pictures (RVM and de Sitter), we discuss both but shall mainly concentrate on unstable de Sitter vacuum decay since this is the main aim of the current work.

In the previous section, we have shown that the EMT of the de Sitter spacetime can be consistently renormalized using the off-shell adiabatic renormalization prescription (ARP), defined by Eq.\eqref{RenormalizedEMTScalar}. Now in pure de Sitter one has $H=$ const. However, let us suppose that this state holds good only as the starting point of the cosmological history (or at least the part that can be dealt with within QFT without yet the participation of quantum gravity \footnote{ We may assume e.g. that it comes from the `low-energy' legacy of string theory around the Planck scale; see a possible connection of this sort in \cite{Mavromatos:2020kzj,Mavromatos:2020crd,PhantomVacuum2021}.}), and then decays. This is a reasonable assumption, since  a realistic version of the Universe requires that $H$ become dynamical, as otherwise we could not reproduce the subsequent FLRW eras until our days.  The transition is possible provided that the huge energy stored in de Sitter background decays into relativistic matter. Thus, the pure de Sitter vacuum is used here only as an initial state where we can renormalize an exact quantum field theory. However, this pure state is ephemeral and immediately starts to decay into an incipient radiation epoch. Inflation occurs within that short lapse of time, in which $H\simeq$ const., where the de Sitter spacetime is only approximate. We shall not enter the microscopic details of the transition as this would entail model-dependent assumptions which are not necessary for our considerations,  and in addition they would introduce new phenomenological parameters. It will suffice to perform a description of the transit between the two epochs on general thermodynamical grounds, but the initial de Sitter vacuum that kicks out the entire cosmic process is  nonetheless treated within quantum field theory.

Once the vacuum energy density in de Sitter spacetime is renormalized through the off-shell ARP, we have at hand a well-defined QFT system endowed with full fledged predictive capability, which is set up in the very early stages of cosmic history. As indicated (and will be shortly demonstrated), an immediate physical consequence of this setup is the existence of a new mechanism of fast inflation naturally triggered by a short period in which the Hubble rate stays approximately constant ($H\simeq H_I=$ const.) and very large  (presumably around a GUT scale or even in the neighborhood of the Planck scale) before it decays into radiation.  Notice that this mechanism is entirely different from Starobinsky inflation\cite{Starobinsky:1980te}, for which it is $\dot{H}$ (the cosmic time derivative of $H$) rather than $H$ itself, the relevant quantity that remains constant during the inflationary period; see \cite{{JSPRev2015}} for a detailed discussion.

The quantized matter fields (bosons and fermions) in a generic GUT constitute the material support for this QFT approach. These fields are acting at the quantum level through loop effects. Therefore, no \textit{ad hoc} classical scalar field must be introduced to play the role of an ``inflaton'' in our framework.

\subsection{Two inflationary scenarios: RVM and  de Sitter}\label{sec:UniScenarios}
In the following, we describe the two scenarios that will be discussed in the current work, which are both based on  a mechanism of $H^4$-inflation.
\newline

 $\bullet$ i) RVM-inflation.
 \newline

It  was studied in detail in \cite{SolaPeracaula:2025yco} and here we briefly review  the basic properties.  In this scenario, inflation is brought about by dynamical contributions imprinted on the background spacetime  by the quantum matter effects. It is characterized by the existence of a short period in the very early Universe where  $H$ remains essentially constant.
The inflation mechanism is driven by powers of $H$ greater than $H^2$, as the latter is only used to describe the FLRW regime. These higher powers appear as quantum effects on top of the classical action.  The simplest power capable of triggering inflation and complying with general covariance is $H^4$ (odd powers are not compatible with covariance\footnote{The reason being that we need an even number of derivatives of the scale factor to contract with the metric tensor. For $H\simeq\,$ const., this is possible only for $H^4, H^6, \dots$  Nevertheless, while powers higher than $H^4$ depend on dimensionful coefficients, $H^4$ has a dimensionless coefficient  and hence is the canonical option.}, so e.g. $H^3$ must be discarded). The RVM form for the VED during the inflationary regime was advanced in Eq.\eqref{eq:VEDinfl}, but it can be conveniently rewritten as follows\footnote{For simplicity, we  illustrate the inflationary mechanism by taking a single but non-minimally coupled quantum scalar field, that is, described by the action \eqref{eq:Sphi}. A generalization for $N_0$ scalar fields is straightforward. If they have the same mass and non-minimal coupling,  one simply includes a multiplicity factor $N_0$ in Eq.\,\eqref{eq:VEDinf2l}. On the other hand,
the effect of quantized fermions can be computed as well, but we prefer to simplify our presentation here in order not to make it too cumbersome; see e.g. \cite{CristianJoanSamira2023}.}:
\begin{equation}\label{eq:VEDinf2l}
\rho_\mathrm{vac}^{\rm RVM}(H)=\frac{3\left( \xi-\frac{1}{6}\right)}{16\pi^2}\, H^4+\frac{3{\nu}(H)}{8\pi}\, \mpl^2H^2\,,
\end{equation}
where
\begin{equation}\label{eq:nuH}
{\nu}(H)\simeq \frac{1}{2\pi}\,\left(\xi-\frac{1}{6}\right)\,\frac{m^2}{\mpl^2} \left(-1+\ln\frac{m^2}{H^2}\right)=\epsilon \left(-1+\ln\frac{m^2}{H^2}\right)\,,
\end{equation}
and $\epsilon$ is as in \eqref{eq:epsilonparameter}.
The solution of RVM-inflation is discussed in full detail in \cite{SolaPeracaula:2025yco}, but it will be useful to review it here since it bares resemblance to that of the de Sitter scenario.
\newline

$\bullet$ ii)  de Sitter inflation.
\newline

Notice that the above formula for the VED in the early universe would also be valid for the de Sitter vacuum,  since it only depends on $H$, which is constant during the short period of inflation.  So, if we would approach unstable de Sitter vacuum within the RVM, the above scenario would be the solution.  However, in the case of de Sitter it is possible to go one step beyond since the field equations can be solved exactly and there is no need to find an adiabatic approximation to it. Therefore, our starting point now is the exact solution to the field equations for de Sitter spacetime that we have found in the previous section. Here the Universe is supposed to begin in a phase where $H=$ const. (which allowed us to derive such an exact solution) and then we let it decay into radiation.  Power $H^4$ is again involved in the VED structure, see Eq.\,\eqref{eq:rvMMoPsi}, and acts as the driving force of inflation. However, as in the RVM case, we shall retain the $H^2$ term  as well in the study of the early universe since we wish to assess the effect from the subleading powers.  Given the fact that $H^2\gg m^2$ during inflation ($m$ being a typical particle mass in the GUT), and that $H_0^2\ll m^2$ for the current Hubble parameter, we can neglect the contribution from the VED at present in all our discussions about inflation. Hence, within a very good approximation, the VED at the inflationary phase can be expressed as indicated in Eq.\,\eqref{eq:VED_Ren_final2}. For convenience, we rephrase it as follows:
\begin{equation}\label{eq:VED_inflation_dS}
\begin{split}
    &\rho_\mathrm{vac}^{\rm infl.}(H)=
    \frac{ H^4}{128\pi^2}\left[\frac{2}{15}+24\left(\xi-\frac{1}{6}\right)-144\left(\xi-\frac{1}{6}\right)^2\right]
    -\frac{m^2H^2}{128\pi^2}\left[\frac{4}{3}+48\left(\xi-\frac{1}{6}\right)  \right]\\
    &+\frac{3 m^2 H^2}{16\pi^2}\left(\xi-\frac{1}{6}\right)\left(\psi\left[\frac{3}{2}-\varsigma(\mu)\right]+\psi\left[\frac{3}{2}+\varsigma(\mu)\right]\right)
    +\frac{m^4}{64\pi^2}\Psi(\mu)\,,
\end{split}
\end{equation}
where we have used Eq.\,\eqref{eq:Psifun}.
Once more it is understood that a multiplicity factor $N_0$  can be included to account for additional scalar field contributions with the same mass and non-minimal coupling (or, in general, a sum $\Sigma_i$ over each independent term with different masses $m_i$ and non-minimal couplings $\xi_i$), but as said we restrict  the presentation to the monocomponent scenario to ease the notation. We shall make use of field multiplicities only in numerical examples.
The above result is to be contrasted with the VED in the early universe within the RVM, which is given by Eq.\,\eqref{eq:VEDinf2l}.
In both cases we have a dominant power $H^4$ and subdominant terms of the type $m^2H^2$ and $m^2H^2\ln(m^2/H^2)$. However, as previously noted, in the de Sitter case the VED does not vanish for $\xi=1/6$ due to the trace anomaly (cf. Sec. \ref{sec:FullEMT}). Furthermore, Eq.\,\eqref{eq:VEDinf2l} vanishes for $H=m$ whereas \eqref{eq:VED_inflation_dS} does not. The last feature was expected since in the de Sitter case we start from an exact solution, whose form in general will differ from the subtracted adiabatic expression in Eq.\,\eqref{RenormalizedEMTScalar}. For the RVM, instead, there is an adiabatic expansion involved in the two terms of Eq.\,\eqref{RenormalizedEMTScalar} and hence it vanishes for $M=m$, if the on-shell result is computed at $4$th order too.

In an analogous manner, and keeping some parallelism with the above RVM formula, we can rewrite the result \eqref{eq:VED_inflation_dS} for the renormalized VED in de Sitter scenario as follows:
\begin{equation}\label{eq:VEDdeSitterl}
\rho^{\rm dS}_\mathrm{vac}(H)=\frac{\alpha} {2\pi^2}\, H^4+\frac{3{\tnu}(H)}{8\pi}\, \mpl^2H^2\,,
\end{equation}
where we have defined
\begin{equation}\label{eq:alphadS}
    \alpha=\frac{1}{480}+\frac38 \left(\xi-\frac16\right)- \frac94 \left(\xi-\frac16\right)^2\,.
\end{equation}
In the inflationary regime, it is hard to define $\Tilde{\nu}(H)$ as an overall coefficient of $H^2$  in Eq.\,\eqref{eq:VED_inflation_dS} due to the non-trivial dependence of the digamma functions on $H$, especially in the case of the last term of that equation. However, such a term is proportional to $\sim m^4$ and can be neglected at high energies during the inflationary regime, where $H^4\gg m^4$. Therefore, in a good approximation in this regime,  we can define
\begin{equation}\label{eq:nutildeH}
\begin{split}
    \Tilde{\nu}(H) &= \frac{m^2}{2\pi m_\mathrm{Pl}^2}\left(\xi-\frac{1}{6}\right)\left(\psi\left[\frac{3}{2}-\varsigma(\mu)\right]+\psi\left[\frac{3}{2}+\varsigma(\mu)\right]\right)-\frac{m^2}{\pi m_\mathrm{Pl}^2}\left[\frac{1}{36}+\left(\xi-\frac{1}{6}\right)  \right] \, .
\end{split}
\end{equation}
Recall that $\tilde{\nu}(H)$ depends on $H$ through the argument $\mu=m/H$ in the digamma functions. The above expression, however,  varies slowly with $H$ and it can be taken approximately constant. We shall come back to this issue in the next sections, where we discuss the solution to the cosmological model and especially after we identify the parameter space available for inflation.

\subsection{Decay of the quantum vacuum into radiation}\label{sec:solutions}

We shall next consider approximate analytical solutions to the decay of the quantum vacuum into relativistic matter in the two scenarios described above.   We proceed in turn.

\vspace{0.3cm}
$\bullet$ i) Approximate analytical solution of Running Vacuum
\newline

For RVM-inflation,  the cosmological equations can be solved in a close analytical form with the help of reasonable approximations\,\cite{SolaPeracaula:2025yco}. For example,  we can assume that ${\nu}(H)$ in Eq.\,\eqref{eq:nuH} is  constant since the log evolution with $H$ is very slow.   In this case, equation \eqref{eq:VEDinf2l} can be considered as a particular case of the VED template
\begin{equation}\label{eq:rvm1}
\rv(H)=\frac{3}{\kappa^2} \left( c_0+\nu H^2+ \frac{H^4}{H_I^2} \right)\,,
\end{equation}
where $\kappa^2=8\pi G_N=8\pi/\mpl^2$ and $H_I$ is a formal parameter at this point which will be related to the Hubble rate at the inflationary time. The structure \eqref{eq:rvm1}  for the VED actually embodies the unified description of the cosmic history in the RVM context from the very early times to our time. In the absence of any power of $H$, the parameter $c_0$ in that equation is related to the cosmological constant simply  as $\CC=3c_0$. However, in the presence of these powers, this is no longer  true as the measured value of the cosmological term at present ($H=H_0$) is $\CC=\kappa^2\rv(H_0)$. Since $\nu$ is small and $H\ll H_I$ in the late Universe, the previous relation between $\CC$ and $c_0$ still holds approximately, but in general $\rv(H)$ evolves with the expansion, and so does the dynamical CC term $\Lambda(H)=\kappa^2\rv(H)$.  In the current Universe, we can neglect the $H^4$ term and we recover the simpler form \eqref{eq:RVMcanonical}. In the early Universe, however,  the $H^4$ power becomes dominant and we may completely neglect $c_0$, which as we have seen is of the order of the measured CC value today, but we can still keep the $H^2$ term as a subdominant correction. The coefficient $\nu$ for the $H^2$ term is a slowly varying function of $H$, see Eq.\,\eqref{eq:nueff2}. During the inflationary period, it is natural to take it at the constant value $H=H_I$, which we call it $\nu_I$. Hence, neglecting low energy terms,
\begin{equation}\label{eq:nueffinfRVM}
\nu_I\equiv\nu(H_I)=\frac{1}{2\pi}\,\left(\xi-\frac16\right)\,\frac{m^2}{\mpl^2}\left(-1+\ln \frac{m^2}{H_I^{2}}\right)\,.
\end{equation}
Thus, during the inflation stage, $\nu$ in Eq.\,\eqref{eq:rvm1} means $\nu(H_I)$ as given above. On the other hand, the scale of inflation ${H}_I$ for the RVM can be simply inferred from comparing equations \eqref{eq:VEDinf2l} and \eqref{eq:rvm1}:
\begin{equation}\label{eq:HIvalue}
H_I=\sqrt{\frac{2\pi}{\xi-\frac{1}{6}} }\ m_\mathrm{Pl}\,.
\end{equation}
Despite that there is no severe restriction on $\xi$ from inflation in the RVM, we can see that the condition $\xi>1/6$ is necessary. Therefore, in particular,  both the minimal coupling $(\xi=0)$ and the conformal coupling ($\xi=1/6$) are excluded. It certainly justifies having performed our calculation for general non-minimal coupling from the very beginning. In the de Sitter scenario, this is even more justified, as we shall see.

The cosmological equations to solve are the following:
\begin{equation}
\begin{split}\label{eq:{eq:friedmann}n}
3H^2&=\kappa^2 (\rho_\mathrm{rad}+\rho_\mathrm{vac}(H)) \; , \\
3H^2+2\dot{H}&=-\kappa^2 (P_\mathrm{vac}(H)+\frac{1}{3}\rho_\mathrm{rad}) \,,
\end{split}
\end{equation}
where $\rho_\mathrm{vac}(H)$ is given by \eqref{eq:rvm1}.
At this stage of cosmic history we just have vacuum energy exchanging energy with relativistic particles (i.e. radiation, with EoS $w_{\rm rad}=1/3$).
As shown in \cite{SolaPeracaula:2025yco} -- see also our discussion later on in Sec.\,\ref{sec:EosVacuum} --- during inflation the RVM vacuum  satisfies the traditional equation of state $\Pv(H)=-\rv(H)$ up to terms which vanish when $H=$ const.   This fact is very useful for easing an approximate analytical solution.  Indeed, the above equations can then be combined as follows:
\begin{equation}\label{eq:Hdiffeq}
\dot H+2H^2=\frac{16\pi G}{3 } \rho_{\rm vac}(H)\,.
\end{equation}
It is advantageous to rewrite this equation using the scale factor rather than the cosmic time as the independent variable. Defining the prime derivative  $^\prime=d/da$, we find:
\begin{equation}\label{eq:diffHdeSitter2}
    a H H' + 2 H^2=2\left(c_0+\nu H^2+\frac{H^4}{H_I^2}\right)\,.
\end{equation}
For the solution of the cosmological equations in the early Universe
we may neglect the small parameter $c_0$ (related to today's measured value of the cosmological term), and we find:
\begin{equation}\label{eq:diffHdeSitter3}
    a H' + 2 (1-\nu)\, H -2\frac{H^3}{H_I^2}=0\,.
\end{equation}
An analytical solution to this equation for $H$ can easily be found in terms of the scale factor, and then also for the corresponding energy densities of vacuum and radiation. Altogether, after a straightforward calculation, the results read as follows:
\begin{align}
H(\hat{a})&=\frac{\Tilde{H}_I}{\sqrt{1+\hat{a}^{4(1-\nu)}}}\,,\label{eq:infaprox}\\
\rho_\mathrm{vac}(\hat{a})&={\rho}_I\frac{1+\nu\hat{a}^{4(1-\nu)}}{[1+\hat{a}^{4(1-\nu)}]^2}\label{eq:infaprox2}\,,\\
\rho_\mathrm{rad}(\hat{a})&={\rho}_I(1-\nu)\frac{\hat{a}^{4(1-\nu)}}{[1+\hat{a}^{4(1-\nu)}]^2}\label{eq:infaprox3}\,,
\end{align}
where we have introduced the total energy density at the early start of the inflationary period: ${\rho}_I=\frac{3}{\kappa^2}\Tilde{H}_I^2$. In the above, we have used the rescaled variables $\hat{a}=a/a_*$ and $\Tilde{H}_I=\sqrt{1-\nu}H_I$, where $a_*$
is related to the equality point $a_{\rm eq}$, defined to be the value of $a$ where the vacuum energy density at the end of the inflationary epoch equals the radiation energy density: $\rv(a_{\rm eq})=\rho_r(a_{\rm eq})$. This point fixes the integration constant and then $a_*$ is defined  as follows:
\begin{equation}
\left(\frac{a_{\rm eq}}{a_*}\right)^{-4(1-\nu)}=1-2\nu\,.
\end{equation}
Notice that in practice the difference between $a_*$ and $a_{\rm eq}$ is very small since $|\nu|\ll1$, hence $a_*\simeq a_{\rm eq}$.
 Furthermore, since $\dot{H}=-2H_I^2(1-\nu) \hat{a}^{4(1-\nu)}/(1+\hat{a}^{4(1-\nu)})^2$, we have $|\dot{H}/H^2|\propto\hat{a}^4\ll1$ for $\hat{a} \ll 1$ (again because $|\nu|\ll1$). It follows that we can safely neglect $\dot{H}\approx 0$ and successive derivatives during inflation; i.e. there is indeed slow roll in this period.

Interestingly, we observe that the initial point $a=0$ is nonsingular in this framework since the Hubble function and the energy densities are well-defined functions taking finite values at that point: $H(0)=\Tilde{H}_I$, $\rho_\mathrm{vac}({0})={\rho}_I=\frac{3}{\kappa^2}\Tilde{H}_I^2$ and $\rho_\mathrm{rad}(0)=0$. In other words, RVM-inflation is characterized by a non-singular de Sitter phase.
Although in previous phenomenological studies \cite{BLS2013,Perico:2013mna,JSPRev2015,Sola:2015csa,BLS2015,Yu2020} it was possible to postulate these equations on semiqualitative grounds, in \cite{SolaPeracaula:2025yco}
the VED structure \eqref{eq:rvm1} was fully substantiated in the context of QFT in curved spacetime.

Let us also note from the above analytical solution that the period of inflation connects smoothly with the radiation-dominated epoch. In fact, for $\hat{a}\gg 1$ (or  $a\gg a_*$) the following asymptotic behavior obtains: $\rho_\mathrm{rad}(\hat{a})\simeq {\rho}_I(1-\nu) \hat{a}^{-4(1-\nu)}\simeq \rho_I a_*^4 \,a^{-4}$ for small $|\nu|\ll1$. This is essentially the standard evolution law of the radiation energy density: $\rho_\mathrm{rad}=\rho_\mathrm{rad}^0{a}^{-4}$. From this fact a useful estimate for $a_*$ ensues which depends on the current cosmological parameters and the inflationary scale:
\begin{equation}\label{eq:astar}
a_{*}\sim
\left(\Omega_{\rm rad}^0\,\frac{\rco}{\rho_I}\right)^{\frac{1}{4}}\,.
\end{equation}
Here $\Omega_{\rm rad}\sim 10^{-4}$ and $\rco\sim 10^{-47}$ GeV$^4$. The value of $\rho_I$ ranges from $M_X^4$ (where $M_X\sim 10^{16}$ GeV is a typical GUT scale)  up to $\rho_I\sim \mpl^4$ for the Planck scale $\mpl\sim 10^{19}$ GeV.
Another remarkable feature that can be read off from equations \eqref{eq:infaprox2} and \eqref{eq:infaprox3} is the following: for $\nu\hat{a}^{4(1-\nu)}\gg1$ the vacuum energy is suppressed by the small factor $\nu$ (i.e. $\rho_\mathrm{vac}/\rho_\mathrm{rad}\sim \nu$) and hence the primordial BBN period can proceed standard, i.e. in accordance with the usual thermal history\cite{{Asimakis:2021yct}}.

\vspace{0.5cm}

$\bullet$ ii) Approximate analytical solution of Unstable de Sitter Vacuum
\newline

The RVM approach is unavoidable in the FLRW case since an exact solution of the field equations is impossible\cite{CristianJoan2020,CristianJoan2022a,CristianJoan2022b,CristianJoanSamira2023,SolaPeracaula:2025yco}. As indicated previously, if we would treat unstable de Sitter vacuum also within the  effective RVM approach, we would start from $H=$ const in the early universe  and let it decay into radiation in exactly the same manner as in the previous scenario. Indeed, the expression \eqref{RenormalizedEMT} for the ZPE would be the same, except for the last term that depends on the time derivatives of $H$.  But this does not affect our inflation mechanism, which is precisely based on having a period where $H\simeq$ const.  Thus, the corresponding VED in the early universe adapts once more to the RVM form \eqref{eq:VEDinfl} and the decaying solution is given anew by equations \eqref{eq:infaprox}-\eqref{eq:infaprox3}. However,  for the de Sitter case we can do much more than that. In fact, because $H=$ const. in the initial state, we were able to solve the model as an exact QFT in the previous sections and produce analytical expressions for $\rho_{\rm vac}(H)$ and $P_{\rm vac}(H)$. This is precisely the approach that we refer to specifically  as `unstable de Sitter vacuum'.  Equipped with the exact and renormalized solution, we let de Sitter decay into radiation using the set of equations \eqref{eq:{eq:friedmann}n}, from which we can obtain a differential equation for $H(a)$.  Expressing the result once more in terms of derivatives with respect to the scale factor, it gives:
\begin{equation}\label{eq:diffHdeSitter}
    a H H' + 2 H^2 =\frac{16\pi G}{3}\rho^{\rm dS}_\mathrm{vac}(H)\,,
\end{equation}
%where the VED in the second term is now given by Eq.\,\eqref{eq:VEDdeSitterl}.
where the VED in the second term is now given by the full expression \,\eqref{eq:VED_inflation_dS}.  This differential equation can only be solved numerically.

Despite that the VED is now more complicated than in scenario i), a reasonable analytical solution is still possible using the simplified form \eqref{eq:VEDdeSitterl} for the VED. We can then approximate Eq.\,\eqref{eq:diffHdeSitter} as follows:
\begin{equation}\label{eq:diffHdeSitterAprox}
    a H H' + 2 H^2 =2\left(\tilde{\nu}_I H^2+ \frac{4\alpha}{3\pi \mpl^2}\,H^4\right)\,,
\end{equation}
which adapts to the  template \eqref{eq:diffHdeSitter2} for $c_0=0$, provided that the coefficient $\tilde{\nu}_I$ is constant. The latter is defined as the value of the coupling \eqref{eq:nutildeH} at $H=H_I$, namely
\begin{equation}\label{eq:nutildeHinfl}
\begin{split}
    \tilde{\nu}_I\equiv\tilde{\nu}(H_I) &= \frac{m^2}{2\pi m_\mathrm{Pl}^2}\left(\xi-\frac{1}{6}\right)\left(\psi\left[\frac{3}{2}-\varsigma(\mu_I)\right]+\psi\left[\frac{3}{2}+\varsigma(\mu_I)\right]\right)-\frac{m^2}{\pi m_\mathrm{Pl}^2}\left[\frac{1}{36}+\left(\xi-\frac{1}{6}\right)  \right]\,,
\end{split}
\end{equation}
with $\mu_I=m/H_I$. As it is clear, equation \eqref{eq:nutildeHinfl} is the de Sitter counterpart of the RVM equation \eqref{eq:nueffinfRVM}.
Therefore, within this approximation we can re-use the same formal analytical solution given for scenario i), the only difference being that $\nu$ in the analytical formulas  \eqref{eq:infaprox}-\eqref{eq:infaprox3} now should mean the above defined coefficient $\tilde{\nu}_I$.
This strategy looks sensible, provided the coefficient $\tnu(H)$ in Eq.\eqref{eq:VEDdeSitterl} remains essentially constant, which is only approximately true since $\tnu(H)$ contains digamma functions of the variable $\mu=m/H(t)$, leading, however, to a very slow time evolution. Thus, we cannot avoid finding also an exact numerical solution to  Eq.\,\eqref{eq:diffHdeSitter} and compare the two results. A similar situation occurs for the RVM approach, see Eq.\eqref{eq:VEDinf2l}  and the corresponding  discussion in \cite{SolaPeracaula:2025yco}. The numerical and analytical solutions of the de Sitter case will be compared in Sec.\ref{sec:AnaliticalNumerical}. But first, we have to identify the parameter space available to trigger inflation  and determine the corresponding value of the inflationary scale $H_I$.  We do this in the next section.

\subsection{Parameter space available for inflation}\label{sec:PhysicalRegion_xi}

In order to produce numerical results, we must first determine the allowed range for the non-minimal coupling $\xi$ such that inflation can occur. Obviously, we have to make sure that the coefficient of $H^4$ is positive.  In the RVM case, see Eq.\,\eqref{eq:VEDinf2l}, the only condition to satisfy (for a single scalar field) is $\xi>1/6$. As noted previously, this excludes, in particular, the minimal coupling ($\xi=0$) and the conformal coupling ($\xi=1/6$) to produce inflation.  However, for the de Sitter inflation, the range of $\xi$ turns out to be much narrower. If we focus on the terms of $\mathcal{O}(H^4)$, a necessary condition to have inflation is that $\alpha>0$ in Eq.\,\eqref{eq:alphadS}.
%\begin{equation}\label{eq:condinflonefield}
  %  \frac{1}{960\pi^2}+\frac{3}{16\pi^2}\left(\xi-\frac{1}{6}\right)-\left(\xi-\frac{1}{6}\right)^2\frac{9}{2\pi^2}>0 \, .
%\end{equation}
This translates into the following (approximate) numerical range for $\xi-1/6$:
\begin{equation}\label{eq:Physregionximinus16}
    -0.00538< \left(\xi-\frac{1}{6}\right) < 0.172
\end{equation}
or, equivalently, for $\xi$:
\begin{equation}\label{eq:inflation_limit}
    0.161<\xi<0.339\,.
\end{equation}
As can be seen, for a single component, the non-minimal coupling  is fairly well determined by the condition of inflation. As was also the case in the RVM, Eq.\,\eqref{eq:inflation_limit} shows that a scalar field with minimal coupling to gravity ($\xi=0$) could not trigger $H^4$- inflation in the de Sitter case either. However, in the present case the conformal coupling $\xi=1/6$ is admitted in the physical region, in contrast to the RVM scenario.

It is useful and enlightening  to understand the origin of the tight constraint \eqref{eq:inflation_limit} imposed in the de Sitter case on $\xi$ at a deeper level. Note that the coefficient of the $H^4$ term in \eqref{eq:alphadS}  for the de Sitter scenario has two additional terms compared to the coefficient of $H^4$ in the RVM treatment, Eq.\,\eqref{eq:VEDinf2l}. The term proportional to $\xi-1/6$ is the same in both scenarios and is actually the only one in the RVM. However, the first term in $\frac{\alpha}{2\pi^2} H^4$, which is  $\frac{1}{960 \pi^2} H^4$, is a direct consequence of the trace anomaly discussed in Sec.\ref{sec:FullEMT}.  In fact, in the very early universe we can neglect the term $\rho_\Lambda(M)$ in the VED formula given by Eq.\eqref{eq:VacuuumTraceform}, and therefore we can write, to a very good approximation,
\begin{equation}\label{eq:VacuuumTraceformInfl}
\rv^{\rm infl.}=\frac{\langle T_{00}^{\delta \phi}\rangle_\mathrm{Ren}}{a^2}= -\frac14  \braket{T^{\delta\phi}}_\mathrm{Ren}=- \frac14\left(-\frac{1}{240\pi^2}\right) H^4=+\frac{1}{960\pi^2} H^4\,,
\end{equation}
which is precisely the aforementioned trace anomaly contribution to $\rv^{\rm infl.}$. Alternatively, the above result can also be obtained from  the first term of Eq.\eqref{eq:minusAnomaly} (divided by $a^2$) after being inserted in the subtraction prescription \,\eqref{eq:subtractionFRandNonDiv}. The reason is clear:  the anomaly originates entirely from finite terms of the $4$th order adiabatic expansion (specifically those virtually independent of the mass scale after an appropriate change of integration variable\,\cite{CristianJoan2022a}) in the conformal limit $\xi=1/6$. Thus, only the first term of \eqref{eq:minusAnomaly} satisfies these conditions.  Finally, the third term in \eqref{eq:alphadS}, the one proportional to the coefficient $(\xi-1/6)^2$,  comes from combining the finite part proportional to the same coefficient in Eq.\,\eqref{eq:unrenorT00} minus the contribution of the second term in Eq.\,\eqref{eq:minusAnomaly} (again divided by $a^2$) which also carries the same coefficient. In all cases, we neglect the mass terms since we consider the regime $H^4\gg m^4$, which is tantamount to having an effective massless limit. Notice that none of the last two mentioned contributions to $H^4$ are present in the RVM treatment since the first two terms in \eqref{eq:minusAnomaly} are independent of the scale and hence they cancel in the ARP procedure\footnote{Indeed, recall that in the RVM approach one computes the on-shell value $\left\langle T_{\mu\nu}^{\delta\phi} \right\rangle (m)$ through the adiabatic expansion (not from an exact result, which in general does not exist, as e.g. in the FLRW case). Hence the mentioned terms, which are independent of the mass scales $m$ and $M$ are cancelled in performing the off-shell adiabatic subtraction \eqref{RenormalizedEMTScalar}, see \cite{CristianJoan2022a} for computational details.}.  This explains the more complex structure of the coefficient of $H^4$ in  Eq.\,\eqref{eq:alphadS} and hence the reason for the much narrower domain of the physically allowed region \eqref{eq:inflation_limit} for inflation in the exact de Sitter scenario.  Finally, we note that if we set the on-shell condition $M=m$, the Bunch-Davies EMT given by \eqref{eq:BDresult} applies. In that case, the overall coefficient of $H^4$ would not be given by the full Eq.\,\eqref{eq:alphadS}, since the linear term in $\xi-1/6$ (the only one contributing in the RVM case) would be missing. The corresponding (approximate) allowed range for $\xi$ would then be $0.136<\xi<0.197$, which is even narrower than \eqref{eq:inflation_limit}.  However, as previously emphasized, the off-shell formulation has the advantage that $M$ can be chosen at the value of $H$ in the corresponding cosmological epoch. Notice e.g. that for the early universe, $H$ can be much larger than $m$.

Let us generalize the inflationary condition in the de Sitter case when we have more than one scalar field with different nonminimal couplings ${\xi}_i$. In the context of a typical GUT with many scalar fields $N_0$ (which may also involve supersymmetric particles, of course), each contributing a coefficient $\alpha_i$ of the same type as in Eq.\,\eqref{eq:alphadS}, the condition for inflation becomes the following:
\begin{equation}\label{eq:condinflationNo}
 A\equiv \sum_i^{N_0}\alpha_i=  \frac{N_0}{480}+\frac{3}{8} \sum_i^{N_0}\bar{\xi}_i-\frac{9}{4}\sum_i^{N_0} \bar{\xi}_i^2>0 \,,
\end{equation}
where $\bar{\xi}_i\equiv \xi_i-1/6$.  In this more general situation, the limit \eqref{eq:inflation_limit} does not necessarily apply to each component and remains only a sufficient condition for\eqref{eq:condinflationNo} to hold.  The value of \eqref{eq:condinflationNo} determines the parameter $H_I$ defined in \eqref{eq:rvm1}  and is therefore very relevant to the mechanism of inflation in the de Sitter scenario.  It is easy to see that
\begin{equation}\label{eq:HIA}
    H_I=\sqrt{\frac{3\pi}{4 G A}}=\sqrt{\frac{3\pi}{4A}}\ \mpl \, ,
\end{equation}
where in the second equality we have expressed the result in terms of the Planck mass, $\mpl$.

Recall that for the decaying de Sitter scenario we can still employ the analytical solution defined by equations \eqref{eq:infaprox}-\eqref{eq:infaprox3}, provided that $\Tilde{\nu}(H)$ in Eq.\eqref{eq:nutildeH} is approximately constant and given by Eq.\eqref{eq:nutildeHinfl}.  This is reasonable since the dependence of $\tilde{\nu}(H)$ on $H$ through $\mu=m/H$ in the digamma functions is mild and is further suppressed  due to the fact that $\xi-1/6$ is restricted in the narrow range \eqref{eq:Physregionximinus16}. Therefore, as a rough estimate within the allowed range of $\xi$, we can approximate
\begin{equation}\label{eq:nutildeapprox}
\begin{split}
    \Tilde{\nu}_I \approx -\frac{m^2}{36\pi m_\mathrm{Pl}^2} \,,
\end{split}
\end{equation}
which is constant,  very small in absolute value ($m^2/\mpl^2\ll1$)) and negative during the inflationary stage.  This parameter has a slight influence on the evolution of the VED in that regime, as it only determines the contribution of the power $\sim H^2$, which is subdominant in front of the leading power $\sim H^4$ during inflation. However, when we exit the inflationary phase, the $\sim H^2$ power takes over and provides the connection with the FLRW radiation epoch.  We refer once more the reader to Fig.\ref{fig:negative_rho}, where we have shown the relative importance of the different contributions to the VED using a numerical scenario within the physically allowed region of parameter space.  It can be easily seen that $\sim H^4$ drives inflation comfortably well and that during this period the effects of $\sim m^2H^2$ and $\sim m^4$ are negative but clearly suppressed.  Eventually, the power $H^2$ becomes dominant in the evolution of the vacuum energy, although it remains always subdominant compared to the radiation density due to the smallness of $\Tilde{\nu}(H)$. As the expansion makes progress, the sign of $\bar{\nu}(H)$ can change since it eventually becomes proportional to $\xi-1/6$. This occurs when the condition $m/H\gg1$ is fulfilled,  since  the formula for $\tilde{\nu}(H)$ that holds in later epochs is different -- see our discussion of the late universe in Sec. \ref{sec:VEDindS}, in particular Eq.\,\eqref{eq:barnueff2}. For example, in the recent universe we have $H\simeq H_0$ and hence the log in that formula becomes overwhelming: $\ln\left(\frac{m^2}{H_0^2}\right)={\cal O}(100)$. Therefore, the value of $\Tilde{\nu}(H)$ in the late universe is positive or negative depending on whether $\xi-1/6>0$ or  $\xi-1/6<0$, respectively. In any case, as explained there, it remains small in absolute value, of order $10^{-3}$ at most.

%%%%%%%%%%%%%%%%%%%%%%%%%%%%%%%%%%%%%%%%%%%%%%%%%%
%%%%%%%%%%%%%%%%%%%%%%%%%%%%%%%%%%%%%%%%%%%%%%%%%%%

%%%%%%%%%%%%%%%%%%%%%%%%%%%%%%%%%%%%%%%%%%%%%%%%%%%%%%%%%%
%\begin{figure}[h!]
%  \begin{center}
%      \resizebox{1.10\textwidth}{!}{\includegraphics{NumSol1.pdf}}
%      %\hspace{0.3cm}
%\caption{\textbf{(a)} Numerical solution of the VED  (solid line) versus the analytical solution \eqref{eq:infaprox2}(dotted line) around the transition time from inflation to the radiation epoch; \textbf{(b)} Numerical solution for both the VED and the radiation energy \eqref{eq:infaprox3} densities during the same transition period. In both cases $\nu=10^{-4}$, and $m$ of order of a GUT scale, the dependence being only logarithmic.}
%\label{Fig:NumSolution}
%  \end{center}
%\end{figure}
%%%%%%%%%%%%%%%%%%%%%%%%%%%%%%%%%%%%

%%%%%%%%%%%%%%%%%%%%%%%%%%%%%%%%%%%%%%%%%%%%%%%%%%%%%%%%%%
%%%%%%%%%%%%%%%%%%%%%%%%%%%%%%%%%%%%%%%%%%%%%%%%%%%%%%%%%%
%\begin{figure}[t]
%  \begin{center}
%      \resizebox{0.53\textwidth}{!}{\includegraphics{wvac_Infl.pdf}}
%      %\hspace{0.3cm}
%\caption{EoS of the running vacuum in the very early universe, as a function of $\hat{a}=a/a_*$. It is shown the transition from  the value at inflation  ($\wv=-1$) into de radiation-dominated epoch, where the vacuum adopts the EoS of radiation: $\wv=1/3$. Compare with Fig. \ref{Fig:EoSvacExtended}. }
%\label{Fig:EoSvacuumInfl}
%  \end{center}
%\end{figure}
%%%%%%%%%%%%%%%%%%%%%%%%%%%%%%%%%%%%%%%%%%%%%%%%%%%%%%%%%%

\subsection{Analytical versus numerical solutions to the cosmological equations}\label{sec:AnaliticalNumerical}
As noted previously, an exact analytical solution of the cosmological equations describing the inflationary period  is not possible, and one must resort to numerical analysis. However, we find that in the unstable de Sitter scenario, it is still possible to obtain a sufficiently accurate analytical solution during the inflationary period using the same analytical formulae \eqref{eq:infaprox}-\eqref{eq:infaprox3} that we used for the RVM, although with $\nu$ given now  by the previously defined $\tilde{\nu}_I=\tilde{\nu}(H_I)$.

We do not expect large deviations between the exact numerical solution and the analytical one, and  moreover they should both lead to the same final state when the Universe exits the inflationary epoch, meaning that they should provide the same smooth connection with the incipient radiation epoch. We shall show that it is indeed  so.
However, at this point, caution may be appropriate. We must ensure that the value of $A$ in Eq.\,\eqref{eq:HIA} can be  sufficiently large to avoid  $H_I$ overshooting the trans-Planckian regime\footnote{Although we do not wish to adopt particular frameworks that implement trans-Planckian values, let us mention  that they can be contemplated from different perspectives, including classical-quantum gravity duality arguments\cite{Sanchez:2020rqj} and  string theory considerations\cite{Silverstein:2008sg}. For example, in string monodromy models of inflation, a field (typically an axion) with sub-Planckian values can drive inflation
along a trans-Planckian excursion after its cycle path is broken by periodically modulated effects e.g. on its potential, thus producing large field inflation (owing to its not coming back to itself as it moves around a circle in the manifold).  A class of de Sitter models can be constructed on this basis, in which the fields may roll over a distance
in field space large compared to the Planck mass. See also \cite{Dorlis:2025gvb}, based on stringy RVM extensions.}.
Even for a sufficiently large scalar field multiplicity  $N_0$, it is not fully warranted that one can escape this regime since the presence of $\bar{\xi}_i^2$  with a negative sign in Eq.\,\eqref{eq:condinflationNo} prevents this expression from being numerically large. This is in contradistinction to the RVM case; see Eq.\,\eqref{eq:HIvalue}, where the term $\bar{\xi}^2$  was absent, and so $\bar{\xi}=\xi-1/6$ can have large values. All that said, this is only a mere precaution, as $H_I$ is just a formal parameter introduced in the definition \eqref{eq:rvm1}. A more physical condition to remain in the safe range is to demand that the maximum value of $\rv$ remains below $\mpl^4$. That maximum is obtained from Eq.\,\eqref{eq:infaprox2} at $a=0$, viz. $\rv(0)=\rho_I= \frac{3}{\kappa^2}\Tilde{H}_I^2\simeq \frac{3}{\kappa^2} {H}_I^2$, where we can neglect the small value of $\nu$ for this consideration. Now $\rv(0)<\mpl^4$ implies  $H_I^2<(\kappa^2/3)\,\mpl^4$, or
\begin{equation}\label{eq:subplankian}
H_I<\sqrt{\frac{8\pi}{3}}\,\mpl\simeq 2.89\,\mpl\sim 3\, \mpl.
\end{equation}
Then from equation \eqref{eq:HIA} the upper bound on $H_I$ translates into the lower bound  $A>9/32$. Assuming $\xi_i$ in the range \eqref{eq:inflation_limit} $\forall i$, Eq.\,\eqref{eq:condinflationNo} implies $N_0\gtrsim 135$, i.e. $N_0\gtrsim{\cal O}(100)$, which is typically fulfilled in GUT's context.
Our operational region for the numerical analysis  satisfies these conditions.
%%%%%%%%%%%%%%%%%%%%%%%%%%%%%%%%%%%
\begin{figure}[t]
  \begin{center}
      %\resizebox{0.6\textwidth}{!}
      %\resizebox{0.8\textwidth}{!}{\includegraphics{Figures/T_inf.pdf}}
      %\hspace{0.3cm}
      \includegraphics[width=0.49\linewidth]{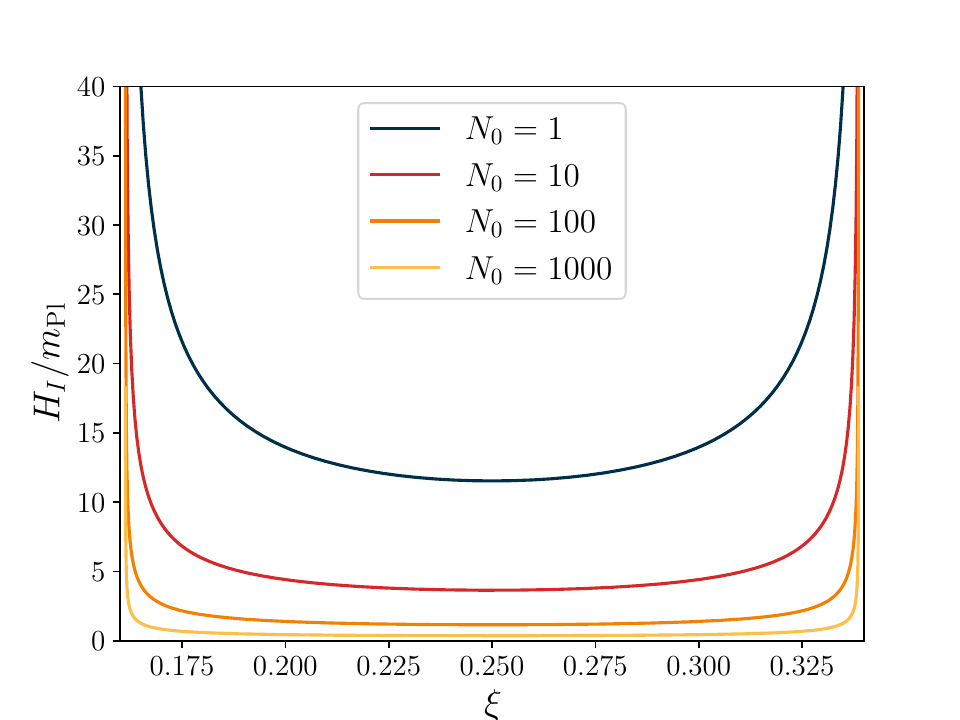}
      \includegraphics[width=0.49\linewidth]{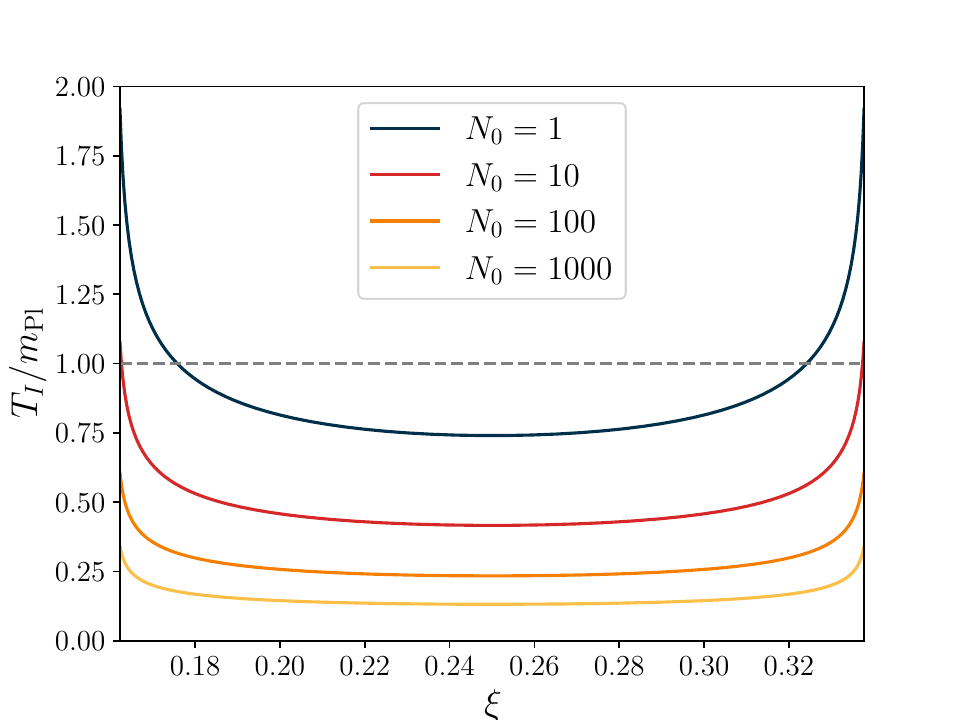}
\caption{Hubble rate $H_I$ from Eq.\,\eqref{eq:HIA} and temperature $T_I$ of radiation immediately after inflation, Eq.\,\eqref{eq:TempdeSitter}, for different multiplicities of the fields, $N_0$, and for $\xi$ (assumed common to all of them) in the allowed region for inflation, as discussed in the text. It is seen that for large enough $N_0$ the condition \eqref{eq:subplankian} is fulfilled all over the permitted range, and $T_I<\mpl$.}
\label{Fig:T_inf}
\end{center}
\end{figure}
%%%%%%%%%%%%%%%%%%%%%%%%%%%%%%%%%%%%%
Furthermore, we can check that physical quantities such as the temperature $T_I$ associated with the total energy density $\rho_I$ remains sub-Planckian in the region of interest. This fact is clearly illustrated in Fig. \ref{Fig:T_inf} for the allowed region \eqref{eq:inflation_limit} of the non-minimal coupling. The mentioned  $T_I$ is of the order of the maximum temperature of the heat-bath of relativistic matter into which the vacuum decays (cf. Sec.\ref{sec:Thermoidynamics} for more details) and is given by
\begin{equation}
    T_I = \left(\frac{30\rho_I}{\pi^2 g_*}\right)^{1/4}\,.
\end{equation}
 Here the characteristic factor $g_*=\mathcal{O}(100)$ provides the number of active degrees of freedom (d.o.f.) at the given temperature (e.g. $g_*=160.75$ in non-supersymmetric $SU(5)$, but in general it is larger for general GUT's.).

We can estimate the above temperature in our case as follows. Using \begin{equation}
    \rho_I=\frac{3H_I^2 m_\mathrm{Pl}^2}{8\pi} \,
\end{equation}
and the relation \eqref{eq:HIA}, we find
\begin{equation}\label{eq:TempdeSitter}
    T_I = \left(\frac{135}{16\pi^2 g_* A}\right)^{1/4}\ \mpl\,.
\end{equation}
Assuming that the parameter $A$ receives the main contribution  from the term proportional to $N_0$ in Eq.\,\eqref{eq:condinflationNo}, which is tantamount to assuming that $\bar{\xi}_i\simeq 0$ for all $i$ (a realistic situation according to the typical range found for the non-minimal couplings), we get
\begin{equation}\label{eq:TempdeSitter2}
    T_I = \left(\frac{135\times 30}{\pi^2 g_* N_0}\right)^{1/4}\ \mpl\simeq 0.2\, \mpl\,,
\end{equation}
where for the numerical evaluation we have taken $g_*=160.75$ and $N_0=1000$. For $N_0=100$ the result would be only slightly bigger, $T_I\simeq 0.3 \mpl$. Roughly speaking, $T_I\sim g_*^{-1/4} \mpl\sim 0.28 \mpl$ within an order of magnitude approximation. Thus, the expected temperature is sub-Planckian by about one order of magnitude.  For larger values of $N_0$ (which, as said, is a common situation in GUT's and especially in their supersymmetric counterparts), we find that the above temperature is granted to lay approximately one order of magnitude below the Planckian regime, and hence we are still entitled to use an approximate field theoretical treatment. %\footnote{A more detailed estimate of this temperature will be provided in Sec.\ref{sec:Thermoidynamics}}.
In Fig.\,\ref{Fig:T_inf} we illustrate the exact numerical dependence of $H_I$ and $T_I$ as a function of $\xi$  for different field multiplicities, assuming for simplicity that all scalar fields have the same $\xi_i=\xi$.
%\begin{figure}[t]
%  \begin{center}
%      %\resizebox{0.6\textwidth}{!}
%      %\resizebox{0.8\textwidth}{!}%{\includegraphics{Figures/T_inf.pdf}}
      %\hspace{0.3cm}
%      \includegraphics[width=0.7\linewidth]%{Figures_new/HI_xi_N0.pdf}
%\caption{Inflationary Hubble rate $H_I$ for %different multiplicities of the fields, $N_0$, and %for $\xi$ in the allowed region for inflation,  as %discussed in the text. It is seen that for large %enough $N_0$ transplanckian values are avoided all %over the permitted range.}
%\label{Fig:H_inf}
%\end{center}
%\end{figure}
\begin{figure}[t]
    \centering
    \includegraphics[width=0.6\linewidth]{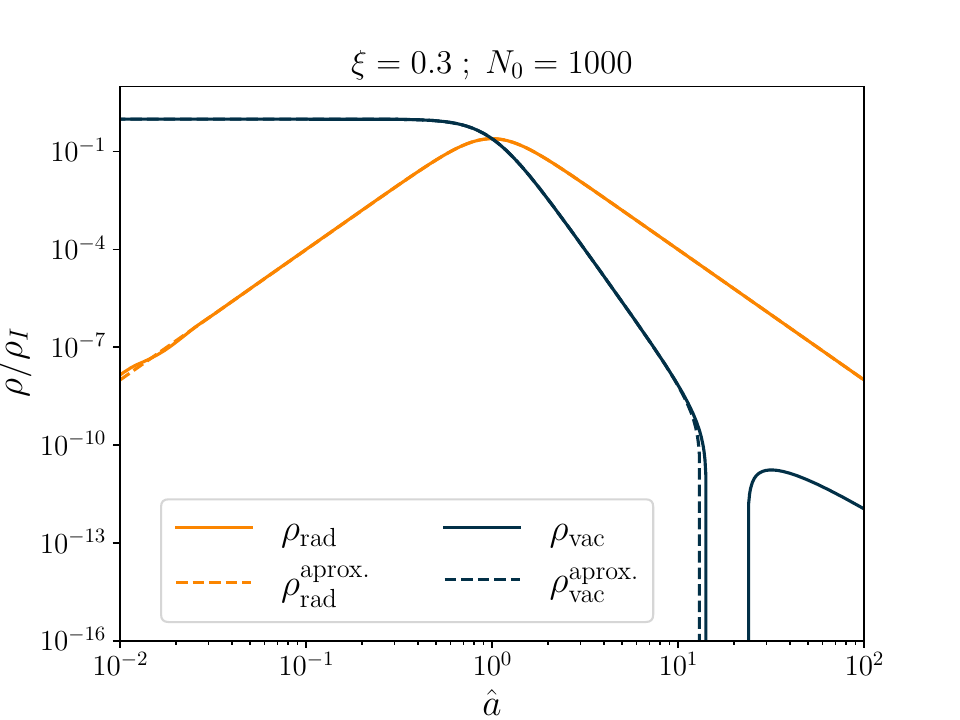}
    \caption{De Sitter decay scenario: energy densities for vacuum and radiation for $\xi>1/6$, as a function of  the scale factor normalized to the transition point, $\hat{a}=a/a_*$. We compare the numerical solution to Eq.\,\eqref{eq:diffHdeSitter} (solid lines) with the (RVM-like) analytical approximation (dashed lines) given in \eqref{eq:infaprox2}-\eqref{eq:infaprox3}. As in previous figures, we use $m=0.001m_\mathrm{Pl}$ for the scalar field mass.}
    \label{fig:ved_greater}
\end{figure}

\begin{figure}[t]
    \centering
    \includegraphics[width=0.6\linewidth]{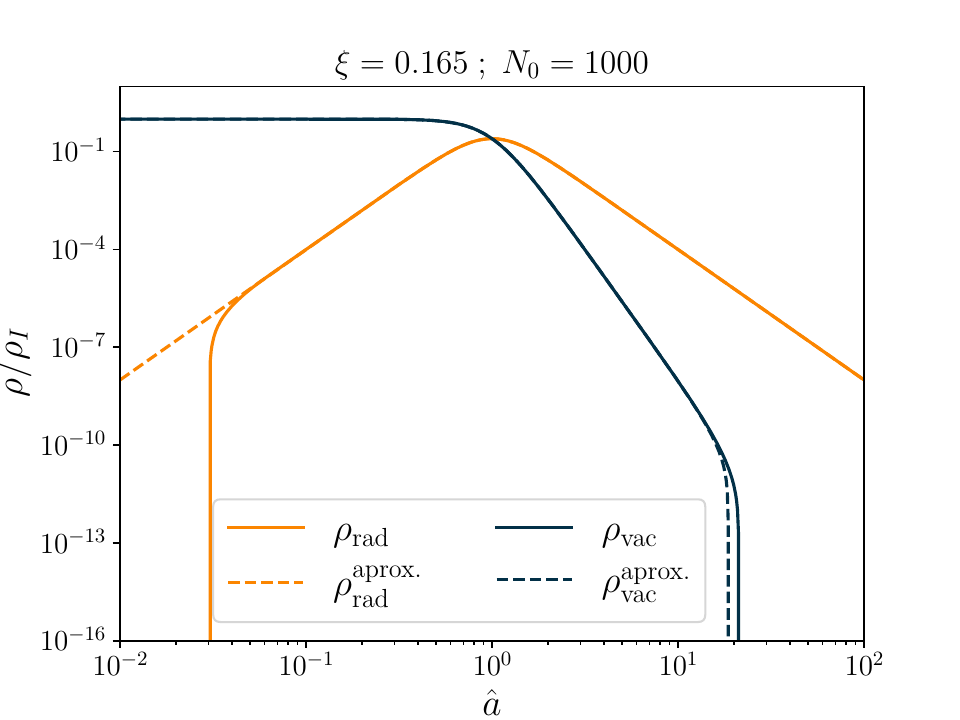}
    \caption{Same as Fig. \ref{fig:ved_greater} but taking $\xi<1/6$.}\label{fig:ved_lower}
\end{figure}
Let us now consider the exact numerical solution to the differential equation \eqref{eq:diffHdeSitter}.  We  have used \texttt{Mathematica}\cite{Mathematica} to numerically determine the Hubble rate $H(a)$ in a specified range of $\hat{a}=a/a_*$ under an appropriate boundary condition inspired by the analytical approximation Eq. \eqref{eq:infaprox}: $H(\hat{a}=1)=\frac{H_I}{\sqrt{2}}$. We have checked that this guarantees $H(\hat{a}\ll1)=H_I$, as it should. % where $H_I$ can be obtained from the quartic coefficient of $\rho_{\rm vac}(H)$ compared to the mentioned analytic solution.
In Figures \ref{fig:ved_greater} and \ref{fig:ved_lower} we show the energy densities of vacuum and radiation for $\xi>1/6$ and $\xi<1/6$, respectively, within the allowed range previously determined. Let us start by analyzing the results obtained in Fig.\ref{fig:ved_greater}. It can be seen  that in the beginning ($\hat{a}=0$) there is no radiation at all while the vacuum dominates and its energy density is the total energy density of the universe, $\rho_I$. Nevertheless, the vacuum energy decays very fast into radiation and the latter quickly dominates the universe. This feature emerging from the numerical analysis is nicely reflected also in the analytical solution  \eqref{eq:infaprox2}-\eqref{eq:infaprox3} when we look at the post-inflationary epoch $\hat{a}\gg1$.  The fast vacuum decay into radiation makes  the inflationary period transit continuously into the standard FLRW radiation-dominated epoch.  This is nothing but the practical implementation of `graceful exit' of the inflationary stage in our context.  There is, therefore,  no conventional `reheating' period in $H^4$-inflation, namely one characterized  e.g. by cold dark matter particles decaying subsequently into relativistic particles. This was already noted in \cite{BLS2013,Perico:2013mna,Sola:2015csa,BLS2015,Yu2020} on phenomenological grounds.  In fact, rather than having a highly non-adiabatic `reheating' event, as typically occurring in inflaton-mediated formulations\, \cite{KolbTurner,LiddleLyth} and in Starobinsky inflation\,\cite{Starobinsky:1980te}, in our case we meet a relatively long non-equilibrium heating up period in which the vacuum instability drives the model progressively to the radiation phase. In addition,  from the analytical formulas, one can see that in the region $\hat{a}\gg 1$ the radiation energy density scales very approximately as $\rho\sim a^{-4}$ since $|\nu|\ll1$, and at the same time the VED is suppressed by the tiny coefficient $\nu$, since $\rv(\hat{a})/\rho_r(\hat{a})\simeq \nu$ for $\hat{a}\gg1$. For de Sitter, $\nu$ is meant to be $\tilde{\nu}_I$, Eq.\,\eqref{eq:nutildeHinfl}, but the behavior is formally similar to the RVM case. Our QFT account of the early cosmic history seems pretty successful; it not only achieves a graceful exit from the inflationary phase into the radiation-dominated era, but also leaves a tiny remnant of vacuum energy during the radiation-dominated epoch, which is highly suppressed in front of the energy density of radiation. This property is of momentous importance in order not to spoil the success of Big Bang Nucleosynthesis (BBN) within  the standard model of cosmology. One can see e.g. in the phenomenological analysis of  Ref.\cite{Asimakis:2021yct} that the BBN bounds on $\nu$ are well within expectations, i.e. $|\nu|<10^{-3}$, and hence fully compatible with the global fits to the cosmological data, see e.g. \cite{SolaPeracaula:2021gxi,SolaPeracaula:2023swx,deCruzPerez:2025dni} and \cite{Sola:2015wwa,Sola:2016jky,Sola:2017znb,SolaPeracaula:2016qlq,Sola:2016zeg,SolaPeracaula:2017esw}.

In Figures \ref{fig:ved_greater} and \ref{fig:ved_lower} we also compare the numerical solution with the corresponding (RVM-like) analytical approximation described in Sec.\,\ref{sec:solutions}, given by the energy densities \eqref{eq:infaprox2} and \eqref{eq:infaprox3}, labeled $\rho^\mathrm{aprox.}_i$.  The two types of solutions are superimposed in these figures.
We find that the bulk qualitative properties of RVM-inflation as described in detail in our previous work\cite{SolaPeracaula:2025yco} are also reproduced in the de Sitter scenario. Indeed, we confirm that the primeval universe is dominated by an approximately constant vacuum energy density corresponding to a large and constant value of $H_I$ and of $\rv\simeq \rho_I$. Because this approximate de Sitter phase subsequently decays into radiation, the vacuum becomes subdominant in front of radiation and we are led to a graceful exit from the inflationary epoch. Worth noticing in these figures, too,  is the fact that the numerical solution for radiation is well described by the approximate analytical solution for most of the inflationary phase and especially for the post-inflationary epoch  (i.e. for $\hat{a}\gg 1$). Within the inflationary region, the numerical and approximate analytical solutions display differences for $\hat{a}\ll 1$, e.g., in Fig. \ref{fig:ved_lower}. Needless to say, the numerical solution prevails in this case. However, both the numerical and analytical solutions  provide a highly consistent description of the transition period from inflation into the FLRW regime, and the standard cosmological expansion is accurately recovered.

\begin{figure}[t!]
    \centering
    \includegraphics[width=0.485\linewidth]{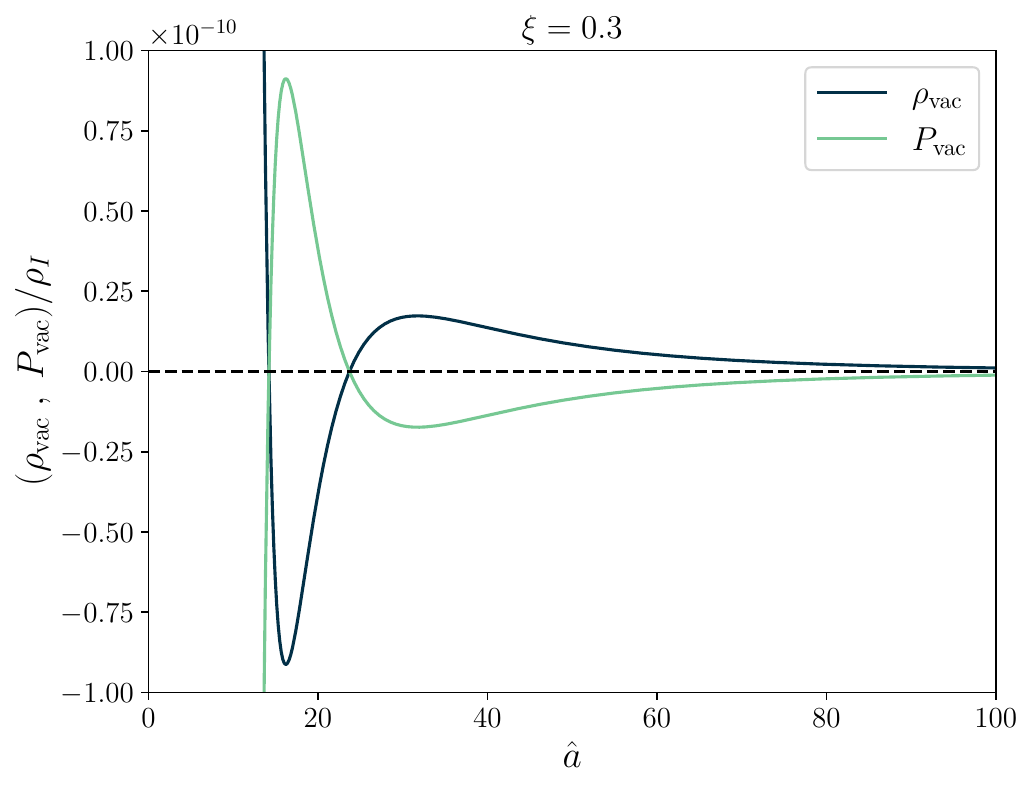}
    \includegraphics[width=0.485\linewidth]{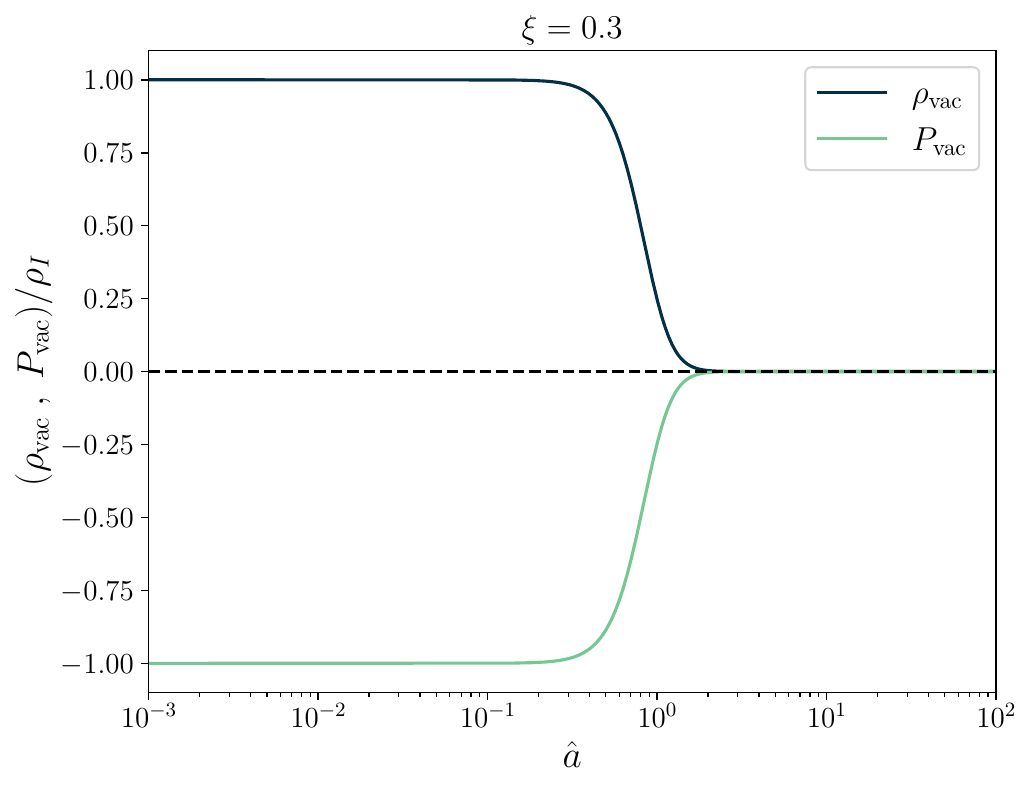}
    \includegraphics[width=0.485\linewidth]{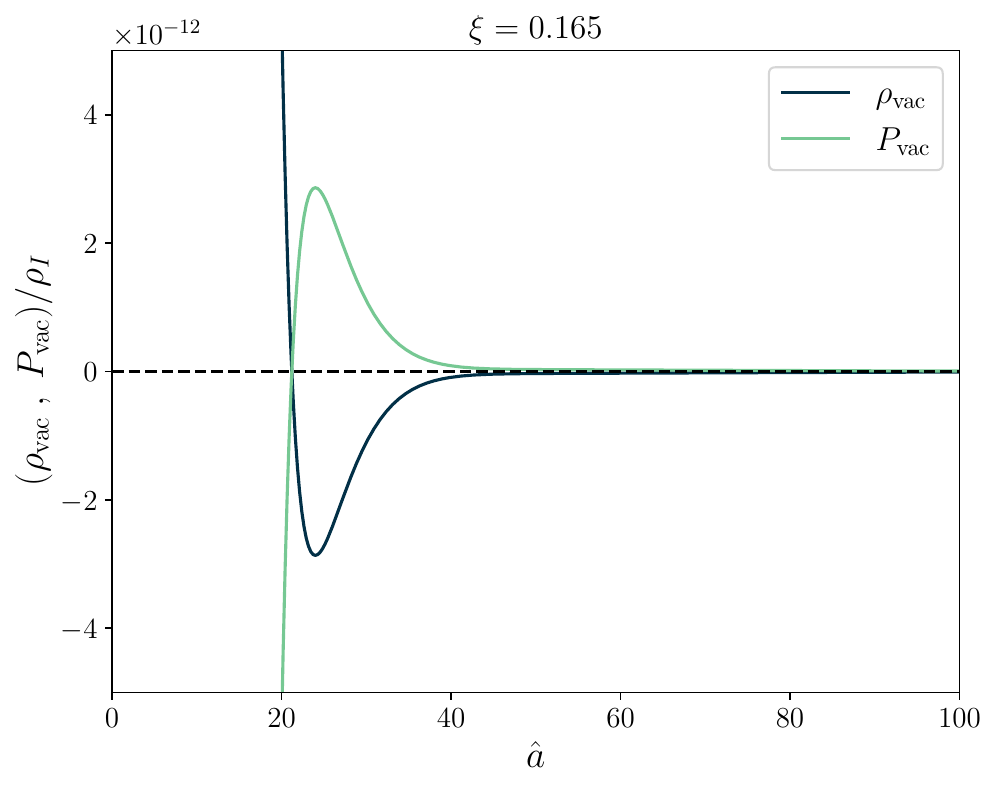}
    \includegraphics[width=0.485\linewidth]{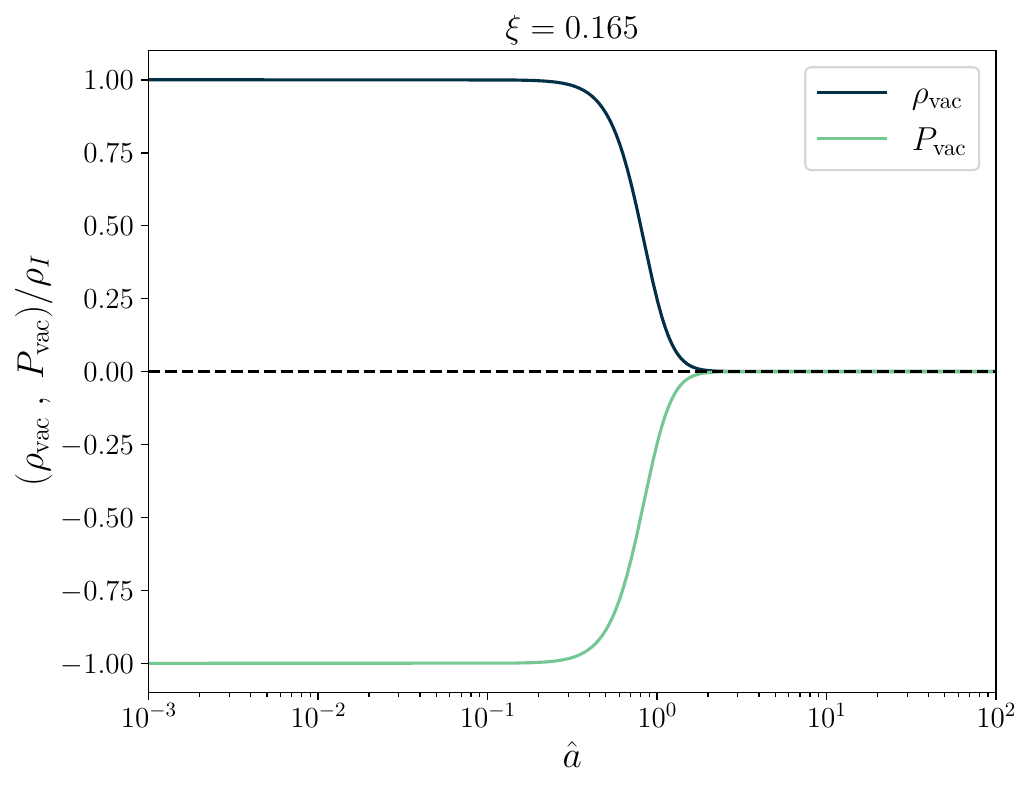}
    \caption{Energy density ($\rv$) and pressure ($\Pv$) of the unstable de Sitter vacuum for the same cases considered in Figs. \ref{fig:ved_greater} and \ref{fig:ved_lower}. Because $\rv$ and $\Pv$ in the last region take much higher values  the features observed on the plots on the left cannot be appreciated in those  on the right.}\label{fig:rhop_inflation}
\end{figure}

From the current analysis and our previous study \cite{SolaPeracaula:2025yco}, we confirm that a smooth transition from the inflationary to the radiation period occurs in the two scenarios under discussion, that is, i) running vacuum (RVM) and ii) unstable de Sitter vacuum. Taking into account the qualitatively  similar behavior of these two models in the very early universe and their extrapolation to the current universe (cf. Sec.\ref{sec:VEDindS}), in both cases we are led to a unified QFT model of inflation and DE embracing the entire cosmological history. The most remarkable difference is that in the de Sitter case the VED, despite it being time-evolving as in the RVM, its EoS is nevertheless unaffected by quantum corrections and remains stuck at the canonical value $\wv=-1$, whereas the EoS of the RVM vacuum is sensitive to quantum effects which produce a departure from $\wv=-1$ in the post-inflationary epoch.  In Fig.\ref{fig:rhop_inflation}, we show the vacuum density and pressure of the de Sitter decay model considering the regions $\hat{a}\gg 1$ (left plots) and $\hat{a}\ll 1$ (right plots). The former plots correspond to a period where the VED is suppressed since inflation was left behind and we entered the radiation-dominated epoch, whereas the latter describe the transition from vacuum dominance to radiation dominance.

%%%%%%%%%%%%%%%%%%%%%%%%%%%%%%%%%%%%%
%%%%%%%%%%%%%%%%%%%%%%%%%%%%%%%%%%%%%%%

%%%%%%%%%%%%%%%%%%%%%%%%%%%%%%%%%%%%%%%%%%%%%%%%%%%%%%%%%%
%%%%%%%%%%%%%%%%%%%%%%%%%%%%%%%%%%%%%%%%%%%%%%%%%%%%%%%%%%
%\begin{figure}[t]
%  \begin{center}
%      \resizebox{0.58\textwidth}{!}{\includegraphics{H4H2.pdf}}
%      %\hspace{0.3cm}
%\caption{Numerical detail (in a normal scale) of the part of Fig. \ref{Fig:NumSolution} around the sharp dip produced by the logarithmic term of the VED in Eq.\,\eqref{eq:VEDinfl} once the early radiation epoch is attained and the $H^4$ power becomes subdominant versus the overall ${\cal O}(H^2)$ contribution. Shown are the full VED (solid line) and the individual contributions from the $H^4$ and $H^2$ terms (dotted lines).}
%\label{Fig:NumDetails}
%  \end{center}
%\end{figure}
%%%%%%%%%%%%%%%%%%%%%%%%%%%%%%%%%%%%%%%%%%%%%%%%%%%%%%%%%%

\subsection{Unified description of inflation and (dynamical) dark energy}\label{sec:unifeid}
All in all, the unified QFT description that we have obtained of the entire cosmological evolution from the very early times to the current universe in both pictures,  i) running vacuum (RVM) and ii) unstable de Sitter vacuum,    involves  an initial inflationary epoch that is dominated by the power $H^4$ of the Hubble rate (which is capable of triggering inflation for a short period where $H\simeq$ const.). Because the predicted structure for the VED involves subleading powers  $H^2$  as well, they take over after the inflationary stage has been finished and thanks to them it is possible for the transit to be carried out in a non-disruptive way from the very early universe to the late universe without an intervening reheating period of the standard type in inflaton models\cite{Kallosh:2025ijd,Martin:2013tda}, as we have shown in previous sections. The predicted late time evolution in each case is given by equations \eqref{eq:RVMcanonical} and \eqref{eq:dSlate1}. Overall, the set of equations \eqref{eq:infaprox}-\eqref{eq:infaprox3} provides a fairly accurate analytical description of the unified solution from the very early times to the radiation-dominated epoch. The subsequent transit to the matter-dominated epoch follows when massive particles become nonrelativistic at late times and the EoS of matter changes from $w=1/3$ to $w=0$. However, the primeval kick-out of the entire cosmological expansion hinges entirely on the initial vacuum dynamics defined by Eq.\,\eqref{eq:rvm1}. And the main point to be stressed here is that this VED form is no longer a phenomenological proposal\cite{BLS2013}, since it has been derived rigorously  in a QFT context by considering the renormalization of the quantum vacuum energy in curved spacetime and has led to the above mentioned scenarios i) and ii) which cover the complete cosmological expansion\footnote{The full analytical interpolation between the inflation epoch all the way down to the current epoch is not possible since the approximate analytical solution \eqref{eq:infaprox}-\eqref{eq:infaprox3} assumes $c_0=0$ in Eq.\eqref{eq:rvm1}, i.e. it neglects today's  cosmological term in the early universe. While it is feasible numerically, of course,  it is not relevant for our discussion.}.

These two unified pictures of the cosmic history can provide a possible explanation for the dark energy (DE) of our universe. In fact, in both cases the foreseen cosmic expansion matches very closely the $\CC$CDM one but is not literally reproduced.  This is actually a bonus since it entails a  remnant of (dynamical!) vacuum energy, which is  predicted to exist in the late-time universe on pure QFT grounds,  playing somehow the role of a  ``fossil energy''\cite{{Fossil2008}} that  hints at the existence of a dramatic period of fast inflation in the very early times. This energy remnant  still undergoes  a mild evolution at present, and it appears in the form of what we call ``dynamical dark energy''. If our picture were to be correct, the observed DE of our universe should just be the residual quantum vacuum energy left over in our time, which could serve as a smoking gun of the huge inflationary event which occurred during the  primeval universe.

It is well-known that dark energy is being scrutinized nowadays using a large variety of {\it ad hoc} parameterizations to describe  the most recent data\cite{DESI:2024mwx,DESI:2024aqx,DESI:2025zgx,DESI:2025fii}. These are useful since they currently hint at potential evidence of its dynamical character, see \cite{Park:2024vrw,deCruzPerez:2025dni}, for example; and  may even provide a model-agnostic reconstruction of the background quantities associated with dark energy -- cf. \cite{Gonzalez-Fuentes:2025lei,Gonzalez-Fuentes:2026rgu} and references therein.  Notwithstanding, while these parameterizations may be able to capture an inherent dynamical feature in it,  they are far from providing a satisfactory explanation of its ultimate nature. In our work, by contrast,  we have advanced a possible theoretical explanation on fundamental grounds, for we have seen that the dynamical DE could be the result of quantum effects stemming from the QFT framework discussed in the present work, see  Sections \ref{sec:VEDinRVM} and \ref{sec:VEDindS}. If so,  there is no a priori need of introducing {\it ad hoc} quintessence or phantom fields to explain its origin.  The upshot is that all these spurious fields may well be banished and might be dispensed with since a more fundamental explanation can be provided, which is able to account for inflation in the very early times and the accelerated expansion of the universe at late times within one and the same framework. Moreover, the fact that the speeding up of the universe in this QFT description is predicted to evolve with the expansion is fully in line with current observations, which are interpreted through the mentioned parameterizations of the DE that suggest its dynamical character.

\jtext{To close this section, the following observation may be in order. We would like to emphasize that  despite the  virtues of the unified framework for DE and Inflation that we are proposing here, further studies will be, of course,  indispensable  in order to assess the reach of its implications and its overall consistency. In  particular, a more detailed study of the cosmic perturbations is mandatory. We should nonetheless  mention that  this subject has already  been treated in part in the literature in different works  exploring various aspects of the cosmic perturbations in the context of the running vacuum model, see e.g. \cite{Grande:2010vg,Grande:2008re,Grande:2007wj}. In particular, in the work\cite{Gomez-Valent:2018nib} it is shown in two gauges (Newtonian and synchronous) that the treatment of cosmic perturbations within the running vacuum framework can help alleviate the so-called $\sigma_8$-tension.   Furthermore, the successful comparison of the running vacuum framework  against the observational data made e.g. in Refs. \cite{Sola:2015wwa,Sola:2016jky,Sola:2017znb,SolaPeracaula:2016qlq,Sola:2016zeg,SolaPeracaula:2017esw} and \cite{SolaPeracaula:2021gxi,SolaPeracaula:2023swx} includes the explicit treatment of the cosmic perturbations, which is convenient for testing the data on large scale structure formation. It is found that it provides a better fit than the standard $\CC$CDM.  At the same time, other theoretical works \cite{Basilakos:2013xpa,Basilakos:2014moa,Basilakos:2015yoa,Basilakos:2019zsf} have explored additional aspects of the running vacuum framework in the early universe with the help of the effective potential method and compare the RVM with Starobinsky inflation, sometimes involving a supergravity context, since there is a stringy formulation of the RVM \cite{Mavromatos:2020kzj,PhantomVacuum2021}. These studies indicate that the RVM may provide a correct description of the cosmological perturbations, a fact which helps to improve the fits to the growth data, see  the devoted study  \cite{Gomez-Valent:2018nib}. As for the primordial perturbations in the very early universe within this unified framework,  further investigation is needed to fully confirm its viability. However,  we should also emphasize  that the two unified models under consideration are very similar in the primordial stages of the early universe and that one of them is the de Sitter model decaying into radiation. Therefore, during the inflationary epoch, while the universe is in the $H\simeq$ const. phase, they both behave alike and we know that under de Sitter spacetime conditions the primordial perturbations are correctly generated. To fully confirm these natural expectations, more devoted studies are needed, which  will be presented elsewhere.}

\section{Some further phenomenological implications}\label{sec:pheno}

After the formal QFT materials presented in  previous chapters, in  which we have addressed the renormalization of the energy and pressure of the quantum vacuum in the RVM and de Sitter scenarios, it seems appropriate to consider some phenomenological implications. In Sec.\,\ref{sec:unifeid}, we have already remarked perhaps one of the most important phenomenological aspects that emerge very clearly from our theoretical framework: the dynamical nature of the vacuum energy, i.e. its evolution with the cosmic expansion. This may provide a possible fundamental explanation for the observed dynamical DE in current observations. In this section, we shall dwell more on these phenomenological aspects. Specifically, we consider in more detail the late-time implications of the vacuum dynamics for the de Sitter scenario and compare it with the RVM case.  Then we address a time-honored cosmological parameter in all studies of the dynamical DE: the equation of state (EoS), in this case of the quantum vacuum.  In addition, we shall also consider an important thermodynamical aspect, namely the calculation of the huge entropy of the universe at present, a result that cannot be accounted for within the standard $\CC$CDM. These are very different phenomenological aspects, but they are all highly relevant. The possible resolution of these disparate problems within the same framework illustrates the reach of our unified proposal.

%\newpage
\subsection{Late-time VED dynamics in  the unstable de Sitter scenario}\label{sec:VEDindS}

In Sec.\ref{sec:VEDinRVM} we have considered the prediction of the RVM for the evolving vacuum energy density in the current universe.  Now we want to find out what the corresponding running of the VED at late times is like for the unstable de Sitter scenario. As we shall discover, the formal resemblance between the RVM and de Sitter is remarkably close, as, in fact, both cosmological frameworks lead to a very similar prediction for the VED evolution in the late universe.

To check this, let us insert the expanded expression of the term $m^4\Psi$ in the limit $m/H \gg1$ from Eq.\eqref{eq:psiexpansion2} into Eq.\,\eqref{eq:rvMMoPsi} and neglect the contributions
$\mathcal{O}(H^4)$ in the late-time universe. We find:
\begin{equation}
\begin{split}
    &\rho_\mathrm{vac}(H)\simeq\rho_\mathrm{vac}(H_0)+\frac{m^2}{64\pi^2}\left[12H^2\left(\xi-\frac{1}{6}\right)\right]\left(\ln\left(\frac{m^2}{H^2}\right)+\frac{4(-1+9\xi)H^2}{3m^2}\right)\\
    &-\frac{m^2}{64\pi^2}\left[12H_0^2\left(\xi-\frac{1}{6}\right)\right]\left(\ln\left(\frac{m^2}{H_0^2}\right)+\frac{4(-1+9\xi)H_0^2}{3m^2}\right)\\
    &+\frac{1}{128\pi^2}\left\{m^2(H^2-H_0^2)\left[-\frac{4}{3}-48\left(\xi-\frac{1}{6}\right)  \right]\right\}+\frac{m^4}{64\pi^2}\left\{\frac{4(-1+9\xi)}{3m^2}(H^2-H_0^2)\right\} \, .
\end{split}
\end{equation}
This can be conveniently rephrased in the RVM form
\begin{equation}\label{eq:dSlate1}
\begin{split}
    \rho_\mathrm{vac}(H)\simeq \rho_\mathrm{vac}(H_0)+\frac{3\bar{\nu}_\mathrm{eff}(H)m_\mathrm{Pl}^2}{8\pi}(H^2-H_0^2) \, ,
\end{split}
\end{equation}
provided we define\footnote{We distinguish the inflationary $\tilde{\nu}_{\rm eff}(H)$ defined in \eqref{eq:nutildeH} with a tilde from the low-energy $\bar{\nu}_{\rm eff}(H)$ defined here.  }
\begin{equation}
\begin{split}\label{eq:barnueff1}
    \bar{\nu}_\mathrm{eff}(H)&=\frac{8\pi}{3}\left\{ \frac{12m^2}{64\pi^2m_\mathrm{Pl}^2}\left(\xi-\frac{1}{6}\right) \ln \frac{m^2}{H^2} + \frac{m^2}{128\pi^2m_\mathrm{Pl}^2}\left[ -\frac{4}{3}-48\left(\xi-\frac{1}{6}\right)\right]+\frac{m^2}{64\pi^2m_\mathrm{Pl}^2} \left[ \frac{4(-1+9\xi)}{3}\right] \right\}\,.
\end{split}
\end{equation}
However, a simple rearrangement in the above expression shows that it boils down to just
\begin{equation}
\begin{split}\label{eq:barnueff2}
    \bar{\nu}_\mathrm{eff}(H)=\frac{1}{2\pi}\left(\xi-\frac{1}{6}\right) \frac{m^2}{m_\mathrm{Pl}^2} \left(\ln \frac{m^2}{H^2}-1 \right)\,.
\end{split}
\end{equation}
For the late universe, we may approximate $H=H_0$ inside logarithms and neglect additive terms which are much smaller than $\ln({}m^2/H_0^2)={\cal O}(100)$. In this limit, therefore, we find that $\bar{\nu}_\mathrm{eff}(H_0)$ given above is exactly coincident with the corresponding low-energy parameter appearing in the RVM case, see Eq.\,\eqref{eq:nueffAprox2}. This is remarkable and tells us that it is subject to the same phenomenological bounds: $\bar{\nu}_\mathrm{eff}(H_0)\lesssim 10^{-4}-10^{-3}$\,\cite{SolaPeracaula:2021gxi,SolaPeracaula:2023swx,deCruzPerez:2025dni}. From the foregoing considerations and applying analogous reasoning as in Sec.\ref{sec:betaVED},  it follows that the low-energy $\beta$-function of the running VED for the de Sitter case takes the same form as in the RVM case. Therefore,
\begin{equation}\label{eq:betavnueffdS}
\beta^{\rm dS}_{\rm vac}(H)=\frac{3\bar{\nu}_\mathrm{eff}}{4\pi}\mpl^2 H^2=\left(\xi-\frac{1}{6}\right)\frac{3 m^2H^2 }{8 \pi^2 }\ln\frac{m^2}{H_0^2}\,.
\end{equation}
In other words, we obtain again the smooth type of  behavior  $\beta^{\rm dS}_{\rm vac}(H)\propto m^2 H^2$  rather than the traditional $\propto m^4$\,\cite{Brown:1992db,Akhmedov:2002ts,Ossola:2003ku,Martin:2012bt}. So, as in the RVM case,  the low-energy running of the VED in de Sitter space is perfectly smooth and causes no fine tuning troubles either.

The results obtained show that the late time evolution of the VED in the two scenarios under consideration, that is, the RVM, which is given by Eq.\eqref{eq:RVMform}, and the unstable de Sitter vacuum, whose VED evolution is given by  \eqref{eq:dSlate1}, is formally the same.  This is rewarding since the two unification scenarios turn out to lead qualitatively to the same picture of the current universe.  However, as we shall see, they can be distinguished quantitatively, in principle,  from the fundamental equation of state (EoS) of the corresponding vacuum, which receives different quantum effects in each case, see Sec. \ref{sec:EosVacuum}. Another difference appears in the allowed range for the non-minimal coupling $\xi$, which is much more restricted in the de Sitter case, see Sec.\,\ref{sec:PhysicalRegion_xi}.

The fact that $\xi-1/6$ can be positive or negative  within the allowed range implies that the parameter $\bar{\nu}_{\rm eff}$ defined in \eqref{eq:barnueff2} can have both signs. Therefore, from the running equation for the VED, Eq.\eqref{eq:dSlate1}, it follows that the latter can either decrease ($\xi-1/6>0$) or increase ($\xi-1/6<0$)  with expansion. These situations correspond to having `effective quintessence' and `effective phantom DE' behavior, respectively. This is so despite the fact that the genuine EoS of the de Sitter vacuum remains invariably stuck at $\wv=-1$, as we have shown.
%%%%%%%%%%%%%%%%%%%%%%%%%%%%
\begin{figure}[t]
  \begin{center}
      \includegraphics[width=0.7\linewidth]{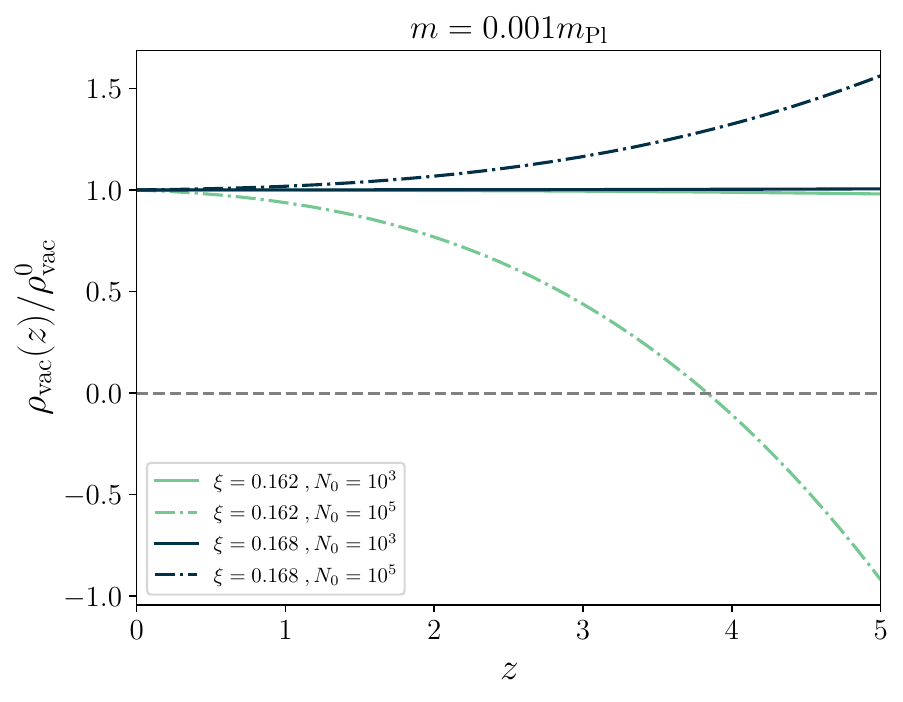}
\caption{Evolution of the VED for the decaying de Sitter scenario in the current universe, Eq.\,\eqref{eq:dSlate1}, normalized to the present density, as a function of redshift. We use allowed values for the parameters in the physical region determined in Section \ref{sec:PhysicalRegion_xi}. For common values $\xi_i=\xi$ for all the fields above or below  $1/6\simeq 0.166$ the vacuum  mimics quintessence or phantom-like DE (more specifically `phantom matter', see the text), respectively.}
\label{fig:VED_low}
\end{center}
\end{figure}
%%%%%%%%%%%%%%%%%%%%%%%%%%%%
These two kinds of effective DE behaviors are illustrated  in Figure \ref{fig:VED_low},

From the foregoing, it follows that the evolution of the VED is not necessarily tied to a departure of the fundamental EoS from $-1$, for the vacuum can display effective EoS behaviors that mimic dynamical DE -- as pointed out long ago in \cite{SolaStefancic,Das:2005yj, Basilakos:2013vya}. Note that for the values of $\xi_i=\xi\ (\forall i)$ displayed in the figure,  a large multiplicity $N_0$ of fields is needed to significantly deviate the VED from a constant  value in the redshift range where the current data are sensitive. In particular, it is remarkable that for some range of values of $\xi$ and $N_0$ it is possible to push the VED into the negative domain. The penetration into this domain cannot be extrapolated to the early universe, of course, since the $H^4$ term is overwhelming there. However, the possible situation with negative DE around our time is peculiar, since it leads to the notion of ``phantom matter'', which  is characterized by negative energy but positive pressure, in contrast to the usual phantom DE.  The notion of phantom matter was introduced phenomenologically in \cite{Grande:2006nn}. However, more recently, it has received important theoretical support,  as it appears in stringy versions of the RVM \,\cite{PhantomVacuum2021}; and also phenomenological support, as it can  help cure the cosmological tensions\cite{Gomez-Valent:2024tdb,Gomez-Valent:2024ejh,Gomez-Valent:2025mfl}.

\subsection{Equation of state of the quantum vacuum}\label{sec:EosVacuum}

In the RVM scenario, the vacuum EoS for the post-inflationary universe (including the radiation- and matter- dominated epochs)  was studied in detail in\,\cite{CristianJoan2022b,SolaPeracaula:2025yco}   and we limit ourselves to report the results. Calculation of the vacuum pressure in the RVM renders
\begin{equation}\label{eq:PressureRVM}
\begin{split}
\Pv(M)=&-\rv(M)+\frac{\left(\xi-\frac{1}{6}\right)}{8\pi^2}\dot{H}\left(m^2-M^2-m^2\ln\frac{m^2}{M^2}\right)\\
&-\frac{3}{8\pi^2}\left(\xi-\frac{1}{6}\right)^2\left(6\dot{H}^2+3H\ddot{H}+\vardot{3}{H}\right)\ln \frac{m^2}{M^2}+ f(\ddot{H}, \vardot{3}{H},...) \, .
\end{split}
\end{equation}
The two explicitly displayed corrections correspond to the second and fourth adiabatic orders.  The correction proportional to $\dot{H}$ in the first line can be relevant, since it is of the order $H^2$ and can influence the late-time universe. Nevertheless, the second line of the above expression can be entirely neglected today. In particular,
the  function $f(\ddot{H}, \vardot{3}{H},...)$   emerges  from higher order adiabatic terms that we need not show explicitly here, see \cite{CristianJoan2022a,CristianJoanSamira2023}. It suffices to say that $f(\ddot{H}, \vardot{3}{H},...)=0$ for $H=$ const.  and that these terms are irrelevant after inflation. The above result shows that the term in parentheses in the pressure equation \eqref{eq:VacuunPressure}, which was identically null for de Sitter spacetime, is non-vanishing for the RVM in FLRW spacetime. This introduces a crucial distinction between these two models. In both cases the VED is dynamical, as we have seen in sections \ref{sec:VEDinRVM}. and \ref{sec:VEDindS}.  However, only for the RVM the fundamental EoS itself is dynamical due to non-vanishing quantum effects. This is a remarkable prediction of QFT concerning the properties of the RVM quantum vacuum. Expressed as a function of the redshift, the dynamical EoS reads\cite{CristianJoan2022b}:
\begin{equation}\label{eq:EoS2}
w_{\rm vac}(z)=\frac{\Pv}{\rv}=-1 +\frac{\nu_{\rm eff}\left(\Omega_{\rm m}^0 (1+z)^3+\frac{4}{3}\Omega_{\rm r}^0 (1+z)^4\right)}{\Omega_{\rm vac}^0+\nu_{\rm eff}\left(-1+\Omega_{\rm m}^0 (1+z)^3 +\Omega_{\rm r}^0(1+z)^4+\Omega_{\rm vac}^0 \right)}\,,
\end{equation}
where  $\nueff$ is defined in \eqref{eq:nueffAprox2} and  $\Omega^0_i=\rho^0_i/\rho^0_c=8\pi G_N\rho^0_i/(3H_0^2)$ are the current cosmological parameters for matter and radiation. It is remarkable that the EoS of the cosmological vacuum in the RVM appears to be a function of the cosmological redshift rather than being stuck at the traditional value $w_{\rm vac}=-1$. The deviation is proportional to $\nueff$ and is therefore caused by the quantum effects of the underlying QFT. In the remote future, i.e. for $z\to-1$, we find $\wv\to-1$, as it should, since then we retake a pure inflationary era.

\begin{figure}[t!]
    \centering
    \includegraphics[width=\linewidth]{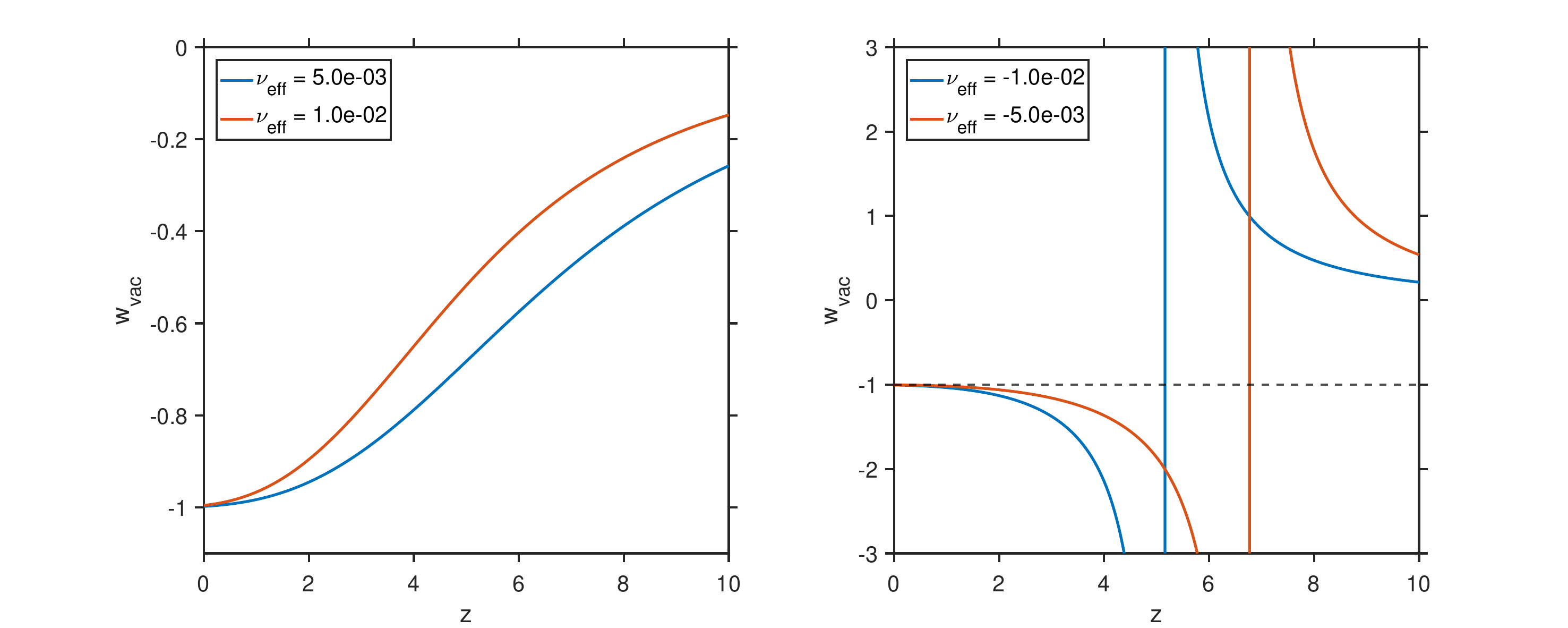}
    \caption{Equation of state parameter $w_{\rm vac}$ in the recent universe within the RVM (Eq. \eqref{eq:EoS2}) for $\nueff > 0$ (left plot) and $\nueff < 0$ (right plot), corresponding to effective quintessence and phantom behaviors, respectively.}
    \label{Fig:EoSplots_RVM}
\end{figure}

In Fig.\,\ref{Fig:EoSplots_RVM} we illustrate these effects in a numerical example.  The plot on the left shows the RVM vacuum mimicking quintessence, whereas the plot on the right shows the RVM vacuum mimicking phantom DE. The vertical asymptote in the last case is related to the fact that the VED vanishes around the redshift $z\simeq 5$. This does not correspond to any singular behavior since at all times the physical quantities (pressure and density) remain finite. As formerly mentioned, typical related situations have been previously reported in the literature on pure phenomenological grounds; see,  e.g. \cite{SolaStefancic,Das:2005yj,Basilakos:2013vya}. However, in the present case, these results emerge for the first time from  QFT calculations in curved spacetime.

%%%%%%%%%%%%%%%%%%%%%%%%%%%%%%%%%%%%%%%

For very low redshift  $z\ll1$ the above expression for the EoS boils down to the simpler form
\begin{equation}\label{EqStateScalar}
w_{\rm vac}(z)\simeq -1+\nu_{\rm eff}\frac{\Omega_{\rm m}^0}{\Omega_{\rm vac}^0}(1+z)^3\,.
\end{equation}
As expected, it stays close to the classical value $\wv^{\rm cl}=-1$, which is usually assumed to be the canonical value for the vacuum state. However, for $\nueff>0$,  the above equation shows in a transparent way that the VED behaves as quintessence ($w_{\rm vac}(z)\gtrsim-1$) in the late universe around our time, whereas for $\nueff<0$ it adopts an effective phantom DE behavior ($w_{\rm vac}(z)\lesssim-1$). Furthermore, the more precise formula \eqref{eq:EoS2} shows that at higher and higher redshift the EoS parameter $\wv\to 0$ or $\wv\to 1/3$ depending on whether the universe sits in the matter- or radiation-dominated epochs, where the redshift evolution is driven by the terms $(1+z)^3$ and $(1+z)^4$, respectively. We mention in passing that the EoS of the RVM vacuum also receives contributions  from the fermion fields\cite{CristianJoanSamira2023}.  However, including fermions leads to a cumbersome discussion, which requires performing a renormalization of the EMT  contributions from the  fermion sector. To keep our presentation more focused, we chose not to discuss these additional contributions in this work.

While in the RVM we get small departures of the  vacuum EoS from the canonical one,  we have shown in Sec. \ref{sec:RenPressure} that the EoS of the unstable de Sitter vacuum stays fixed at $\wv=-1$ throughout the cosmic history. Thus, at a fundamental level the rigid EoS in the de Sitter case is different from the dynamical one in the RVM case. Phenomenologically, however, this does not make a big difference, for we have shown in the previous section that they both lead to the same law for the late time running vacuum evolution, and this feature is what determines the `effective DE behavior' of vacuum, which is  the only one accessible to current observations. This is a remarkable feature of our unified approach. The net result is that when either of these two models is analyzed through standard parameterizations of the DE they can both lead to effective quintessence or phantom DE behavior\cite{SolaStefancic,Das:2005yj, Basilakos:2013vya}. Recently, this feature has also been studied for a variety of DE models in light of the latest observations\,\cite{deCruzPerez:2025dni}.

Let us close this section by mentioning that in the old versions of the RVM (see \cite{JSPRev2013,JSPRev2015} and references therein), the relation $\wv=-1$ was just assumed. The existence of quantum corrections to this relation was unveiled much later, when the RVM structure was derived from explicit QFT calculations in curved spacetime\cite{CristianJoan2020,CristianJoan2022a,CristianJoan2022b,CristianJoanSamira2023}. Thus, the current results show that the unstable de Sitter scenario somehow behaves as the old RVM, in that $\wv=-1$ holds exactly for the whole cosmic history, while in contrast the modern (QFT) RVM formulation has an evolving EoS, $\wv=\wv(z)$.

\subsection{Particle and entropy production in $H^4$-inflation}\label{sec:Thermoidynamics}

In the two $H^4$ inflationary scenarios i) and ii) defined in Sec.\,\ref{sec:UniScenarios} being  studied here, inflation is associated with a period where $H\simeq H_I$ remains approximately constant during inflation. It is well-known that when we can have this kind of situation in some particular framework, we are then in a position to solve the horizon problem. Indeed, a light pulse that began in
the remote past at $t=t_1\gtrsim t_i$ (shortly after inflation started at $t_i$) will have traveled until the end of inflation,
$t_f$, the physical distance\begin{equation}\label{eq:horizonInfinite}
d_{H}(a_f)= a_f\int_{a_1}^{a_f}\frac{d
a'}{a'^2\,H}\simeq\frac{a_f}{H_I}\left(\frac{1}{a_1}-\frac{1}{a_f}\right)\simeq
\left(\frac{a_f}{a_1}\right)\,H_I^{-1}\,,
\end{equation}
where $H\simeq H_I$ in this period, and we use the fact that  $a_f\equiv a(t_f) \sim e^{H_I t_f}$ is  exponentially greater than
$a_1\sim  e^{H_I t_1}$ at the end of inflation ($t_f\gg t_1$). By the same token,  the
above integral (the particle horizon) can be as large as desired.
As a result, all entropy production can be causally produced, in contrast to the standard $\CC$CDM  model for which $d_H(a)\sim a^{3/2}$ in the matter-dominated epoch, and $d_H(a)\sim a^{2}$ in the radiation-dominated epoch. Hence in both cases  $d_H(a)/a\rightarrow 0$ for $a\to 0$ (namely the observers become fully isolated in the remote past).

The horizon problem in the $\CC$CDM is intimately connected with the famous entropy problem, which challenges the standard thermal history account of cosmic history\,\cite{KolbTurner}.  In this way, we are naturally led to thermodynamical considerations on the cosmological evolution. Thermodynamical methods in the context of cosmology were pioneered in the  well-known seminal works \cite{Prigogine1986,Prigogine:1988jax,Prigogine:1989zz}. Such studies involve, of course, the notion of particle and entropy production. As indicated previously, this phenomenological approach has also been applied in the RVM context and other scenarios, cf. \cite{Sola:2015csa,BLS2015,Yu2020,Lima:2025nhh}. However, here we make a further step since we use QFT methods to compute the evolution of the vacuum energy density and combine them with the thermodynamical description of the matter and entropy production.

According to the Second Law of Thermodynamics, the total entropy flow of an isolated system always increases until it reaches equilibrium, i.e. $\nabla_\mu s^\mu\geq0$ (equality applies only at equilibrium). Here, $s^\mu=s\,U^\mu$ is the entropy flow, with  $s=n\sigma$  the entropy density in the comoving frame, $n$  the number density of particles and $\sigma$ the specific particle entropy (the entropy of an individual particle).
Since there is a production of particles from vacuum decay, the particle flux $n^\mu=n U^\mu$ is not conserved and we have $ \nabla_\mu n^\mu=n\Gamma$, where $\Gamma$ is the particle  production rate.  In a more explicit form, in the FLRW metric,
\begin{eqnarray}\label{eq:dotn}
\dot{n}+3H n=n\Gamma\,.\label{balanceLaw}
\end{eqnarray}
On the other hand, the corresponding energy density  of particles is conventionally expressed as follows:
\begin{eqnarray}
\dot{\rho}_{\rm rad}+3H (\rho_{\rm rad}+p_{\rm rad})=\beta n\Gamma\,,\label{AnomalousConservLaw}
\end{eqnarray}
where $\beta>0$ is a dynamical quantity with energy dimensions. From Eq.\,\eqref{eq:dotn}  we immediately find  that the particle  production rate can be written as expected:
$\Gamma={\dot N}/{N}$, where $N=n a^3$ is the total number of particles in the comoving volume. If we reach equilibrium and the number of particles is conserved, $\Gamma=0$ and we recover the standard conservation laws for the number of particles and for the energy density: $\dot{n}+3Hn=0$ and $\dot{\rho}_{\rm rad}+3H (\rho_{\rm rad}+p_{\rm rad})=0$.
Change of particle number and variation of entropy are connected. In fact, the relative cosmic time variation of the total entropy $S$  in the comoving volume $a^3$ is related to the variation in the number of particles in that volume and to the relative variation of the specific entropy per particle.  Using $S=N\sigma$, we have
\begin{eqnarray}
\frac{\dot S}{S}=\frac{\dot N}{N}+\frac{\dot\sigma}{\sigma}
=\Gamma+\frac{\dot\sigma}{\sigma}\,,\label{dotentropy1}
\end{eqnarray}
or, equivalently,
\begin{eqnarray}
\frac{dS}{dt}=S\left(\frac{\dot\sigma}{\sigma}+\Gamma\right)=N\dot{\sigma}
+N\sigma\Gamma\,.\label{dotentropy2}
\end{eqnarray}
Clearly, particle and entropy production are correlated.  In the frequent particular case in which the entropy per particle remains constant, i.e. $\dot{\sigma}=0$ (adiabatic process),  we have
\begin{equation}\label{eq:sigmaconstant}
\frac{\dot S}{S}=\frac{\dot N}{N}=\Gamma\,.
\end{equation}
Assuming also that the particle production from vacuum is adiabatic, such that some basic thermodynamic equilibrium relations are preserved, one finds that the quantity $\beta$ introduced previously in Eq.\,(\ref{AnomalousConservLaw}) becomes determined as follows\cite{Calvao:1991wg,Lima1996,Yu2020}:
\begin{equation}\label{betaequation2}
  \beta=\frac{\rho_{\rm rad}+p_{\rm rad}}{n}\,.
\end{equation}
In addition, since particle production stems from the interaction between vacuum and matter, the local energy density conservation law can also be expressed as
\begin{eqnarray}
\dot\rho_{\rm rad}+3H(\rho_{\rm rad}+p_{\rm rad})=-\dot{\rho}_{\rm vac}\,,\label{energyconservation2}
\end{eqnarray}
provided the vacuum satisfies the canonical EoS $\Pv=-\rv$. We have shown that this is strictly true in the exact de Sitter scenario, even in the presence of quantum effects,  whereas it is only approximately correct in the RVM.  But even in the latter case, that canonical vacuum condition  is very accurately fulfilled during the inflationary period when the vacuum decays into radiation.  Therefore, we assume that it holds in both cases during this period. It then follows from the consistence of the above equations that
\begin{eqnarray}
\Gamma=\frac{-\dot{\rho}_{\rm vac}}{n\,\beta}=\frac{-\dot{\rho}_{\rm vac}}{\rho_{\rm rad}+p_{\rm rad}}=\frac{-\dot{\rho}_{\rm vac}}{(1+w_{\rm rad})\rho_{\rm rad}}=-\frac34 \frac{\dot{\rho}_{\rm vac}}{\rho_{\rm rad}}\,,
\label{relation1}
\end{eqnarray}
where $w_{\rm rad}=p_{\rm rad}/\rho_{\rm rad}=1/3$ is the EoS of relativistic matter. This result is reasonable: if the entropy per particle is constant ($\dot\sigma=0$), then, when the vacuum decays ($\dot\rho_{\rm vac}<0$),  its energy is fully invested in the creation  of new particles ($\Gamma>0$),  whereas if $\dot\rho_{\rm vac}>0$ the particles disappear into vacuum ($\Gamma<0$).

 From Eq.~(\ref{relation1}) combined with the analytical densities in the inflationary regime,  Eqs.~(\ref{eq:infaprox2}) and (\ref{eq:infaprox3}), which are valid to high accuracy both for the de Sitter and RVM scenarios (with corresponding values of $\nu$), we find the expression for the particle production rate:
%\begin{eqnarray}\label{17}
%\Gamma(\hat a)=\frac{3 H_I \sqrt{1-\nu} \left[2-\nu+ \nu(1-2\nu)^{-1} \hat a^{4(1-\nu)}\right]}{\left[1+ (1-2\nu)^{-1}\hat a^{4(1-\nu)}\right] \sqrt{\alpha  (1-2\nu)^{-1}\hat a^{4(1-\nu)}+\alpha }}>0,
%\end{eqnarray}
\begin{eqnarray}\label{17}
\Gamma(\hat a)=3 H_I\, \frac{2-\nu+ \nu\, \hat a^{4(1-\nu)}}{\left[1+ \hat a^{4(1-\nu)}\right] \sqrt{1+\hat a^{4(1-\nu)}}}>0\,.
\end{eqnarray}
The particle production rate is positive and remains approximately constant in the initial stages  $\ha\ll1$, where $\Gamma\simeq 6 H_I$.  Such a large value of the primeval $\Gamma$ leads to a massive production of particles from  vacuum decay during inflation.  For $\ha\gg 1$, instead,  it behaves much more tamed, just as
\begin{eqnarray}\label{17b}
\Gamma(\hat a\gg1)\simeq3\nu\, \frac{H_I}{\sqrt{1+\hat a^{4(1-\nu)}}}=3\nu H(\hat{a}),
\end{eqnarray}
where we have used \eqref{eq:infaprox}. Because $H$ decreases and  $|\nu|\ll1$, $\Gamma$ eventually  becomes much smaller than $H_I$, but in the asymptotic limit there is still a particle production rate, and hence for $\nu>0$ the  vacuum continues (mildly) decaying into particles even in the post-inflationary era up to our days.

What about the entropy produced in the early times and how does it compare to the current observations? As previously noticed, the above thermodynamical  account is essentially valid for both unstable de Sitter vacuum and the RVM.  Therefore, after exiting the inflationary period we may use the analytical formulas for the radiation  to compute analytically the radiation entropy of the produced relativistic particles. We know that the corresponding energy density increases as the fourth power of the temperature,
\begin{eqnarray}\label{eq:BBformula}
\rho_{\rm rad}=\frac{\pi^2}{30}g_\ast T_{\rm rad}^4\,,
\end{eqnarray}
where again factor $g_\ast$ accounts for the total number of effectively massless d.o.f. at a given temperature\,\cite{KolbTurner} (e.g. $g_*=160.75$ in non-supersymmetric $SU(5)$).\\

If we consider the de Sitter scenario, the value of $\nu=\nu(H)$ in the primeval period is a slow function of $H$ which, as noted, can be estimated by setting  $H=H_I$ in Eq.\,\eqref{eq:nutildeH}. We find that it is very small, of the order of $|\nu|\sim 10^{-5}$ at most,  even after including a multiplicity factor $N_0=1000$.  This result can also  be estimated using the simpler formula \eqref{eq:nutildeapprox}.  Recall that at later times the value of $\nu$ is different and is given by Eq.\,\eqref{eq:barnueff2} for $H\simeq H_0$. Numerically, the late value can be much larger owing to the large logarithm $\ln(m^2/H_0^2)$ in the current universe. As a result, $\Tilde{\nu}_{\rm eff}$ can reach $10^{-4}-10^{-3}$ in order of magnitude, and this can have beneficial consequences for the observed phenomenology\,\cite{ SolaPeracaula:2021gxi,SolaPeracaula:2023swx,deCruzPerez:2025dni}. Regarding the transition scale factor from inflation to radiation, $a_*$, it is given by Eq.\,\eqref{eq:astar}. Numerically, in the case of the de Sitter scenario around the Planck scale, we find\footnote{Just for comparison, recall that the equality point between radiation and nonrelativistic matter occurs much later in the cosmic expansion, viz. around
$a_{EQ} \sim 3 \times 10^{-4}$ ($z\simeq 3400$). As expected,  $a_{\rm eq}\simeq a_*\lll a_{EQ}$.}
\begin{equation}\label{eq:astar2}
a_{*}\sim
\left(10^{-4}\,\frac{10^{-47}}{10^{76}}\right)^{1/4}\simeq 2\times10^{-32}\,.
\end{equation}
Here we used $\rho_I\simeq\frac{3}{\kappa^2} {H}_I^2\lesssim\mpl^4\sim 10^{76}$ GeV$^4$ in accordance with our discussion in Sec.\,\ref{sec:AnaliticalNumerical}. We shall use this estimate for  $a_{*}$ shortly.\\

\begin{figure}
    \centering
    \includegraphics[width=0.49\linewidth]{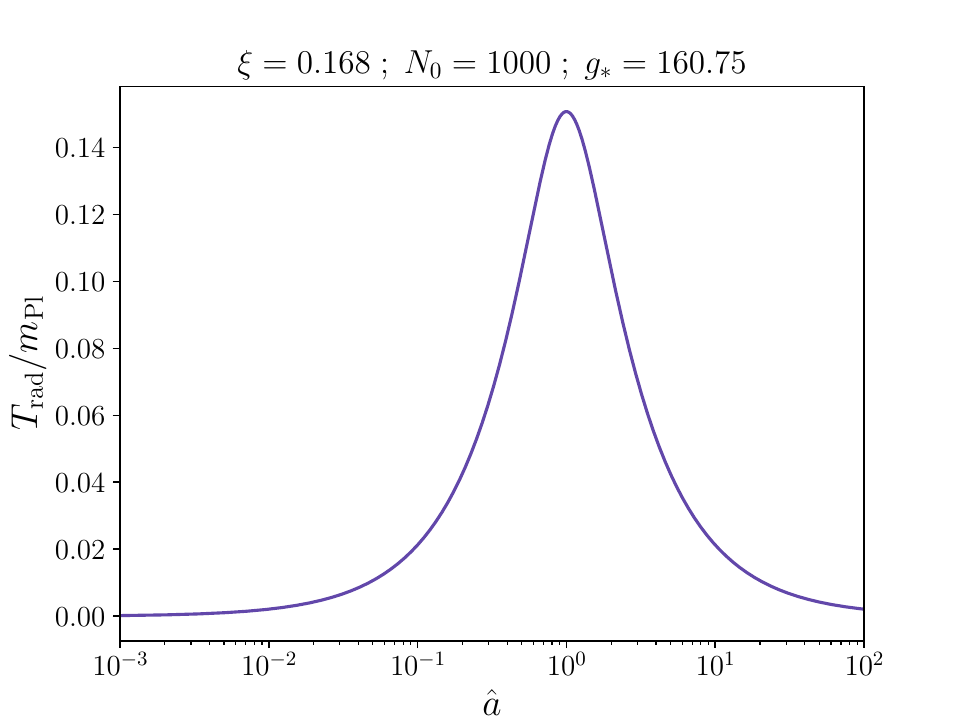}
    \includegraphics[width=0.49\linewidth]{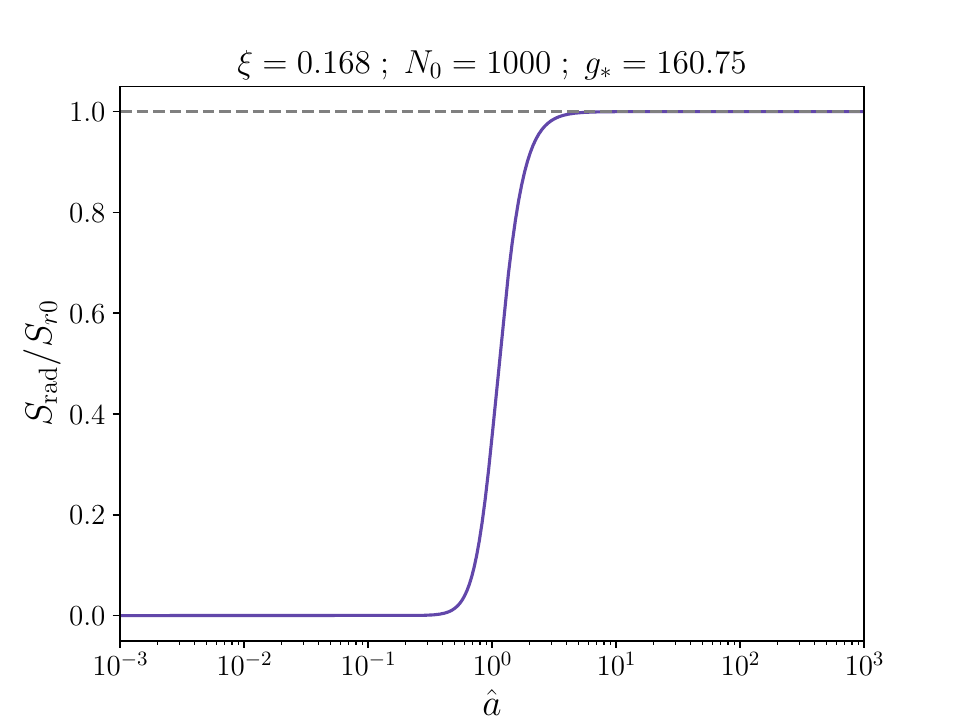}
    \caption{Evolution of temperature and entropy of the heat-bath of radiation as the inflationary stage ends. We illustrate the result for $\xi=0.168$, $N_0=1000$, $g_{*}=160.75$ and $m=0.001\mpl$ using the exact formulas. The particle entropy is initially very small but it rockets fast until a saturation value. }
    \label{fig:temp_entropy}
\end{figure}

As we have seen above, the particle production rate associated with vacuum decay is extremely large, $\Gamma\simeq 6 H_I$,  during the inflationary period, although the vacuum is still dominant until we approach the transition point $\hat{a}\sim 1$.  The relativistic particles from vacuum decay constitute a heat-bath of radiation whose temperature can be computed by equating the radiation density $\rho_{\rm rad}(a)$ given by \eqref{eq:infaprox3} to the black-body formula \eqref{eq:BBformula}.  This yields
\begin{equation}\label{eq:TempRad}
T_\mathrm{rad}(\hat{a})=T_I\,(1-\nu)^{1/4} \frac{\hat{a}^{(1-\nu)}}{\left[1+\hat{a}^{4(1-\nu)}\right]^{1/2}}\,,
\end{equation}
where  $T_I$ is related to $\rho_I$ through $\rho_I=\frac{\pi^2}{30}\,g_\ast\,T_I^4$.  In the left plot of Fig. \ref{fig:temp_entropy}  we show this function.
The maximum shown in that figure can be computed easily  and  is achieved exactly at the transition  point $\hat{a}=1$ ($a=a_*$), attaining the value
\begin{equation}\label{eq:Tmax}
T_\mathrm{max}=\frac{T_I}{\sqrt{2}}\,(1-\nu)^{1/4} = \frac{1}{\sqrt{2}}\left[\frac{30\,\rho_I\,(1-\nu)}{\pi^2 g_*}\right]^{1/4}=\left[\frac{15\times 135 (1-\nu)}{2 \pi^2 g_* N_0}\right]^{1/4}\,\mpl\simeq 0.16 \,\mpl\,,
\end{equation}
for $N_0=1000$ and $g_*=160.75$, upon neglecting  terms of order $|\nu|\ll 1$. The above result for the maximum temperature is indeed consistent with our first estimate made in Sec.\ref{sec:AnaliticalNumerical}.  It follows that the maximum  temperature is one order of magnitude below the Planck mass, which is meaningful. As a matter of fact the temperature remains much smaller than that in most of the inflationary regime, as can be appraised in Fig. \ref{fig:temp_entropy}. The temperature raises up to the maximum at $\hat{a}=1$ and then drops again to much smaller values.  Notice that for $\hat{a}\gg1$ (i.e. $a\gg a_{*}$, corresponding to a region  deep into the radiation-dominated epoch) the scaling of the temperature \eqref{eq:TempRad} with the scale factor goes as
\begin{equation}
T_\mathrm{rad}\, a^{1-\nu}={\rm const.}
\end{equation}
Recalling that $|\nu|\ll 1$,  we recover in very good approximation the canonical scaling law of the adiabatic regime: $T_\mathrm{rad}\propto 1/a$ up to small scaling corrections of order $\nu$.

Let us consider now the evolution of the radiation entropy associated to vacuum decay into relativistic particles. The corresponding (comoving) radiation entropy is given by
$S_{\rm rad}= \left(4\rho_{\rm rad}/3T_{\rm rad}\right) a^3$\cite{KolbTurner}. It follows that
\begin{eqnarray}\label{eq:SrRVM}
S_{\rm rad}(\hat{a})=\frac{2\pi^2}{45}\,g_\ast T_{\rm rad}^3a^3=\frac{2\pi^2}{45}\,g_{*}\,T_I^3\,\astar^3\,(1-\nu)^{3/4}\,\frac{\ha^{6-3\nu}}{\Big[1+\ha^{4(1-\nu)}\Big]^{3/2}}\,.
\end{eqnarray}
The above radiation entropy is displayed in Fig. \ref{fig:temp_entropy} (right plot) side by side with the corresponding radiation temperature (on the left). We can see that during the heating-up period the comoving entropy rockets approximately as the sixth power of the scale factor, $S\sim \hat{a}^{(6-3\nu)}\sim \hat{a}^6$, until it finally reaches an approximate saturation plateau in the radiation-dominated phase:
\begin{equation}\label{eq:SrSaturation}
  S_{\rm rad}(\ha\gg1)\simeq \frac{2\pi^2}{45}\,g_{*}\,T_I^3\astar^3(1-\nu)^{3/4}\ha^{3\nu}
  \equiv S_{r0} \ha^{3\nu}\,.
  \end{equation}
It is not exactly a flat plateau for $\nu\neq 0$, but since $\nu$ is very small, the ulterior evolution of the entropy is much more tempered.
Equation \eqref{eq:SrSaturation} stands for  the (approximate)  asymptotic comoving entropy\,\footnote{We do not consider here the entropy contribution from the horizon, which is associated with the Generalized Second Law; for the RVM it has been computed in \cite{Yu2020}.}. For  $\nu=0$ the quantity $g_{*}T_{\rm rad}^3 a^3$ becomes
conserved during the adiabatic phase  and hence it must   equal the current value
$g_{s,0}\,T_{\gamma 0}^{3}\,a_0^{3}$, in which $T_{\gamma 0}\simeq 2.725\,$K
(CMB temperature now) and  $g_{s,0}=2+6\times
(7/8)\left(T_{\nu,0}/T_{\gamma 0}\right)^3\simeq 3.91$ is the entropy factor
for the light d.o.f. today, computed from the ratio of the present neutrino and photon temperatures. Thus, the huge entropy enclosed in our horizon today, $H_0^{-1}$, is \footnote{The physical volume of the universe is related to the comoving volume as $V = a^3 L^3$, where $L^3$
is the coordinate volume at present.  For simplicity, we have previously assumed the coordinate
volume $L^3 = 1$, but for the present consideration we take the horizon $L=H_0^{-1}$, and hence the physical volume at present ($a=1$) is $V_0=\left(H_0^{-1}\right)^3$, which leads to Eq.\,\eqref{eq:S0}.}
\begin{equation}\label{eq:S0}
S_{0}=
\frac{2\pi^2}{45}\,g_{s,0}\,T_{\gamma 0}^3\,\left(H_0^{-1}\right)^3\simeq
2.3 h^{-3} 10^{87}\sim 10^{88} \ \ \ \ \ \ (h\simeq 0.7).
\end{equation}
This huge number cannot be accounted for in the standard model of cosmology without violating causality, this being the origin of the entropy and horizon problems\cite{KolbTurner,LiddleLyth}. In the  unified framework that we have considered here (both within the RVM and unstable de Sitter vacuum), the large (and causally generated) entropy at
the end of the inflation period is bulk-transferred to the radiation phase, and then it is preserved
by the standard (adiabatic) evolution, up to a small $\nu$-dependent correction. Thus, the observed entropy was causally produced in our remote past, and the result does not depend
on the details of the underlying GUT.

Only within a rough order of magnitude, it is easy to convince oneself that the numbers fit pretty well. In fact, we can see from the above formulas that in order to match the desired total amount of entropy in our unified scenario, we need to  fulfill the condition  $g_*T_I^3 a_*^3\sim g_{s,0} T_{\gamma 0}^3 a_0^3$ (with $a_0=1$ at present). This relation connects the inflationary epoch and the current epoch. Hence it is a prime and nontrivial consistency condition to be fulfilled, as  $T_{\gamma 0}\simeq 2.7\,K\sim 2.3\times10^{-13}$ GeV is a measured quantity at present, whereas $T_I$ and $a_*$ are primordial parameters that belong to the very early universe and which we have previously estimated in our theoretical framework. The matching is therefore relevant and constitutes a significant result. Let us check it within an order of magnitude using, e.g. our unstable de Sitter scenario, see \cite{SolaPeracaula:2025yco} for the RVM case.
From the estimate \eqref{eq:astar2} for $a_*$ and the typical value of the $g_*$ factor for a GUT that we have used previously, we find the temperature at which the radiation entropy was produced and subsequently leveled off:
\begin{equation}\label{ee:TIestimate}
T_I\sim \left(\frac{g_{s,0}}{g_*}\right)^{1/3} \frac{T_{\gamma 0}}{a_*}\sim\left(\frac{3.91}{160.75}\right)^{1/3} \frac{2.3\times 10^{-13}}{2.0\times 10^{-32}}\ {\rm GeV} \simeq 0.2\, \mpl\,,
\end{equation}
where $\mpl\simeq 1.22\times 10^{19}$ GeV. The estimated temperature $T_I$ turns out to be about one order of magnitude below the Planck mass. This fits fairly well, in order of magnitude, with our original estimates for $T_I$, see Eqs.\,\eqref{eq:TempdeSitter2} and \eqref{eq:Tmax}, demonstrating the numerical consistency of our analysis. Needless to say, the values of $a_*$  and $g_*$ are sensitive to the details of the particular  GUT under consideration near the Planck scale.  However, the sensitivity is mild and it is rewarding to see that the orders of magnitude fit reasonably well. Therefore, we conclude that the huge entropy \eqref{eq:S0} of the current universe can be explained within our overarching QFT model of the cosmic evolution. This is an encouraging result, demonstrating once more the internal consistence of our unified framework for the cosmic evolution from the very early times until today.

We should emphasize that the above discussion on inflation has been derived within a QFT formulation that goes well beyond previous phenomenological considerations on these matters. Our approach provides a theoretical basis for a possible solution to these cosmological problems within fundamental physics. As noted in \cite{Sola:2015csa}, the clues to the cosmological problems of the present may  have profound roots in the past. But for this to be useful we need an overarching picture of the cosmological evolution that links the different cosmic epochs within one and the same theoretical paradigm, starting from the very early universe until the present time.  The unified QFT approach that we have presented here seems to meet this important condition.

%\newpage

\section{Discussion and conclusions}\label{sec:conclusions}

In this paper, we have emphasized the notion of quantum vacuum in the context of fundamental physics. This applies, in particular, to quantum field theory (QFT) in curved spacetime and hence it suggests that the vacuum energy density (VED)  should also play a relevant role in the fundamental description of the cosmological evolution. From this point of view, the quantum dynamics of vacuum should naturally be connected to what we call (dynamical) dark energy (DE) in the present universe and even to the mechanism of inflation in the early universe. Our QFT approach is semiclassical in that only the quantized matter fields determine the quantum vacuum energy. This part is safe and established, whereas quantum gravity (QG) considerations can wait for a final quantum theory of the gravitational field, which does not exist at present.  Two QFT scenarios along these lines have been addressed here, both of them aimed at a unification purpose in our description of the cosmic history, to wit:
scenario i) is the running vacuum model (RVM)\cite{JSPRev2022, JSPRev2013,JSPRev2015}, which we have revisited in light of the recent works\cite{CristianJoan2020,CristianJoan2022a,CristianJoan2022b,CristianJoanSamira2023,SolaPeracaula:2025yco}; scenario ii) is  unstable de Sitter vacuum, which we have analyzed in detail here for the first time within the off-shell adiabatic renormalization scheme proposed in \cite{CristianJoan2020,CristianJoan2022a}. In both cases, we have addressed the renormalized energy-momentum tensor (EMT) of a quantized scalar field non-minimally coupled to the FLRW background and derived the corresponding VED. These scenarios consistently lead to a dynamical solution for the cosmic vacuum, with vast implications for the description of the cosmological history. In particular, we have demonstrated the consequences of vacuum dynamics for the current universe and also for the very early universe. In the former, the evolution of the quantum vacuum leads to a possible QFT explanation for DE as dynamical VED without invoking quintessence and the like, whereas in the latter we find a new mechanism of inflation without appealing to {\it ad hoc} inflaton fields. \jtext{Our study depicts a possible unified framework for DE and inflation which might be  possible on fundamental grounds, as we summarize below.  However, as it is natural  for a comprehensive work of this sort,  further investigations are mandatory in order to confirm the full viability of this unified approach, which is presented here as a promising proposal in its preliminary stage.}

The RVM quantum field approach leads to an off-shell adiabatic expansion in an even number of derivatives of the scale factor, which is actually demanded by general covariance\cite{JSPRev2022}. The method is particularly useful for  solving QFT in slowly evolving cosmological backgrounds, since, in general, an exact field treatment does not exist in these cases. Such is indeed the case for FLRW spacetime, so the solution to the field equations is only  approximate and can be expressed as an adiabatic WKB series. In contrast, for the de Sitter spacetime, an exact solution is possible in the early universe. De Sitter spacetime can be suitable as an initial condition in cosmic evolution; although it cannot be, of course,  a realistic picture for the subsequent expansion history since inflation would never end. The origin of such an initial condition might ultimately be connected to QG or even to stringy versions of the RVM\,\cite{Mavromatos:2020kzj}, despite that these theoretical realms fall well beyond the scope of the semiclassical QFT approach that we have followed here; see, however,  \cite{NickJoan_PR} for a comprehensive exposition wherein  these ideas are explored in the stringy domain as well.

After reviewing the RVM scenario, we have focused on our main computation,  de Sitter spacetime, which is an exactly soluble QFT.  For the mentioned reasons, we treat it as an unstable background that decays early on (during inflation) into a radiation state which can be conceived as the primeval radiation epoch of the standard FLRW cosmology. The cosmological equations of the decay process can no longer be solved exactly, so  semianalytical and numerical methods are mandatory at this point. The semianalytical approximation is very helpful for understanding the physics of the exact numerical solution.
Furthermore, since we are in a QFT context,  renormalization is necessary. We use off-shell adiabatic renormalization of the EMT, first introduced in \cite{CristianJoan2020,CristianJoan2022a}. In both unification scenarios, the renormalized theory leads to a unified account  of the cosmological evolution, which differs very little from the standard $\CC$CDM cosmological model in the late universe but provides a significant completion of the cosmic history at very early times; in particular, it provides a specific $H^4$-mechanism for the inflationary phase and its connection with the radiation-dominated epoch, where the VED is  driven by the subleading power $H^2$. During the radiation phase, the VED is negligible  compared to the radiation density, so primordial BBN and the standard thermal history are unaffected. \jtext{The net outcome is  a possible unified QFT framework that might provide a natural link  between inflation and the FLRW regime. No such link exists in the  $\CC$CDM model, in which inflation is not even entertained and  a rigid $\CC$ is just imposed \textit{ad hoc}. So the fact that a unification scheme can be devised  within renormalizable QFT in curved spacetime looks promising, in principle,  and it may be  a good starting point for further investigations aimed at verifying its full consistency.}

The aforementioned prescription for off-shell adiabatic renormalization depends on a floating scale $M$\cite{CristianJoan2020,CristianJoan2022a}. The appearance of such a scale is characteristic of the renormalization group approach and is to be expected in any renormalization scheme in QFT owing to the intrinsic breaking of conformal invariance by quantum effects. We find that in the on-shell situation we recover the classical Bunch-Davies result obtained from  point-splitting renormalization \cite{Bunch:1978yq}, and also with other methods  \cite{Dowker:1975tf}, which constitutes a nontrivial check of our calculation.  However, the presence of the scale $M$ in our case gives room for further physical exploration in the cosmological arena. This approach to cosmic expansion within a QFT context is in the very spirit of the RVM\cite{JSPRev2022} and is unprecedented in the literature to the best of our knowledge. The scale $M$ plays the role of a renormalization point and, therefore, following the usual practice in ordinary gauge theories, physical contact is made by eventually fixing $M$ to the typical value of the energy of the process being explored. In particle physics, for example, this is accomplished through the energy scale defined by the center-of-mass energy of a scattering event, or the mass of the particle in a decay. However, in the cosmological context, the ``process'' is just the universe evolution itself, and hence we can explore the expansion history by fixing $M$ at the value of the Hubble rate $H$ at each cosmological epoch\footnote{Incidentally, this setting has also been recently tested with success in the lattice\,\cite{Dai:2024vjc}.}.  As indicated previously, upon renormalizing the EMT according to the above procedure, the VED takes on the form of an expansion in an even number of time derivatives of the scale factor,  which can conveniently be rephrased in terms of powers of $H$ and its time derivatives, $\rho_{\rm vac}=\rho_{\rm vac}(H, \dot{H},\ddot{H}, \dots)$.  The obtained expression for the VED can then be used to explore the entire history of the universe from the very early times where inflation occurs, going through the radiation- and matter-dominated epochs until reaching the current DE epoch. In this framework, what we call the dark energy density can be thought of as the remnant tail of the huge vacuum energy that brought about the exponential inflation in primordial times and is still decaying very slowly at present. The DE shows up in the manner of an unsuspected  ``fossil energy''\cite{{Fossil2008}} reminiscent of our highly energetic past.  In fact, the two QFT scenarios under consideration i) and ii) consistently predict that the DE is dynamical and is ultimately connected  to the vacuum energy of primordial inflation.

\jtext{As a result of the former considerations, perhaps the principal message of our work  is that the measured  VED, $\rho_{\rm vac}^0 =\rho_{\rm vac}(H_0)$, may  not be a constant throughout the cosmic history, and hence that  the cosmological `constant' $\Lambda=8\pi G\,\rho_{\rm vac}$ may not really be a fundamental constant of nature. In our context, instead, we find that they are effective (``running'') quantities,  therefore suggesting that the VED is dynamical, $\rv=\rv(H)$, and the associated cosmological term too, $\Lambda=\Lambda(H)$. }  This means that the observed value applies only today, i.e. at $H=H_0$. Nevertheless, the evolution of the VED between two nearby epochs in recent history is very mild and is given by $\delta \rho_{\rm vac}(H)\propto \nu_{\rm eff} \mpl^2 H^2$ (similarly, $\delta\CC(H)\propto \nueff H^2$), where $|\nu_{\rm eff}| \ll 1$ is essentially the $\beta$-function coefficient of its running. Despite the fact that the running laws for scenarios i) and ii)  are different at high energies, they turn out to coincide  exactly at low energies (the current universe). This is a remarkable result of our QFT computation. Hence there is a perfect convergence of the de Sitter picture into the RVM picture in the late universe.  The common coefficient $\nu_{\rm eff}$ is a computable quantity that depends on the number of bosonic and fermionic fields responsible for the quantum effects\cite{CristianJoan2020,CristianJoan2022a,CristianJoan2022b,CristianJoanSamira2023}; and, what is more, $\nueff$ can be measured observationally.

\jtext{While further theoretical work is, of course,  necessary to verify the full consistency  of these unification scenarios, their phenomenological implications  for the late universe have actually been repeatedly tested in the literature since more than 10 years ago.}  For example, in a number of phenomenological studies, the RVM has been successfully confronted with a large number of cosmological observations, and it has been shown to seriously compete with the corresponding global fits obtained from a  $\CC$CDM description of the same data, see \cite{Sola:2015wwa,Sola:2016jky,Sola:2017znb,SolaPeracaula:2016qlq,Sola:2016zeg,SolaPeracaula:2017esw} and \cite{SolaPeracaula:2021gxi,SolaPeracaula:2023swx}. This fact has recently been re-validated in\,\cite{deCruzPerez:2025dni}.  The preferred fitting values for the parameter $\nu_{\rm eff}$ fall in the ballpark of $\sim 10^{-4}-10^{-3}$.  It is worth mentioning that these analyses also demonstrate that the $H_0$-tension and the growth tension can be significantly alleviated in some cases\,\cite{SolaPeracaula:2021gxi,SolaPeracaula:2023swx}. Besides, it goes without saying that the fact that the running law for the VED in the de Sitter case is formally identical to that of the RVM at low energies implies, of course, that the previous phenomenological considerations translate verbatim into the de Sitter scenario as well.

In a deeper theoretical vein, perhaps the most noticeable property of the unified QFT models of cosmic evolution that we have  put forward here is the fact that the VED running is free of the undesired quartic mass contributions $\sim m^4$ from the quantum matter fields with  non-vanishing rest mass\cite{CristianJoan2020,CristianJoan2022a}. If these quartic terms were present in our approach, they would recreate the need for extreme fine-tuning, which is one of the most unpalatable aspects of the cosmological constant problem\cite{Weinberg89} -- see also \cite{JSPRev2022} for a devoted discussion.  Obviously, avoiding them  is another important theoretical achievement of the current framework.

The implications of the new mechanism of inflation in the very early universe, which is common to both scenarios i) and ii), are also significant.
Despite exhibiting non-negligible differences in the structure of the VED at high energies, the leading power $H^4$ is in both cases the driving force in those early times. Such power is irrelevant for the present universe, but plays a major role during a short period where it remains approximately constant ($H\simeq$ const.) and triggers exponential expansion. Due to this period of $H$ constancy, the new mechanism of inflation differs substantially from Starobinsky's inflation, where $H$ is never constant\cite{Starobinsky:1980te}. However, the  $H^4$- mechanism leads to a graceful exit from the inflationary phase into  the radiation-dominated era until the current epoch, the last two periods being dominated by the ordinary subpower $H^2$. As an additional bonus, the unified scenarios of cosmic evolution investigated here can overcome the flatness and horizon problems as well as the entropy problem. In fact, we have checked that the large entropy observed today can be generated during the short period of $H^4$-inflation in a way consistent with causality. This result demonstrates once more the internal consistency of our approach since it successfully connects the two opposite ends of the cosmic history within one and the same unification framework.

No less important to stand out is the fact that in either scenario we do not require the presence of spurious fields such as the `inflaton'  (nor quintessence or phantom DE at low energies, as we will comment in a moment) since inflation is brought about by pure QFT  effects on the dynamical background. The two unification frameworks are similar in the main properties, but exhibit some differences. In particular, the range of physically allowed values for the non-minimal coupling  $\xi$ is significantly more restricted in the de Sitter case than in the RVM. Another important distinction is that the fundamental EoS of the RVM is evolving in the late universe, whereas the fundamental EoS of the unstable de Sitter vacuum remains stuck to the canonical value $\wv=-1$.  However, as we further discuss below, at the phenomenological level, we do not actually measure the fundamental EoS of the quantum vacuum but the effective one associated with the cosmic evolution of the DE.

The two  QFT models that we have analyzed in this work lead to what may be called the ``renormalized $\CC$CDM'', that is,  a ``renormalization group improved''  $\CC$CDM model\cite{JSPRev2015,JSPRev2022} of the cosmic evolution, which differs very slightly from the standard $\CC$CDM in the present epoch, but with the important property that the parameters of the model acquire a nontrivial QFT status. This nuance is not just a theoretical nicety. As a practical bonus, the  renormalization effects which modify the ``tree-level'' values of the original parameters can be computed explicitly and  help cure the cosmological tensions \, \cite{SolaPeracaula:2021gxi,SolaPeracaula:2023swx}.

The renormalization framework used in our calculations also leads to an even milder (logarithmic) running of the `gravitational constant', which we express schematically  as \,$\delta G/G\sim {\cal O}(\ln H)$ -- details are provided in Appendix \ref{sec:RunninGandVED}. Although it does not alter the main quantitative results that we have obtained for the VED, it is useful to test the consistency of the running laws for $\rv(H)$ and $G(H)$. Such a logarithmic evolution of $G(H)$ is compatible with the standard tests on the variation of the fundamental constants of Nature \cite{Fritzsch:2012qc,Sola:2016our,Fritzsch:2016ewd}.   Some of these features had been considered in the old formulation of the RVM using semiqualitative arguments based on the RG\cite{ShapSol,Guberina:2002wt,Shapiro:2004ch,Babic:2004ev}.  However, the full QFT treatment of the subject was not performed until the off-shell adiabatic procedure was formally introduced in the aforementioned references\cite{CristianJoan2020,CristianJoan2022a}.

We should also mention that the analysis presented here could depend to some extent on the scale setting procedure. In general scenarios, in which there can also be exchange of energy between vacuum and matter and where other fundamental `` constants'' are varying with the cosmological expansion\cite{Fritzsch:2012qc}, one may expect that the scale setting can be slightly different, see the phenomenological analyses of this more general situation in\cite{SolaPeracaula:2021gxi,SolaPeracaula:2023swx}. We cannot know this a priori and hence it is ultimately a matter of phenomenological test. However, given a minimal scenario such as the one considered here, in which matter is not interacting with vacuum and only $\rv$ and $G$ can evolve with the expansion, the setting $M=H$ appears to be fully consistent, as demonstrated in detail  in Appendix \ref{sec:RunninGandVED}, where we also comment that the lattice simulations of the quantum vacuum in Ref.\cite{Dai:2024vjc} also favor $M=H$.

\jtext{Let us finally address a very important implication for the current universe, which we have hinted at above but on which we believe is worthwhile to put more emphasis.  Theoretical consistency for a proposal is, of course, very important, but physics is no less keen of phenomenological implications.  The following property might actually provide  smoking gun evidence for the unification scenarios of the cosmic expansion that we have been discussing in this work.  This important aspect is connected to the possibility that the running vacuum energy can mimic quintessence or phantom DE.}  In fact,  the background cosmology for these models is almost indistinguishable from the $\CC$CDM in the late universe, but the EoS of scenario i) (the RVM) deviates from the canonical value $\wv=-1$ in a potentially measurable way. In contrast, the EoS of scenario ii) (the decaying de Sitter vacuum) remains canonical. This feature makes the two models distinguishable at a fundamental level.  More specifically, in the RVM the vacuum EoS evolves with cosmic  expansion, $w=w(z)$, and can mimic quintessence  ($w_{\rm vac}(z)\gtrsim  -1$) or phantom DE ($w_{\rm vac}(z)\lesssim-1$) around the present time, depending on the sign of the running parameter $\nu_{\rm eff}$ ($>0$ or $<0$, respectively, cf. Fig.\ref{Fig:EoSplots_RVM}).  On the other hand,  the fact that the EoS of the decaying de Sitter vacuum  remains stuck at $\wv=-1$ does not preclude this scenario from mimicking dynamical DE as well, for the mentioned evolution of the VED in this model (cf. Fig.\ref{fig:VED_low}) causes a `mirage effect' of dynamical EoS\,\cite{Basilakos:2013vya} -- irrespective of its fundamental EoS. In fact, a decrease or increase of the VED with the cosmic expansion points to effective quintessence or phantom DE, respectively, when viewed through the lens of typical DE parameterizations\cite{SolaStefancic,Das:2005yj}. This is the mirage effect we were referring to.  These parameterizations may grasp the dynamical behavior of the DE, but do not have the capacity of measuring the fundamental EoS,  just the effective one.  \jtext{Therefore, they cannot exclude the possibility that the measured effect may be  (dynamical) quantum vacuum energy speeding up the cosmic evolution.  While a full confirmation would require  the participation of several observables at a time,  one should remain open to this possibility.  If so, the fashionable evidence on dynamical DE recently collected by DESI measurements\cite{DESI:2024mwx,DESI:2024aqx,DESI:2025zgx,DESI:2025fii} might ultimately  have a fundamental QFT explanation. For example, the observed  dynamical DE could just be running vacuum energy in the context of a unified fundamental theory of the cosmological evolution encompassing the expansion history from the very early times to the present universe, i.e. along the lines of the present proposal. As previously indicated, this is only a possibility, but an attractive one  if we take into account that it involves fundamental physics.}
\jtext{Further studies are necessary, of course, e.g. to confirm the full viability of the inflationary scenario through a detailed treatment of cosmic perturbations and its comparison with the observational constraints. This is left for future work. However, we believe that the unified framework under consideration is sufficiently rich as to eventually make possible to understand the main properties of the cosmological expansion in terms of fundamental physics.}

Finally, the fact that this might be possible places once more the issue of the quantum vacuum in the forefront of modern cosmology and indicates that rather than being blamed as the source of inextricable problems such as the cosmological constant problem, it could actually be the clue to its resolution and even provide an explanation of the practical problems of observational cosmology.

%\newpage
\vspace{1.5cm}

%\begin{scriptsize}
{\bf Acknowledgments}:  JSP is funded by projects  PID2022-136224NB-C21 (MICIU), 2021-SGR-249 (Generalitat de Catalunya) and CEX2024-001451-M (ICCUB). AGF is funded by the grant FPU24/01241 (MICIU). JSP, CMP and AGF acknowledge networking support by the COST Association Action CA21136 ``{\it Addressing observational tensions in cosmology
with systematics and fundamental physics (CosmoVerse)}''.  CMP acknowledges financial support from the María Goyri Programme through a Professor Lector contract at the University of Girona. JSP also acknowledges participation in the COST Action CA23130 ``{\it Bridging
high and low energies in search of quantum gravity (BridgeQG)}''. JSP thanks N. E. Mavromatos for discussions.
%\end{scriptsize}

\newpage

%\vspace{0.75cm}

\appendix
\counterwithin*{equation}{section}
\renewcommand\theequation{\thesection\arabic{equation}}

\section{Useful formulae}

\subsection{Some dimensional regularization integrals}\label{sec:DRintegrals}

While our renormalization framework is based on off-shell adiabatic subtraction, dimensional regularization (DR) can be used as an auxiliary tool and should not be associated with minimal subtraction, which we do not use at all in our calculations.  We denote the 't Hooft mass unit characteristic of  the DR procedure  as $\Tilde{\mu}$\cite{Collins84} (not to be confused with our definition $\mu\equiv m/H$, which is a dimensionless parameter, whereas $\Tilde{\mu}$ has natural dimension 1). Since we perform the renormalization of the QFT when the classical background is de Sitter spacetime, for which $H=$ const., it is convenient to use the correspondence
\begin{equation}\label{eq:measureDRdeSitter}
d^3k\xrightarrow{}(a\Tilde{\mu})^{3-N} d^Nk=\frac{1}{|\tau|^3} \left(\frac{\Tilde{\mu}}{H}\right)^{3-N}\, d^Nz\,
\end{equation}
when we generalize the $3$-dimensional momentum integrals to $N=3-2\varepsilon$ dimensions,  with the understanding that the limit $\varepsilon\to 0$ is to be taken in the final results after the poles at the value $N=3$ have been canceled. We choose $a\Tilde{\mu}$ instead of the usual $\Tilde{\mu}$ for convenience. This does not have any impact in the renormalized expressions, since those cannot depend on this DR parameter. The last expression in \eqref{eq:measureDRdeSitter} ensues from using Eq.\eqref{eq:attau} and the definition of the $z$ variable $z=-k\tau=k |\tau|$, where $\tau$ is the conformal time.
It follows that, in our de Sitter QFT context,  UV-divergent integrals of the form $\int k^2dk ( \dots)$, in which the dots involve a spherically symmetric function of the integration variable,  can be dimensionally regularized through the prescription
\begin{equation}\label{eq:integralsDRdeSitter}
\begin{split}
\int k^2dk \ ( \dots)&=\frac{(2\pi)^3}{4\pi} \int \frac{d^3k}{(2\pi)^3}\ ( \dots)\to
\frac{(2\pi)^3}{\Omega_3}\frac{1}{|\tau|^3} \left(\frac{\Tilde{\mu}}{H}\right)^{3-N}\, \int \frac{d^Nz}{(2\pi)^N}\ ( \dots)\\
&=\frac{\Omega_N}{\Omega_3}\frac{(2\pi)^3}{(2\pi)^N|\tau|^3} \left(\frac{\Tilde{\mu}}{H}\right)^{3-N} \, \int dz z^{N-1}\ ( \dots)\,,
\end{split}
\end{equation}
where $\Omega_N= 2\pi^{N/2}/\Gamma(N/2)$ is the solid angle subtended by $N$-dimensional space. For $N=3$, $\Omega_3=4\pi$.
As an example, consider an integral of the type $\int k^2 dk \ \omega_k^{-n}$, with $\omega_k=\sqrt{k^2+a^2m^2}$. We find
\begin{equation}\label{eq:integralexample}
\int k^2 dk\frac{1}{\omega_k^n}\to
\frac{(2\pi)^3}{\Omega_3}\frac{1}{|\tau|^3} \left(\frac{\Tilde{\mu}}{H}\right)^{3-N}\,\int \frac{d^Nz}{(2\pi)^N} \frac{1}{\omega_k^n}=\frac{(2\pi)^3}{\Omega_3} |\tau|^{n-3} \left(\frac{\Tilde{\mu}}{H}\right)^{3-N}\,\int \frac{d^Nz}{(2\pi)^N} \frac{1}{\left(z^2+\mu^2\right)^{n/2}}\,,
\end{equation}
where in the last step we have used the definition $\mu=m/H$, which should not be confused with the aforementioned $\tilde{\mu}$.
The remaining integral with respect to $z$ can be computed in DR.  The following master formulas are useful for our calculations:
\begin{equation}\label{eq:IN}
\begin{split}
    I_N(n,\mu^2)&=\left(\frac{\Tilde{\mu}}{H}\right)^{3-N}\int \frac{d^Nz}{(2\pi)^N} \frac{1}{(z^2+\mu^2)^{n/2}}=\left(\frac{\Tilde{\mu}}{H}\right)^{3-N}\frac{1}{(4\pi)^{N/2}} \frac{\Gamma\left(\frac{n-N}{2}\right)}{\Gamma(n/2)} \left(  \frac{1}{\mu^2}\right)^{\frac{n-N}{2}} \\
    &=\frac{1}{(4\pi)^{n/2}} \frac{\Gamma \left( \frac{n-3}{2}+\varepsilon \right)}{\Gamma(n/2)}\left(\frac{\mu^2}{4\pi}  \right)^{\frac{3-n}{2}}\left(\frac{\mu^2H^2}{4\pi \Tilde{\mu}^2}  \right)^{-\varepsilon}=\frac{1}{(4\pi)^{n/2}} \frac{\Gamma \left( \frac{n-3}{2}+\varepsilon \right)}{\Gamma(n/2)}\left(\frac{\mu^2}{4\pi}  \right)^{\frac{3-n}{2}}\left(\frac{m^2}{4\pi \Tilde{\mu}^2}  \right)^{-\varepsilon}\,,
    \end{split}
\end{equation}
\begin{equation}\label{eq:IN2}
\begin{split}
    L_N(n,k,\mu^2)&=\left(\frac{\Tilde{\mu}}{H}\right)^{3-N}\int \frac{d^Nz}{(2\pi)^N}\frac{z^k}{(z^2+\mu^2)^{n/2}}=\left(\frac{\Tilde{\mu}}{H}\right)^{3-N}\frac{\mu^{N-n+k}}{(4\pi)^{N/2}}\ \frac{\Gamma(\frac{n-N-k}{2})\Gamma(\frac{N+k}{2})}{\Gamma(\frac{N}{2})\Gamma(\frac{n}{2})}\,.
    %\left(\frac{1}{\mu^2}\right)^{\frac{n-N-k}{2}}\,.
\end{split}
\end{equation}
The following expansions for $\varepsilon\to 0$ are useful to extract the divergences from the poles of the Gamma function at the negative integer numbers, including zero:
\begin{equation}\label{eq:Gamma012}
\begin{split}
    &\Gamma_0\equiv\frac{\Gamma(\varepsilon)}{(4\pi)^{-\varepsilon}}=\frac{1}{\varepsilon}-\gamma_E+\ln 4\pi + \mathcal{O}(\varepsilon) \,,\\
    &\Gamma_1\equiv\frac{\Gamma(-1+\varepsilon)}{(4\pi)^{-\varepsilon}}=-\frac{1}{\varepsilon}-1+\gamma_E-\ln 4\pi + \mathcal{O}(\varepsilon) \,,\\
    &\Gamma_2\equiv\frac{\Gamma(-2+\varepsilon)}{(4\pi)^{-\varepsilon}}=\frac{1}{2\varepsilon}+\frac{1}{4}(3-2\gamma_E+2\ln 4\pi)+ \mathcal{O}(\varepsilon)\,,
\end{split}
\end{equation}
where $\gamma_E$ is Euler's constant.
\subsection{Asymptotic expansions of the Hankel functions}\label{sec:AppendixHankel}
The leading asymptotic behavior of the Hankel functions reads as:
\begin{equation}
\begin{split}
 &\mathbb{H}_\varsigma^{(1)}(|z|\xrightarrow{}\infty) \sim \sqrt{\frac{2}{\pi z}}e^{i\left(z-\frac{1}{2}\varsigma \pi -\frac{\pi}{4} \right)} \sim \sqrt{\frac{2}{\pi z}}e^{iz} e^{\frac{\pi}{2}\Im \varsigma} \,, \\
  &  \mathbb{H}_\varsigma^{(2)}(|z|\xrightarrow{}\infty) \sim  \sqrt{\frac{2}{\pi z}}e^{-i\left(z-\frac{1}{2}\varsigma \pi -\frac{\pi}{4} \right)}\sim \sqrt{\frac{2}{\pi z}}e^{-i z} e^{-\frac{\pi}{2}\Im \varsigma}\,,
\end{split}
\end{equation}
where in the asymptotic limit we can discard the phases but keep the term with $\varsigma$ since it can be imaginary\footnote{Note that in the main text we keep this convention of phases for Hankel functions\cite{AbramowitzStegun} and define the modes accordingly. The used convention is simply for convenience, it has no impact on our final results.}.
As explained in the text, if we use  $z=-k\tau=k\abs{\tau}>0$ ($-\infty<\tau<0$) as the argument of the Hankel function,  then only $\mathbb{H}_\varsigma^{(1)}(z)$ satisfies the Bunch-Davies asymptotic condition at short distances, i.e. it reproduces the usual behavior of a positive frequency solution in Minkowski space: $\mathbb{H}_\varsigma^{(1)}(|z|\xrightarrow{}\infty)  \sim \sqrt{\frac{2}{\pi z}}e^{i z}=\sqrt{\frac{2}{\pi z}}e^{-i k\tau}$.  This is easily verified since in the asymptotic limit $i\partial_\tau\mathbb{H}_\varsigma^{(1)}=\omega \mathbb{H}_\varsigma^{(1)}$ with $\omega=k>0$ (the mass of the particle being neglected for large $k$).
Thus, we will focus only on $\mathbb{H}_\varsigma^{(1)}(z)$\footnote{In alternative approaches using $k\tau=-z$  as the argument of the Hankel functions, it would be $\mathbb{H}_\varsigma^{(2)}(z)$ the solution which satisfies the Bunch-Davies vacuum condition.}. Notice that $\mathbb{H}_\varsigma^{(2)}=\mathbb{H}_\varsigma^{(1)*}$ for $\varsigma$ real. In general, the conjugate of the Hankel function satisfies the property
\begin{equation}\label{eq:hankelrel}
    {\mathbb{H}_\varsigma^{(1)}}^*=e^{\pi \Im \varsigma}\mathbb{H}_\varsigma^{(2)}\,,
\end{equation}
which is valid for $\varsigma$ real or pure imaginary.

We need to go beyond the leading asymptotic term mentioned above. Specifically, a more detailed asymptotic expansion in the powers of $1/z$ is required for the modulus square of the Hankel function and its derivative, as well as for the crossed terms.  The asymptotic results ($z\to\infty$) up to the necessary order can be obtained with the help of \texttt{Mathematica} and read as follows:
\begin{equation}\label{eq:expansionHz}
\begin{split}
    \abs{\mathbb{H}_\varsigma^{(1)}}^2 &= e^{\pi \Im\varsigma}\left[\frac{2}{\pi z}+\frac{-1+4\varsigma^2}{4\pi z^3}+\frac{3(9-40\varsigma^2+16\varsigma^4)}{64\pi z^5} +\mathcal{O}\left(\frac{1}{z^7}\right)\right]\,, \\
    \abs{{\mathbb{H}_\varsigma^{(1)}}'}^2 &=e^{\pi \Im\varsigma}\left[\frac{2}{\pi z}+\frac{3-4\varsigma^2}{4\pi z^3}-\frac{45-184\varsigma^2+16\varsigma^4}{64\pi z^5}+\mathcal{O}\left(\frac{1}{z^7}\right)\right]\,, \\
    \mathbb{H}_\varsigma^{(1)} {\mathbb{H}_\varsigma^{(1)*}}'+\mathbb{H}_\varsigma^{(1)*}{\mathbb{H}_\varsigma^{(1)}}' &=e^{\pi \Im\varsigma}\left[-\frac{2}{\pi z^2}+\frac{3-12\varsigma^2}{4\pi z^4}+\mathcal{O}\left(\frac{1}{z^6}\right)\right]\,.
\end{split}
\end{equation}
The integrals from the exact expression given in the main text (cf. Eq.\,\eqref{eq:T00mexact}) involve terms that are quartic, quadratic and logarithmically divergent, as it becomes clear from these expansions.

An alternative asymptotic expansion of the Hankel functions is also needed. In fact, while the previous expansion in powers of $1/z$ is useful to check the cancellation of UV differences in the ARP procedure,  we also need the asymptotic expansion of Hankel functions in inverse powers of $\omega_z=\sqrt{z^2+\mu^2}$, where $\mu\equiv m/H$.  In this way, we can keep track also of the finite parts and we do not meet any infrared divergence at $z=0$. Specifically, we need an expansion in powers of $1/\omega_z$ in the limit $\omega_z\gg 1$.  Note that this limit is consistent with the adiabatic expansion, which originates from the WKB approximation. The latter is valid for large $\omega_k$, but this limit amounts to large $\omega_z$ as well --- cf. Eq.\,\eqref{eq:defwz}.
The leading terms of the alternative asymptotic  expansion  that we seek  can be found with the help of \texttt{Mathematica}\cite{Mathematica}. The result is the following:
\begin{equation}\label{eq:expansionHwz1}
\begin{split}
   & \abs{\mathbb{H}_\varsigma^{(1)}}^2 = e^{\pi \Im\varsigma}\left[\frac{2}{\pi \omega_z}+\frac{4\varsigma^2+4\mu^2-1}{4\pi \omega_z^3}+\frac{3(9-8\mu^2+16\mu^4+8(-5+4\mu^2)\varsigma^2+16\varsigma^4)}{64\pi \omega_z^5} + \mathcal{O}\left( \frac{1}{\omega_z^7} \right) \right]\\
     &\hspace{1.2cm}= e^{\pi \Im\varsigma}\left[\frac{2}{\pi \omega_z}+\frac{2(1-6\xi)}{\pi \omega_z^3}+\frac{3\mu^2+36\xi(-1+6\xi)}{2\pi \omega_z^5} + \mathcal{O}\left( \frac{1}{\omega_z^7} \right)\right] \,,
       \end{split}
\end{equation}
     \begin{equation}\label{eq:expansionHwz2}
\begin{split}
    \abs{{\mathbb{H}_\varsigma^{(1)}}'}^2 &=e^{\pi \Im\varsigma}\left[\frac{2}{\pi\omega_z}+\frac{3+4\mu^2-4\varsigma^2}{4\pi\omega_z^3}-\frac{45-48\mu^2-184\varsigma^2+16\varsigma^4-(72-96\varsigma^2)\mu^2}{64\pi \omega_z^5}+\mathcal{O}\left(\frac{1}{\omega_z^6} \right)\right] \\
     &=e^{\pi \Im\varsigma}\left[\frac{2}{\pi\omega_z}+\frac{-3+4\mu^2+24\xi}{2\pi\omega_z^3}+\frac{9+4\mu^4-42\xi-72\xi^2+8\mu^2(-1+3\xi)}{2\pi \omega_z^5}+\mathcal{O}\left(\frac{1}{\omega_z^6} \right)\right] \,,
     \end{split}
\end{equation}
\begin{equation}\label{eq:expansionHwz3}
\begin{split}
    \mathbb{H}_\varsigma^{(1)} {\mathbb{H}_\varsigma^{(1)*}}'+\mathbb{H}_\varsigma^{(1)*}{\mathbb{H}_\varsigma^{(1)}}' &= e^{\pi \Im\varsigma}\left[-\frac{2}{\pi\omega_z^2}+\frac{3-8\mu^2-12\varsigma^2}{4\pi\omega_z^4}+\mathcal{O}\left(\frac{1}{\omega_z^6}\right)\right] \\
     &= e^{\pi \Im\varsigma}\left[-\frac{2}{\pi\omega_z^2}+\frac{-6+\mu^2+36\xi}{\pi\omega_z^4}+\mathcal{O}\left(\frac{1}{\omega_z^6}\right)\right]\,,
\end{split}
\end{equation}
where in the alternative expression presented for each one of these expansions we have made use of the definition of $\varsigma$ given in Eq.\,\eqref{eq:nudefinition}.
These formulas can be useful to explore the type of leading contributions expected in our calculation. They are precisely adapted for using the dimensional regularization integrals presented in Appendix \ref{sec:DRintegrals}.  It is easy to check that for $\mu^2\ll 1$ the expansions \eqref{eq:expansionHwz1}-\eqref{eq:expansionHwz3} boil down to the simpler ones \eqref{eq:expansionHz}, as they should. While these expansions are useful to check the cancellation of UV-divergences in the renormalization program and to obtain preliminary results on the leading terms expected in our calculation, our final results have been obtained using the exact integral formulae of Appendix \ref{sec:exactintegrals}.

\subsection{Special integrals involving products of Hankel functions}\label{sec:exactintegrals}
The integrals we need that involve Hankel functions of order $\varsigma$ are of the following kind:
\begin{equation}\label{eq:WeberSchafheitlin}
\begin{split}
    \mathcal{I}(\lambda, \varsigma)&=e^{- \pi \Im \varsigma}\int_0^\infty dzz^\lambda\abs{\mathbb{H}_\varsigma^{(1)}}^2 \, , \\
    \mathcal{J}(\lambda, \varsigma)&=e^{- \pi \Im \varsigma}\int_0^\infty dzz^\lambda\left(\mathbb{H}_\varsigma^{(1)}{\mathbb{H}_\varsigma^{(1)*}}'+\mathbb{H}_\varsigma^{(1)*}{\mathbb{H}_\varsigma^{(1)}}' \right) \, , \\
    \mathcal{K}(\lambda, \varsigma)&=e^{- \pi \Im \varsigma}\int_0^\infty dzz^\lambda\abs{{\mathbb{H}_\varsigma^{(1)}}'}^2 \, ,
\end{split}
\end{equation}
where for the latter, one can use the recurrence relations
\begin{equation}
\begin{split}
    {\mathbb{H}_\varsigma^{(1)}}'(z)&=\frac{1}{2}\left(\mathbb{H}_{\varsigma-1}^{(1)}(z)-\mathbb{H}_{\varsigma+1}^{(1)}(z)\right) \, ,\\
    {\mathbb{H}_\varsigma^{(1)*}}'(z)&=e^{\pi \Im \varsigma}{\mathbb{H}_\varsigma^{(2)}}'(z)=\frac{e^{ \pi \Im \varsigma}}{2}\left(\mathbb{H}_{\varsigma-1}^{(2)}(z)-\mathbb{H}_{\varsigma+1}^{(2)}(z) \right)=\frac{1}{2}\left(\mathbb{H}_{\varsigma-1}^{(1)*}(z)-\mathbb{H}_{\varsigma+1}^{(1)*}(z)\right)\, ,
\end{split}
\end{equation}
and hence,
\begin{equation}
\begin{split}
    \mathcal{K}(\lambda,\varsigma)&=e^{- \pi \Im \varsigma}\int_0^\infty dz z^\lambda \abs{{\mathbb{H}_\varsigma^{(1)}}'(z)}^2=\frac{1}{4}e^{- \pi \Im \varsigma}\int_0^\infty dz z^\lambda \left[ \abs{{\mathbb{H}_{\varsigma-1}^{(1)}}}^2 + \abs{{\mathbb{H}_{\varsigma+1}^{(1)}}}^2 \right.\\
    &-\left.e^{ \pi \Im \varsigma}\left({\mathbb{H}_{\varsigma-1}^{(1)}}{\mathbb{H}_{\varsigma+1}^{(2)}}+{\mathbb{H}_{\varsigma+1}^{(1)}}{\mathbb{H}_{\varsigma-1}^{(2)}}\right)\right] \\
    &=\frac{1}{4}\mathcal{I}(\lambda, \varsigma-1)+\frac{1}{4}\mathcal{I}(\lambda, \varsigma+1)-\frac{1}{4}\mathcal{K}_c(\lambda, \varsigma) \,,
\end{split}
\end{equation}
where we have defined
\begin{equation}
    \mathcal{K}_c(\lambda, \varsigma) = \int_0^\infty dz z^\lambda \left({\mathbb{H}_{\varsigma-1}^{(1)}}{\mathbb{H}_{\varsigma+1}^{(2)}}+{\mathbb{H}_{\varsigma+1}^{(1)}}{\mathbb{H}_{\varsigma-1}^{(2)}}\right)\,.
\end{equation}
The results we find are the following\footnote{The integrals \eqref{eq:WeberSchafheitlin} are standard and fall into the category of the so-called Weber–Schafheitlin integrals -- see\,\cite{Watson1966} (chapter 13.4). They can also be found in advanced
handbooks\cite{AbramowitzStegun}, but can also be dealt with using  \texttt{Mathematica}\cite{Mathematica} or \texttt{Maple}\cite{Maple}. We have cross-checked them using both algebraic tools after significant elaboration.}:
\begin{equation}\label{eq:WeberSchafheitlin2}
\begin{split}
    &\mathcal{I}(\lambda, \varsigma)= \frac{\cos \pi\varsigma}{2\pi^{5/2}}(\lambda-1)\Gamma\left(\frac{1+\lambda}{2}-\varsigma\right) \Gamma\left(\frac{1+\lambda}{2}+\varsigma\right)\Gamma\left(-\frac{\lambda}{2}\right)\Gamma\left(\frac{\lambda-1}{2}\right) \, ,\\
   &\mathcal{J}(\lambda,\varsigma)=\frac{8\cos\pi\varsigma}{(4\varsigma^2-\lambda^2)\pi^{3/2}\Gamma\left(-\frac{\lambda}{2}\right)}\Gamma\left(\frac{1-\lambda}{2}\right) \Gamma\left(1+\frac{\lambda}{2}-\varsigma\right)\Gamma\left(1+\frac{\lambda}{2}+\varsigma\right) \, , \\
    &\mathcal{K}_c(\lambda,\varsigma)=\frac{2\sec{\frac{\pi\lambda}{2}} \cos\pi\varsigma }{\pi^{3/2}} \frac{\Gamma\left(-\frac{\lambda}{2}\right)\Gamma\left(\frac{1-\lambda}{2}\right)}{\Gamma\left(\frac{3-\lambda}{2}\right)\Gamma\left(-\frac{1+\lambda}{2}\right)} \Gamma\left(\frac{1+\lambda}{2}-\varsigma\right)\Gamma\left(\frac{1+\lambda}{2}+\varsigma\right) \, .
\end{split}
\end{equation}
For our calculation, we  need the results for the parameter values $\lambda=2,3,4$ which must be treated as limiting values.
Substituting directly these values of $\lambda$ in the previous equations results in divergent quantities.
More details on the connection of these formulas with the dimensional regularization are given in Appendix \ref{sec:calculationHankelIntegrals}. We will need the previous integrals in the limit $\lambda=(2,3,4)-2\varepsilon$ for $\varepsilon\to 0$. These limits can be performed more safely with the help of \texttt{Mathematica}, and we find the following results:
\begin{equation}\label{eq:WeberSchafheitlin3}
\begin{split}
    \mathcal{I}(4-2\varepsilon, \varsigma)&=\frac{3(m^2/H^2+12\xi)}{8\pi \varepsilon}\left(-2+\frac{m^2}{H^2}+12\xi\right)-\frac{12\xi+m^2/H^2}{16\pi}\left(-2+\frac{m^2}{H^2}+12\xi\right)\left[7+6\left(-\ln 4\right.\right.\\
    &\left.\left.+\psi\left[\frac{5}{2}-\varsigma\right]+\psi\left[\frac{5}{2}+\varsigma\right]\right)\right]\\
    \mathcal{I}(2-2\varepsilon, \varsigma)&=-\frac{-2+m^2/H^2+12\xi}{2\pi\varepsilon}+\frac{-2+m^2/H^2+12\xi}{2\pi}\left(1-\ln4+\psi\left[\frac{3}{2}-\varsigma\right]+\psi\left[\frac{3}{2}+\varsigma\right]\right)\\
    \mathcal{J}(3-2\varepsilon, \varsigma)&=\frac{3(-2+m^2/H^2+12\xi)}{2\pi\varepsilon}+\frac{1}{2\pi}\left(-2+\frac{m^2}{H^2}+12\xi\right)\left\{-5+6\ln 2+ \frac{9}{m^2/H^2+12\xi}\right.\\
    &-\left.3\left(\psi\left[\frac{5}{2}-\varsigma\right]+\psi\left[\frac{5}{2}+\varsigma\right]\right)\right\}\\
    \mathcal{K}_c(4-2\varepsilon, \varsigma)&=\frac{5(m^2/H^2+12\xi)}{4\pi\varepsilon}\left(-2+\frac{m^2}{H^2}+12\xi\right)-\frac{12\xi+m^2/H^2}{8\pi}\left(-2+\frac{m^2}{H^2}+12\xi\right)\left\{9\right.\\
    &+\left.10\left(-\ln 4 + \psi\left[\frac{5}{2}-\varsigma\right]+\psi\left[\frac{5}{2}+\varsigma\right] \right) \right\}\,,
\end{split}
\end{equation}
where $\psi(z)=d\ln\Gamma(z)/dz=\Gamma'(z)/\Gamma(z)$ is the standard digamma function. It is useful for our calculations to recall its asymptotic behavior:
\begin{equation}
\psi(z)\simeq \ln z - \frac{1}{2z}-\sum _{n=1}^{\infty}\frac{B_{2n}}{2n}\,z^{-2n}\ \ \ \ \ (\textrm{for}\ |z|\to \infty)\,,
\end{equation}
with $B_k$  the $k$th Bernoulli number.

%\newpage

\section{Cancellation of UV-divergences in the renormalized ZPE}\label{sec:FiniteEMTdeSitter}

Here we show that the adiabatically renormalized ZPE of de Sitter spacetime is free from UV-divergences, i.e. it is a finite quantity. To perform this check, one has to first expand the divergent terms from the adiabatic expansion up to the asymptotic order $z^{-3}$. Putting $M=m$ in \eqref{eq:defwz}, we find
\begin{equation}\label{eq:expansionomegaonshell}
\begin{split}
    \omega_z(m) &\simeq z+\frac{m^2}{H^2} \frac{1}{2z}-\frac{m^4}{H^4} \frac{1}{8z^3}\,, \\
    \frac{1}{\omega_z(m)}&\simeq \frac{1}{z}-\frac{m^2}{H^2}\frac{1}{2z^3}\,,\\
    \frac{1}{\omega_z^3(m)}&\simeq \frac{1}{z^3}\,.
\end{split}
\end{equation}
Substituting these relations in Eq.\,\eqref{eq:divpartonshell} we arrive at
\begin{equation}\label{eq:T00divonshell}
    \braket{T_{00}^{\delta\phi}}^{(0-4)}_{\mathrm{Div}}(m)=\frac{H^2}{8\pi^2\abs{\tau}^2}\int dz \left\{ 2 z^3+\frac{m^2}{H^2}z-\frac{m^4}{H^4} \frac{1}{4z}+\left(\xi-\frac{1}{6}\right)\left(-6z-3\frac{m^2}{H^2}\frac{1}{z} \right)+\cdots\right\}\,,
\end{equation}
where the dots indicate finite terms.
As anticipated in the main text, the very same result follows from taking the divergent part of \eqref{eq:divpartonshell} and performing similar asymptotic expansions as in \eqref{eq:expansionomegaonshell} but using now $\omega_z(M)$ instead of $\omega_z(m)$. After a straightforward calculation one finds
\begin{equation}\label{eq:DivMm3}
\begin{split}
    \braket{T_{00}^{\delta\phi}}^{(0-4)}_{\mathrm{Div}}(M)&=\frac{H^2}{8\pi^2\abs{\tau}^2}\int dz z^2 \left[ 2\omega_z+\frac{\Delta^2}{H^2\omega_z}-\frac{1}{4}\frac{\Delta^4}{H^4\omega_z^3}+\left(\xi-\frac{1}{6}\right)\left(-\frac{6}{\omega_z}-6\frac{M^2}{H^2\omega_z^3} -3\frac{\Delta^2}{H^2\omega_z^3} \right) \right]\\
    &=\frac{H^2}{8\pi^2\abs{\tau}^2}\int dz \left[2z^3 +\frac{M^2}{H^2}z-\frac{M^4}{H^4}\frac{1}{4z}+\frac{\Delta^2}{H^2}z-\frac{\Delta^2M^2}{H^4}\frac{1}{2z}-\frac{1}{4}\frac{\Delta^4}{H^4}\frac{1}{z}\right.\\
    &+\left.\left(\xi-\frac{1}{6}\right)\left( -6z +3\frac{M^2}{H^2}\frac{1}{z}-6\frac{M^2}{H^2}\frac{1}{z}-3\frac{\Delta^2}{H^2}\frac{1}{z}\right) +\cdots \right] \\
    &=\frac{H^2}{8\pi^2\abs{\tau}^2}\int dz \left[2z^3 +\frac{M^2}{H^2}z-\frac{M^4}{H^4}\frac{1}{4z}+\frac{m^2-M^2}{H^2}z-\frac{m^2M^2-M^4}{H^4}\frac{1}{2z}-\frac{1}{4}\frac{(m^2-M^2)^2}{H^4}\frac{1}{z}\right.\\
    &+\left.\left(\xi-\frac{1}{6}\right)\left( -6z +3\frac{M^2}{H^2}\frac{1}{z}-6\frac{M^2}{H^2}\frac{1}{z}-3\frac{m^2-M^2}{H^2}\frac{1}{z}\right) +\cdots \right]\\
    &=\frac{H^2}{8\pi^2\abs{\tau}^2}\int dz \left\{ 2 z^3+\frac{m^2}{H^2}z-\frac{m^4}{H^4} \frac{1}{4z}+\left(\xi-\frac{1}{6}\right)\left(-6z-3\frac{m^2}{H^2}\frac{1}{z} \right)+\cdots\right\} \\
&=\braket{T_{00}^{\delta\phi}}^{(0-4)}_{\mathrm{Div}}(m)+\cdots\,,
\end{split}
\end{equation}
where again the dots indicate finite terms. This result confirms our claim of finiteness in Eq.\,\eqref{eq:finite}. Note also that in the last step all $M$-dependent terms that are shown explicitly cancel out and the result becomes exactly equal to that in \eqref{eq:T00divonshell}.  However, we should emphasize once more that this is only true for the divergent parts, since $\braket{T_{00}^{\delta\phi}}^{(0-4)}_{\mathrm{Div}}(m)$ and $\braket{T_{00}^{\delta\phi}}^{(0-4)}_{\mathrm{Div}}(M)$ are not exactly equal, they actually differ in finite terms which are relevant and depend on the scale $M$. These differences are computed in Sec.\ref{sec:RenZPEandVED}.

The important result that can be derived right next is the following. By operating the ARP subtraction \eqref{RenormalizedEMTScalar} the divergent terms from the exact on-shell solution and those from the fourth-order adiabatic expansion cancel each other out in a precise way. To demonstrate it,  we need the asymptotic form of the divergent part of $\braket{T_{00}^{\delta\phi}}^{\rm dS}(m)$. The latter can be obtained from the asymptotic expansion ($z\to\infty$) of the Hankel functions in powers of $1/z$ which is provided in the Appendix \ref{sec:AppendixHankel}. We find
\begin{equation}
\begin{split}
    \braket{T_{00}^{\delta\phi}}^{\rm dS}_\mathrm{Div}(m)&=\frac{H^2}{16\pi^2 \abs{\tau}^2}\int dz z^2 \left[ \frac{1}{2z}+\frac{-1+4\varsigma^2}{16z^3}+2z+\frac{3-4\varsigma^2}{4z}-\frac{45-184\varsigma^2+16\varsigma^4}{64z^3} \right. \\
    &-\frac{1}{z}-\frac{3(-1+4\varsigma^2)}{8z^3} + 2z + \frac{-1+4\varsigma^2}{4z}+\frac{3(9-40\varsigma^2+16\varsigma^4)}{64z^3}+\frac{m^2}{H^2} \frac{2}{z}+\frac{m^2}{H^2}\frac{(-1+4\varsigma^2)}{4z^3} \\
    &-\left.6\left(\xi-\frac{1}{6}\right) \left( \frac{4}{z}+\frac{-1+4\varsigma^2}{2z^3}-\frac{2}{z}-\frac{3(-1+4\varsigma^2)}{4z^3}\right)+\cdots\right] \\
    &=\frac{H^2}{8\pi^2 \abs{\tau}^2}\int dz \left\{ 2z^3+\frac{m^2}{H^2}z+\left( \frac{1}{64}-\frac{1}{8}\varsigma^2+\frac{1}{4}\varsigma^4 \right)\frac{1}{z}+\frac{m^2}{H^2} \frac{(-1+4\varsigma^2)}{8z}\right. \\
    -&\left.6\left(\xi-\frac{1}{6}\right)\left[ z+\left( \frac{1}{8}-\frac{\varsigma^2}{2} \right)\frac{1}{z} \right] +\cdots \right\}\,.
\end{split}
\end{equation}
We may now perform the subtraction of the divergent parts of the expressions involved in \eqref{RenormalizedEMTScalar} and verify explicitly the exact cancellation of UV-divergences:
\begin{equation}
\braket{T_{00}^{\delta\phi}}^{\rm dS}_\mathrm{Div}(m)-\braket{T_{00}^{\delta\phi}}^{(0-4)}_{\mathrm{Div}}(M)=\braket{T_{00}^{\delta\phi}}^{\rm dS}_\mathrm{Div}(m)-\braket{T_{00}^{\delta\phi}}^{(0-4)}_{\mathrm{Div}}(m)+\cdots\\
\end{equation}
We have used the fact that the divergent parts of $\braket{T_{00}^{\delta\phi}}^{(0-4)}_{\mathrm{Div}}(m)$ and $\braket{T_{00}^{\delta\phi}}^{(0-4)}_{\mathrm{Div}}(M)$ are the very same, as proven in \eqref{eq:DivMm3}. By selecting the manifestly divergent terms in the former subtraction we are left with:
\begin{equation}
\begin{split}
    &\frac{H^2}{8\pi^2 \abs{\tau}^2}\int dz \left\{ 2z^3+\frac{m^2}{H^2}z+\left( \frac{1}{64}-\frac{1}{8}\varsigma^2+\frac{1}{4}\varsigma^4 \right)\frac{1}{z}+\frac{m^2}{H^2} \frac{(-1+4\varsigma^2)}{8z}-6\left(\xi-\frac{1}{6}\right)\left[ z+\left( \frac{1}{8}-\frac{\varsigma^2}{2} \right)\frac{1}{z} \right] \right.\\
    &-\left.2 z^3-\frac{m^2}{H^2}z+\frac{m^4}{H^4} \frac{1}{4z}-\left(\xi-\frac{1}{6}\right)\left(-6z-3\frac{m^2}{H^2}\frac{1}{z} \right) \right\}\\
    &=\frac{H^2}{8\pi^2 \abs{\tau}^2}\int dz \left\{ \left[  \frac{1}{64}-\frac{1}{8}\varsigma^2+\frac{1}{4}\varsigma^4 +\frac{m^2}{H^2} \frac{(-1+4\varsigma^2)}{8}+\frac{m^4}{H^4}\frac{1}{4}\right] \frac{1}{z}+\left(\xi-\frac{1}{6}\right)\left[-6\left(\frac{1}{8}-\frac{\varsigma^2}{2} \right)+3\frac{m^2}{H^2} \right]\frac{1}{z} \right\}  \\
    &=\frac{H^2}{8\pi^2 \abs{\tau}^2}\int \frac{dz}{z}\left[ (1-6\xi)^2+\left(\xi-\frac{1}{6}\right)(6-36\xi) \right]=0\,.
\end{split}
\end{equation}

Notice that in the last step we borrowed the definition of the parameter $\varsigma$ (the order of Hankel functions) given in Eq.\eqref{eq:nudefinition}. The results derived in this appendix are particularly transparent and are independent of those obtained using the exact formulas of the Appendix \ref{sec:exactintegrals}  and therefore reconfirm our claim expressed in the text that the quantities \eqref{eq:DivMm2} are perfectly finite. While the exact formulas are necessary to collect all the relevant finite pieces of the renormalized result, the cancellation of divergences can be cross-checked in a much simpler way with the method presented in this appendix. Overall, this demonstrates the consistency of the off-shell adiabatic renormalization procedure being used in this work to renormalize the EMT of de Sitter spacetime.

%\newpage

\section{Renormalized VED and trace of the energy-momentum tensor}\label{sec:calculationHankelIntegrals}
In this appendix, we provide some details about the derivation of the formulas put forward in Sec.\ref{sec:RenZPEandVED} and Sec.\ref{sec:RenPressure}.
\
\begin{equation}
\begin{split}
    \braket{T_{00}^{\delta\phi}}^{\rm dS} (m)&=\frac{1}{16\pi a^2\abs{\tau}}e^{- \pi \Im \varsigma}\int dk k^2 g(\varsigma, \mu, \xi, k|\tau|)\,,
    %&= \frac{1}{16\pi a^2\abs{\tau}} \frac{(2\pi)^3}{4\pi}e^{- \pi \Im \varsigma}\int \frac{d^3k}{(2\pi)^3}\left\{\left( \frac{9}{4}+\frac{m^2}{H^2}-12\xi\right)\abs{\mathbb{H}_\varsigma^{(1)}}^2+k^2\abs{\tau}^2\abs{\mathbb{H}_\varsigma^{(1)}}^2+k^2\abs{\tau}^2 \abs{{\mathbb{H}_\varsigma^{(1)}}'}^2 \right. \\
    %&+\left.k\abs{\tau}\left( \frac{3}{2}-6\xi\right) \left( \mathbb{H}_\varsigma^{(1)} {\mathbb{H}_\varsigma^{(1)*}}'+\mathbb{H}_\varsigma^{(1)*}{\mathbb{H}_\varsigma^{(1)}}' \right)\right\} \,.
\end{split}
\end{equation}
where
\begin{equation}
\begin{split}
g(\varsigma,\mu,\xi, k|\tau|)=& \left( \frac{9}{4}+\frac{m^2}{H^2}-12\xi\right)\abs{\mathbb{H}_\varsigma^{(1)}}^2+k^2\abs{\tau}^2\abs{\mathbb{H}_\varsigma^{(1)}}^2+k^2\abs{\tau}^2 \abs{{\mathbb{H}_\varsigma^{(1)}}'}^2  \\
&+k\abs{\tau}\left( \frac{3}{2}-6\xi\right) \left( \mathbb{H}_\varsigma^{(1)} {\mathbb{H}_\varsigma^{(1)*}}'+\mathbb{H}_\varsigma^{(1)*}{\mathbb{H}_\varsigma^{(1)}}' \right) \, .
\end{split}
\end{equation}

In order to compute the divergent integrals, we first restore a factor $\Omega_3=4\pi$ to write them in terms of $d^3k$ and apply the prescriptions \eqref{eq:measureDRdeSitter} and \eqref{eq:integralsDRdeSitter}.
Subsequently we regularize the above result by going to $N$ dimensions and expressing also the results in terms of the variable $z=-k\tau$ (where $\tau$ is the conformal time), we obtain:

\begin{equation}\label{eq:t00appendix}
\begin{split}
    \braket{T_{00}^{\delta\phi}}^{\rm dS} (m)&= \frac{2\pi^2}{16\pi a^2\abs{\tau}} (a\Tilde{\mu})^{3-N}e^{- \pi \Im \varsigma}\int \frac{d^Nk}{(2\pi)^N}g(\varsigma,\mu,\xi, k|\tau|)\\
    &= \frac{\pi}{8a^2\abs{\tau}}\left(\frac{\Tilde{\mu}}{H\abs{\tau}}\right)^{3-N}\frac{\Omega_N}{(2\pi)^N}e^{- \pi \Im \varsigma}\int dk k^{N-1}g(\varsigma,\mu,\xi, k|\tau|)\\
    &=\frac{\pi}{8a^2\abs{\tau}}\left(\frac{\Tilde{\mu}}{H\abs{\tau}}\right)^{3-N}\frac{2\pi^{N/2}}{\Gamma(N/2)(2\pi)^N}\frac{1}{\abs{\tau}^N}e^{- \pi \Im \varsigma}\int dz z^{N-1}g(\varsigma,\mu,\xi,z)\\
    &=\frac{\pi H^2}{4\tau^2}\frac{1}{\Gamma(N/2)}\frac{1}{(4\pi)^{N/2}}\left(\frac{\Tilde{\mu}}{H}\right)^{3-N}e^{- \pi \Im \varsigma}\int dz z^{N-1}g(\varsigma,\mu,\xi, z)\\
    &= \frac{H^2}{32\tau^2\sqrt{\pi}}\frac{1}{\Gamma(N/2)}\left(\frac{H^2}{4\pi\Tilde{\mu}^2}\right)^{-\frac{3-N}{2}} e^{- \pi \Im \varsigma}\int dz z^{N-1}g(\varsigma,\mu,\xi, z)\\
    &= \frac{H^2}{32\sqrt{\pi} \abs{\tau}^2\Gamma\left(\frac{3-2\varepsilon}{2}\right)} \left(\frac{H^2}{4\pi\Tilde{\mu}^2}\right)^{-\varepsilon} \left\{\left( \frac{9}{4}+\frac{m^2}{H^2}-12\xi\right) \mathcal{I}(2-2\varepsilon,\varsigma)+\mathcal{I}(4-2\varepsilon,\varsigma)\right.\\
    &+\left.\frac{1}{4}\mathcal{I}(4-2\varepsilon,\varsigma-1)
    +\frac{1}{4}\mathcal{I}(4-2\varepsilon,\varsigma+1)-\frac{1}{4}\mathcal{K}_c(4-2\varepsilon,\varsigma) +\left( \frac{3}{2}-6\xi\right)\mathcal{J}(3-2\varepsilon,\varsigma)\right\} \,.
\end{split}
\end{equation}
We have used the parametrization $N=3-2\varepsilon$. In the last equality, we have expressed the result in terms of the  special integrals involving products of Hankel functions  defined in Appendix \ref{sec:exactintegrals}. We proceed now with the trace. We find:
\begin{equation}
    \braket{T^{\delta\phi}}^{\rm dS}(m)=\frac{H^4}{8\pi}e^{- \pi \Im \varsigma}\int dz z^2 f(\varsigma, \mu, \xi, z)\,,
\end{equation}
where
\begin{equation}
\begin{split}
    f(\varsigma, \mu, \xi, z) &=\left[(6\xi-1)z^2+\frac{9}{4}-\frac{51}{2}\xi+\frac{m^2}{H^2}(-2+6\xi)+72\xi^2\right]\abs{\mathbb{H}_\varsigma^{(1)}}^2 \\
    &-(6\xi-1) z^2\abs{{\mathbb{H}_\varsigma^{(1)}}'}^2-\frac{3(6\xi-1)}{2}z\left( \mathbb{H}_\varsigma^{(1)} {\mathbb{H}_\varsigma^{(1)*}}'+\mathbb{H}_\varsigma^{(1)*}{\mathbb{H}_\varsigma^{(1)}}' \right) \,.
    \end{split}
\end{equation}
In an analogous way to the $00$th component:
\begin{equation*}
\begin{split}
    \braket{T^{\delta\phi}}^{\rm dS}(m)&=\frac{H^4}{8\pi} e^{- \pi \Im \varsigma}\frac{(2\pi)^3}{4\pi}\int \frac{d^3z}{(2\pi)^3} f(\varsigma, \mu, \xi, z)=\frac{H^4}{8\pi} \frac{(2\pi)^3}{4\pi} \left(\frac{\Tilde{\mu}}{H}\right)^{3-N}e^{- \pi \Im \varsigma}\int \frac{d^Nz}{(2\pi)^N} f(\varsigma, \mu, \xi, z)\\
    &=\frac{H^4 \pi}{4} \frac{2\pi^{N/2}}{\Gamma(N/2)} \left(\frac{\Tilde{\mu}}{H}\right)^{3-N} \frac{1}{(2\pi)^N} e^{- \pi \Im \varsigma}\int z^{N-1} dz f(\varsigma, \mu, \xi, z)\\
    &= \frac{1}{16\sqrt{\pi}} \frac{1}{\Gamma(N/2)}\left( \frac{H^2}{4\pi\Tilde{\mu}^2}\right)^{\frac{N-3}{2}} e^{- \pi \Im \varsigma} \int dz z^{N-1} f(\varsigma, \mu, \xi, z)
    \end{split}
\end{equation*}
\begin{equation}\label{eq:traceappendix}
\begin{split}
    &=\frac{H^4}{16\sqrt{\pi}}\frac{1}{\Gamma\left(\frac{3-2\varepsilon}{2}\right)} \left(\frac{H^2}{4\pi\Tilde{\mu}^2}\right)^{-\varepsilon}\left\{ (6\xi-1) \mathcal{I}(4-2\varepsilon,\varsigma)+\left(\frac{9}{4}-\frac{51}{2}\xi+\frac{m^2}{H^2}(-2+6\xi)+72\xi^2\right) \mathcal{I}(2-2\varepsilon,\varsigma) \right. \\
    &-\left.(6\xi-1)\left(\frac{1}{4}\mathcal{I}(4-2\varepsilon,\varsigma-1)+\frac{1}{4}\mathcal{I}(4-2\varepsilon,\varsigma+1)-\frac{1}{4}\mathcal{K}_c(4-2\varepsilon,\varsigma) \right)-\frac{3(6\xi-1)}{2}\mathcal{J}(3-2\varepsilon,\varsigma)\right\} \, .
\end{split}
\end{equation}
The next step is to expand the above expressions in the limit $N=3-2\epsilon\to 3$ for $\varepsilon\to 0$. This can be done using the formulae \eqref{eq:WeberSchafheitlin3}. Since it involves a substantial amount of algebra, we have performed these calculations with the help of \texttt{Mathematica}.  Substituting these results into the renormalization prescription \eqref{RenormalizedEMTScalar}, we eventually find the renormalized ZPE and renormalized trace of the EMT within our off-shell ARP procedure.  The final results are presented in the main text, see equations \eqref{eq:RenormZPEdeSitter}
and \eqref{eq:trace_ren_exact}.

%\newpage

\section{Running of $G$ and $\rv$:  consistency of the scale setting $M=H$ in the off-shell adiabatic renormalization}\label{sec:RunninGandVED}

Our novel ARP renormalization of the EMT is based  on the off-shell adiabatic subtraction procedure defined in Eq.\,\eqref{RenormalizedEMTScalar}. It may be useful to provide an additional digression on this important aspect of our calculation. The subtracted term in that equation is the vacuum EMT up to fourth adiabatic order and is evaluated at a floating scale $M$. A particular case is $M=m$ (on-shell setting, corresponding to the mass of the particle). However, as suggested in \cite{CristianJoan2020,CristianJoan2022a}, in cosmology it is more convenient to use an off-shell prescription given the fact that the characteristic vacuum energy density during most of the cosmological evolution is certainly much smaller than the average mass of any particle. The only exception occurs during the inflationary stage, where $H$ can be higher than the mass of any involved GUT particle, see e.g. Fig.\,\ref{Fig:T_inf}. But even in this case, the off-shell treatment is useful, since it explores the opposite end of the energy scales involved, namely energies above the mass shell of the particles.  After we perform the full QFT calculation with such a generic scale $M$, we eventually set it to $M=H$ in order to track the physical evolution of the EMT (in particular the VED) in the course of the cosmic expansion, as proposed in the references mentioned.  This looks reasonable since $H$ is the natural/canonical scale of the FLRW metric and there is no other scale in cosmological spacetime that is simpler and more natural. Other scales that have been explored in the past do not offer a comparable level of simplicity and adequacy, see e.g.\cite{ShapSol,Guberina:2002wt,Shapiro:2004ch,Babic:2004ev}.
A formal proof about the exact scale setting procedure to be used is not possible since it may depend on different circumstances (see below). Notice that even in more robust and well tested theories such as in the standard model of elementary particle physics, the usual setting $\mu=M_Z$ (mass of the $Z$ gauge boson) employed for the computation of cross-sections at LEP energies (a well-known former CERN collider) is ultimately an educated guess, which however proved extremely fruitful. This is e.g. how it was possible to test the running of the fine structure constant from $\mu=m_e$ to $\mu=M_Z$. In cosmology, however, the situation may be more delicate, but it is natural to proceed with the same RG philosophy, having in mind that ultimately the setting can be tested phenomenologically, see e.g. \cite{Sola:2015wwa,Sola:2016jky,Sola:2017znb,SolaPeracaula:2016qlq,Sola:2016zeg,SolaPeracaula:2017esw} and \cite{SolaPeracaula:2021gxi,SolaPeracaula:2023swx,deCruzPerez:2025dni} for a number of successful phenomenological analyzes over the years.

Now, a good way to strengthen the viability of the scale setting procedure that is being used in a particular calculation is to check its internal consistency, especially if several quantities are running  at the same time and if a formal relation (e.g. a Bianchi identity) is to be preserved. For example, in our case we have focused on the VED, i.e. $\rv=\rv(H)$. However, in general the running of Newton's coupling $G=G(H)$ also occurs, and these two quantities are indeed connected with the Bianchi identity.

In what follows, for better contextualization, we first provide some discussion of the running couplings in the gravitational context under off-shell renormalization. Then we focus on the corresponding running of $G$ with $H$, which is interesting in itself, but at the same time it will afford a nontrivial internal check that $M=H$ is a consistent scale setting for $G(M)$ and $\rv(M)$.  Despite the fact that the quantitative effect of the $G$-running is not important for the particular calculations presented in this work, as we shall argue, the internal consistency between the two running laws constitutes a precious test that helps to reinforce our off-shell renormalization method.

We start by recalling the relation between the vacuum EMT, $\langle T_{\mu\nu}\rangle$, and the effective action of vacuum\cite{BirrellDavies82,ParkerToms09,Fulling89}, $W$, which describes the quantum  matter vacuum effects of QFT in curved spacetime:
\begin{equation}\label{eq:DefW}
\langle T_{\mu\nu}\rangle=-\frac{2}{\sqrt{-g}} \,\frac{\delta W}{\delta g^{\mu\nu}}\,.
\end{equation}
Taking into account the standard definition of effective action, the relation of $W$ to the effective vacuum Lagrangian, $L_W$, is simply
\begin{equation}\label{eq:EAW}
\begin{split}
W= &\frac{i}{2}Tr \ln (-G_F^{-1})\equiv \int d^4 x \sqrt{-g}\, L_W\,,
\end{split}
\end{equation}
where $G_F^{-1}$ is the inverse Green's function in curved spacetime. The above expressions involve unrenormalized quantities. To generate finite results, we need to define the renormalized Lagrangian. The corresponding off-shell adiabatically renormalized vacuum effective Lagrangian is given by \cite{CristianJoan2022a,SolaPeracaula:2025yco}:
\begin{equation}\label{eq:LWrenormalized}
L_W^{\rm ren}(M )= L_W (m)-L_W^{(0-4)}(M)\,,
\end{equation}
where  $L_W^{(0-4)}(M)$ is computed up to fourth adiabatic order.  This subtraction prescription, which is  performed at the level of the effective vacuum Lagrangian, is the exact analog of the off-shell ARP definition for the renormalized vacuum EMT, Eq.\,\eqref{RenormalizedEMTScalar}, and it suffices to make $L_W^{\rm ren}(M)$ finite.  Indeed, the explicit calculation of the latter has been performed in detail in \cite{CristianJoan2022a} (cf. Appendix C) using an off-shell generalization of the DeWitt-Schwinger expansion, yielding the following finite result:
\begin{equation}\label{eq:LWrenM}
\begin{split}
L_W^{\rm ren}(M)=\delta \rho_\Lambda(M)-\frac{1}{2}\delta\MPl^2(M) R-\delta \alpha_Q(M) \frac{{Q^\lambda}_\lambda}{3}-\delta \alpha_2(M) R^2+\cdots
\end{split}
\end{equation}
The dots stand for higher order adiabatic contributions, and
\begin{equation}\label{eq:deltacouplings}
\begin{split}
&\delta\rL(M)=\frac{1}{8\left(4\pi\right)^2}\left(M^4-4m^2M^2+3m^4-2m^4 \ln \frac{m^2}{M^2}\right),\\
&\delta\MPl^2(M) =\frac{\left(\xi-\frac{1}{6}\right)}{(4\pi)^2}\left(M^2-m^2+m^2\ln \frac{m^2}{M^2}\right),\\
&\delta \alpha_Q(M)=-\frac{1}{2(4\pi)^2}\ln\frac{m^2}{M^2},\\
&\delta{\alpha}_2(M)=\frac{\left(\xi-\frac{1}{6}\right)^2}{4(4\pi)^2}\ln\frac{m^2}{M^2}.
\end{split}
\end{equation}
The quantity $\delta\MPl^2(M)$ above defines the scaling evolution of the reduced Planck mass squared $\MPl^2(M)\equiv\frac{G^{-1}(M)}{8\pi}$.
 All these quantities  are finite renormalization effects which are generated in the  subtraction \eqref{eq:LWrenormalized}. The above equations are fully consistent with the results \eqref{SubtractionrL}-\eqref{Subtractionalpha} previously obtained in Sec.\ref{sec:GeneralizedEqs} from the renormalized Einstein's equations and the renormalized vacuum EMT, with $\alpha_2=\alpha/2$.  The perfect agreement between the two procedures confirms the correct structure of the vacuum effective Lagrangian \eqref{eq:LWrenM}, which we shall use for our next considerations.

In itself $L_W^{\rm ren}(M)$ is obviously not RG-invariant since it is $M$-dependent. However, the full effective Lagrangian involving also the extended classical part (EH part plus HD geometric terms) must be RG-invariant, of course. The classical part of the Lagrangian reads
\begin{equation}\label{eq:LEHHD}
\begin{split}
L_G^{\rm cl.}=L_{EH}+L_{HD}=&-\rho_\Lambda +\frac{1}{2}\MPl^2R+\alpha_Q \frac{{Q^\lambda}_\lambda}{3}+\alpha_2 R^2\\
=&-\rho_\Lambda+\frac{1}{2}\MPl^2  R+\alpha_1 C^2+\alpha_2 R^2+\alpha_3 E+\alpha_4\Box R\,,
\end{split}
\end{equation}
where
\begin{equation}\label{eq:traceQ1}
\frac{1}{3}{Q^\lambda}_\lambda \equiv-\frac{1}{120}C^2+\frac{1}{360}E+\frac{1}{6}\left(\xi-\frac{1}{5}\right)\Box R\,,
\end{equation}
in which $E$ is the Euler density (whose action defines the Gauss-Bonnet term) and $C^2$ is the square of the Weyl tensor.
Thus, the full effective Lagrangian reads
\begin{equation}\label{eq:Full-Leff1}
\begin{split}
L_{\rm eff}&=L_G^{\rm cl.}(M)+L_W^{\rm ren}(M)=-\rL(M)+\frac{1}{2}\MPl^2(M)  R+\alpha_1(M) C^2+\alpha_2(M) R^2+\alpha_3(M) E\\
&+\alpha_4(M)\Box R+\delta \rL(M)-\frac{1}{2}\delta\MPl^2(M) R-\delta \alpha_Q(M) \frac{{Q^\lambda}_\lambda}{3}-\delta \alpha_2(M) R^2+\cdots\\
%\begin{equation}\label{eq:Full-Leff}
&=\left[-\rL(M)+\delta\rL(M)\right]+\frac12\left[\MPl^2(M)-\delta\MPl^2(M)\right] R+\left[\alpha_1 (M) +\frac{1}{120} \delta\alpha_Q(M)\right] C^2\\
&+\left[\alpha_3 (M)-\frac{1}{360}\delta\alpha_Q(M)\right] E+\left[\alpha_4 (M)-\frac{1}{6}\left(\xi-\frac15\right) \delta\alpha_Q(M)\right]\Box R\\
&+\left[\alpha_2 (M)-\delta\alpha_2(M)\right] R^2+\cdots\\
\end{split}
%\end{equation}
\end{equation}
where the dots stand for higher order adiabatic contributions which need not be considered here.

The above Lagrangian is RG-invariant. The paired terms in square brackets are scale independent, and hence their derivatives with respect to the scale $M$ must be zero. Defining as usual the $\beta$-functions
\begin{equation}\label{eq:BetaFunction}
  \beta_i=M \frac{\partial \lambda_i(M)}{\partial M}
\end{equation}
for each of the parameters  $\lambda_i=\rL,\MPl^2,\alpha_1,...,\alpha_4$, and using the relations \eqref{eq:deltacouplings},  we find the relevant results for $\rL$ and $G^{-1}$ in the low energy action,
\begin{equation}\label{eq:BetaFunctionrL}
\beta_{\rL} (M)=\frac{1}{2(4\pi)^2}(M^2-m^2)^2
\end{equation}
\begin{equation}\label{eq:BetaFunctionMPl}
\beta_{\MPl^2} (M)=\frac{\left(\xi-\frac{1}{6}\right)}{8\pi^2} (M^2-m^2)\,,
\end{equation}
and the corresponding results for the HD terms:
\begin{equation}\label{eq:BetaFunctions12}
\begin{split}
\beta_{\alpha_1}=-\frac{1}{120(4\pi)^2}\ \ \ \ \ \
 \beta_{\alpha_2}=-\frac{\left(\xi-\frac{1}{6}\right)^2}{2(4\pi)^2}
 \end{split}
\end{equation}
 \begin{equation}\label{eq:BetaFunctions34}
\begin{split}
  \beta_{\alpha_3}=\frac{1}{360 (4\pi)^2}\ \ \ \ \ \ \
   \beta_{\alpha_4}=\frac{\xi-\frac15}{6(4\pi)^2}\,.
\end{split}
\end{equation}
Despite that we shall no longer consider the $\beta$-function coefficients \eqref{eq:BetaFunctions12}-\eqref{eq:BetaFunctions34} for the HD terms, we can check that they are correctly identified and coincide with standard results in the old literature\,\cite{Nelson:1984sy}.

Of particular interest for us here are the $\beta$-functions in the first two equations \eqref{eq:BetaFunctionrL}-\eqref{eq:BetaFunctionMPl}, since they involve the parameter $\rL$ and $G^{-1}$ of the low-energy EH action.  The first of them leads to a RGE that coincides with Eq.\,\eqref{eq:RGErL}, as it should, thus confirming that the effective action approach is consistent with the vacuum EMT computation.  This result reduces to the well-known expression\cite{Brown:1992db}
\begin{equation}\label{eq:brLMSS}
\beta_{\rL}^{\rm MSS}=\frac{m^4}{32\pi^2}
\end{equation}
in the minimal subtraction scheme (MSS) in DR and for $M^2\ll m^2$, as it should for a scale choice such as $M=H$.
Integrating the RGE  associated with \eqref{eq:BetaFunctionrL} between the scales $M_0$ and $M$ we retrieve  Eq.\eqref{SubtractionrL}. We must emphasize that Eq.\,\eqref{eq:brLMSS} does not characterize the running of the VED, but just that of a coupling in the EH action -- see \cite{JSPRev2022} for additional discussion. The running of the vacuum energy density $\rv(M)$ is, instead, governed by a completely different equation, which we discussed in Sec \, \ref{sec:betaVED} and carries a much softer behavior of the type $\beta_{\rm vac}\propto m^2 H^2$, see Eq.\,\eqref{eq:DerivTotal3}.

On the other hand, Eq.\,\eqref{eq:BetaFunctionMPl} leads to the following RGE for the inverse Newton's coupling:
\begin{equation}\label{eq:RGinvG}
M\frac{d}{d M}\left(\frac{1}{16\pi G}\right)=\frac{(\xi-\frac16)}{(4\pi)^2}\,(M^2-m^2)\,.
\end{equation}
It can also  be integrated  to compute the scaling evolution of $G$ between  $M_0$ and $M$ and one retrieves Eq.\eqref{SubtractionG}.  The result can be conveniently expressed as follows:
\begin{equation}\label{eq:RGENewton}
G(M)=\frac{G(M_0)}{1+\frac{\left(\xi-\frac{1}{6}\right)}{2\pi}G(M_0)\left(M^2-M_0^2-m^2\ln \frac{M^2}{M_0^2}\right)}\,.
\end{equation}

Within the off-shell renormalization procedure, the running of $G$ with the expansion can  be obtained from \eqref{eq:RGENewton} by evaluating the change of $G$ between the two cosmic epochs $M_0=H_0$ and $M=H$. If we confine to the late universe, where the higher powers of $H$ can be neglected, this yields the result
 \begin{equation}\label{eq:runGH}
G(H)=\frac{G_N}{1-\frac{\left(\xi-\frac{1}{6}\right)}{2\pi}\frac{m^2}{m^2_{\rm Pl}}\ln \frac{H^2}{H_0^2}}=\frac{G_N}{1-\epsilon\ln \frac{H^2}{H_0^2}}\,,
\end{equation}
with
\begin{equation}\label{eq:epsilonparameter0}
\epsilon\equiv\frac{1}{2\pi}\,\left(\xi-\frac{1}{6}\right)\,\frac{m^2}{\mpl^2}\,.
\end{equation}
In the above formula, $G_N$ defines the local gravity value usually associated with the inverse Planck mass squared:  $ G(H_0)=G_N=1/m^2_{\rm Pl}$ (in natural units). Furthermore,  we have neglected $H^2-H_0^2$ versus the logarithmic term $m^2\ln \frac{H^2}{H_0^2}$, as the ratio between the former and the latter is of order $H_0^2/m^2\ll 1$, for $H$ close to $H_0$.
It follows that the running of $G$ with $H$ is very mild, since it is logarithmic and depends on the very tiny coefficient $\epsilon$ in front of it. Recall that $m^2\ll \mpl^2$ even for GUT particles at a characteristic unification scale of $M_X\sim 10^{16}$ GeV.  In the de Sitter scenario there is a further suppression factor owing to the fact that the range of $\xi$ values is close to $1/6$ (cf. Sec.\,\ref{sec:PhysicalRegion_xi}).

From the foregoing, it is clear that the quantitative impact of the running of $G(H)$ on that of $\rv(H)$ is small. However, it can be of interest in general scenarios concerned with the variation of the fundamental constants of Nature, see e.g. \cite{Fritzsch:2012qc,Sola:2016our,Fritzsch:2016ewd}. Irrespective of this fact, here we can use the two running laws  for $G(H)$ and $\rv(H)$ to  check the consistency of the scale setting $M=H$, which is common in them.
The consistency check consists in re-deriving the result \eqref {eq:runGH}-\eqref{eq:epsilonparameter0} using explicitly the running law for the VED. We use the RVM form of this law at low energies; recall from Sec.\ref{sec:VEDindS} that both scenarios, the RVM and de Sitter share essentially the same formula for the VED evolution in the late universe.

To proceed, we consider the generalized Einstein's equations \eqref{eq:MEEs}.  On the \textit{r.h.s.} we may include also the contribution from the  background matter field $\phi_b$ defined in Sec.\,\ref{sec:Quantization}. Thus, we have
\begin{equation}\label{eq:MEEsTphi}
\frac{1}{8\pi G(M)} G_{\mu \nu}+\rho_\Lambda(M) g_{\mu \nu}+\alpha(M)\ \leftidx{^{(1)}}{\!H}_{\mu \nu}= \langle T_{\mu \nu}^{\delta \phi}\rangle_{\rm ren}(M)+T_{\mu\nu}^{\phi_b}\,.
\end{equation}
In the absence of interaction between matter and vacuum, $T_{\mu\nu}^{\phi_b}$ is covariantly conserved: $\nabla^\mu T_{\mu\nu}^{\phi_b}=0 $, as can be checked using the explicit form of the EMT in Eq.\,\eqref{EMTScalarField} and the equation of motion for $\phi_b$ (viz. the free KG equation).

Next, we multiply both sides of Eq.\,\eqref{eq:MEEsTphi} by $8\pi G(M)$ and subsequently compute the covariant derivative of the resulting expression.
In doing this, we use the fact that the involved HD tensor is covariantly conserved ($(\nabla^\mu\, \leftidx{^{(1)}}{\!H}_{\mu\nu}=0$); and, of course, also  the Bianchi identity of the Einstein tensor ($\nabla^\mu G_{\mu\nu}=0$). In this way, we find
\begin{equation}\label{eq:nablaEinsteineqs}
\begin{split}
\nabla^\mu\left(G(M) \rho_\Lambda (M) \right) g_{\mu\nu}+&\nabla^\mu\left(G(M) \alpha(M) \right) H_{\mu\nu}^{(1)}\\
&=\nabla^\mu\left( G(M) \braket{T_{\mu\nu}^{\delta\phi}}_{\rm ren}(M) \right)+\nabla^\mu\left(G(M)\right) T_{\mu\nu}^{\phi}\,.
\end{split}
\end{equation}
Finally, by  taking  the $\nu=0$ component of the above equation, and after some calculations, it can be worked out to produce the following expression:
\begin{equation}
\begin{split}\label{MixedConservation}
\dot{G}(M)\left(\rho_{\phi_b}+\rv\right)&+G(M)\dot{\rho}_{\rm vac}+3HG(M)\left(\rv(M)+P_{\rm vac}(M)\right)\\
&=\left( \alpha(M)\dot{G}(M)+G(M)\dot{\alpha}(M)\right) \frac{\leftidx{^{(1)}}{\!H}_{00}}{a^2}\,,
\end{split}
\end{equation}
where dots stand, as usual, for cosmic time differentiation.

Since we perform our analysis in the post-inflationary epoch and, in particular, in the current universe, we may disregard the HD contribution  on the \textit{r.h.s.}  of Eq.\,\eqref{MixedConservation}. Therefore, we are left with the simpler form
\begin{equation}\label{ModifiedConsLaw}
\dot{G}(H)\left(\rho_{\phi_b}+\rv(H)\right)+G(H)\dot{\rho}_{\rm vac}(H)+3HG(H)\left(\rv(H)+P_{\rm vac}(H)\right)=0\,,
\end{equation}
where we have implemented at this point the scale setting $M=H$ This equation can be conveniently cast as follows,
\begin{equation}\label{eq:NonConserVEDandG}
\dot{\rho}_{\rm vac}+3H\left(\rv+P_{\rm vac}\right)=-\frac{\dot{G}}{G}\,(\rho_{\phi_b}+\rv)=- \frac{\dot{G}}{G}\,\frac{3H^2}{8\pi G}\,,
\end{equation}
where in the last step, we used Friedmann's equation. The expression obtained shows that the vacuum is not conserved, since the Bianchi identity requires a time evolution of $G$ as compensation, i.e. $\dot{G}\neq 0$. In fact, we are after the precise form of $G(t)=G(H(t))$.  In general, there may also be exchange of vacuum with matter, but as indicated, we are considering here the simplest situation where matter is self-conserved ($\nabla^\mu T_{\mu\nu}^{\phi_b}=0 $).

Now the \textit{l.h.s.} of Eq.\,\eqref{eq:NonConserVEDandG} can be explicitly computed using  the expressions for the VED and the vacuum pressure discussed in Sections \ref{sec:VEDinRVM} and \ref{sec:EosVacuum}. Since we are considering the late universe we can neglect the higher order terms ${\cal O}\left(H^4\right)$ and hence Eq.\,\eqref{eq:NonConserVEDandG} can be transformed into the following differential equation for $G$:
\begin{equation}\label{ModifiedConsLaw2}
\begin{split}
& \frac{d}{dt}\left[\rv^0+\frac{3\nueff (H)}{8\pi}(H^2-H_0^2)\, \mpl^2\right]+3H\frac{\left(\xi-\frac{1}{6}\right)}{8\pi^2}\dot{H}m^2\left(1-\ln \frac{m^2}{H^2}\right)=-\frac{3H^2 \dot{G}(t)}{8\pi G^2(t)}\,,
\end{split}
\end{equation}
or, equivalently,
\begin{equation}\label{ModifiedConsLaw2b}
\begin{split}
&\frac{H^2 \dot{G}(t)}{G^2(t)}+\left[\dot{\nu}_{\rm eff} (H)(H^2-H_0^2)+ 2H\dot{H}\nueff(H)\right]\mpl^2 +H\frac{\left(\xi-\frac{1}{6}\right)}{\pi}\dot{H}\,m^2\left(1-\ln \frac{m^2}{H^2}\right)=0\,.
\end{split}
\end{equation}
Here, $G(t)=G(H(t))$  is the function that we wish to determine by solving the above equation.  The expression for $\nueff(H)$ and its time derivative  $\dot{\nu}_{\rm eff}(H)=\frac{d}{dt}{\nu}_{\rm eff}(H)$  can be obtained from Eq.\eqref{eq:nueff2}.  After some straightforward calculations, the above expression becomes greatly simplified, and we reach the following compact result:
\begin{equation}\label{MixedConservationApprox3}
\frac{d{G}}{G^2}=\frac{\left(\xi-\frac{1}{6}\right)}{\pi}m^2\frac{dH}{H}\,,
\end{equation}
in which we have replaced the time derivatives $\dot{G}=dG/dt$ and $\dot{H}=dH/dt$  by just the differentials $dG$ and $dH$  since $dt$ cancels on both sides. The reader can easily recognize that Eq.\eqref{MixedConservationApprox3} matches Eq.\,\eqref{eq:RGinvG} for $M=H$ in the limit $H^2\ll m^2$.
Finally, integrating  Eq.\,\eqref{MixedConservationApprox3} by simple quadrature from the present time $(H_0,G(H_0)$, where $G(H_0)=G_N\equiv 1/\mpl^2$, up to an arbitrary time around our past $(H,G(H))$,  we reach the final result
\begin{equation}\label{eq:runGH2}
G(H)=\frac{G_N}{1-\frac{\left(\xi-\frac{1}{6}\right)}{2\pi}\frac{m^2}{m^2_{\rm Pl}}\ln \frac{H^2}{H_0^2}}\,,
\end{equation}
which is just Eq.\,\eqref{eq:runGH} within the same kind of approximations.
In this way, we have verified the consistency of the scale setting $M=H$ following a completely different logical path, now based directly on local conservation laws.

To further complete our discussion of the scale setting procedure, the following complementary observations may be in order:

\begin{itemize}

\item In Ref.\cite{JSPRev2015} and later on in \cite{CristianJoan2020,CristianJoan2022a} various arguments for a possible generalization of the canonical scale setting $M=H$ were proposed in the sense that $M^2$ could be a linear combination of $H^2$ and $\dot{H}$, since they are dimensionally consistent: $M^2=\alpha H^2+\beta \dot{H}$, for dimensionless coefficients $\alpha,\beta$. In this case, the setting $H=M$ would correspond to the particular case $\alpha=1$, $\beta=0$. This scenario has been tested phenomenologically in, e.g. \cite{Sola:2015wwa,Sola:2016jky,Sola:2016zeg} and \cite{SolaPeracaula:2021gxi,SolaPeracaula:2023swx}.  The presence of $\dot{H}$, however, does not affect at all our mechanism of inflation (for which $H=$const.) and it only generalizes the form of the VED at low energies.
Notice that the possibility of such a generalization can be attributed e.g. to the fact that in general the running of the VED may be linked not only to the running of $G$ through the Bianchi identity, as argued above,  but also to the possible exchange between vacuum energy and matter in the present universe (see the mentioned references), so one should be open to more general scenarios.  Now since we cannot know
a priori which are the precise mechanisms of energy exchange between the vacuum, matter and a possible (simultaneous) variation of ``fundamental constants'' such as $G, \alpha$ etc.\cite{Fritzsch:2012qc}, it is not possible to know which is the ``exact'' scale setting procedure. In this sense, it is ultimately a matter of phenomenological test. However, given a minimal scenario such as the one depicted above, in which matter is not interacting with vacuum and only $\rv$ and $G$ can evolve with the expansion, the setting $M=H$ appears to be fully consistent (see also the last point in these considerations).

\item In older versions of the RVM \cite{ShapSol,Guberina:2002wt,Shapiro:2004ch,Babic:2004ev,Grande:2010vg,Grande:2011xf,JSPRev2013}, prior to the formal developments presented in more recent years\cite{CristianJoan2020,CristianJoan2022a,CristianJoan2022b,CristianJoanSamira2023}, different reasons were also given to favor the scale setting $M=H$ over others more phenomenological (such as $M=\rho_c^{1/4}$). We refer the reader to these older references for details.

\item The presence of a cutoff scale is also used in the context of the RG formulation within the QG framework\cite{Reuter:1996cp,Reuter:1993kw}.  In both type of approaches, semiclassical and QG, the kind of reasoning is based on the same RG principle: the breaking of conformal invariance by the quantum theory requires the presence of a floating scale to perform renormalization. Thus, at the end of the calculation, a scale setting is needed to contact physics. Typically, the cutoff scale in the QG context has been called $k$ and different proposals were suggested to fix it, e.g. $k=1/t$ (where $t$ is the cosmic time) and $k=H$ (as in our case).

\item A recent and intriguing example is found in the lattice QG approach of \cite{Dai:2024vjc} . In it, the authors test the canonical version of the RVM formula \eqref{eq:RVMform} as well as the possible generalization mentioned in the first point above which involves both terms $H^2$ and $\dot{H}$. Using the method of Euclidean dynamical triangulations, these authors study the quantum vacuum in the lattice and find, on the one hand,  that the lattice simulations favor a quadratic running with the scale, and on the other  that the scale can be identified
with the Hubble rate ($M=H$). In other words, they favor strongly the canonical realization of the RVM, Eq.\eqref{eq:RVMform}. It is also remarkable that the preferred range of values that they find in their lattice simulations for the dimensionless parameter $\nu$ lies in the ballpark of $10^{-4}-10^{-3}$, see Eq. (83) in that paper.  This  is well in the typical range of values of our $\nueff$ found from the direct fits to the cosmological data performed in the aforementioned references quoted in the first point above, and also from the old theoretical estimate using GUT's\cite{Fossil2008}. At the end of the day it turns out that very different approaches (based on the one hand on QFT models of the vacuum energy and their phenomenological implications, and on the other on numerical simulations in lattice QG) seem to resonate in three important conclusions: i) the behavior of the running vacuum is quadratic in the scale $M$; ii) the scale of the running is identified to be $M=H$, and  iii) The quantitative determination of the parameter $\nueff$ responsible for the running of the VED lies most likely in the range $10^{-4}$ to $10^{-3}$.  We find the consistency of the results obtained from such disparate kind of analyzes quite intriguing and certainly encouraging, since they seem to support the theoretical framework presented here.

\end{itemize}

\end{document}